# THE DIPOLE–QUADRUPOLE THEORY OF SURFACE ENHANCED RAMAN SCATTERING

**A.M. Polubotko**

A.F. Ioffe Physico-Technical Institute, Politechnicheskaya 26 194021

Saint Petersburg RUSSIA

E-mail: alex.marina@mail.ioffe.ru Tel: (812) 274-77-29,

Fax: (812) 297-10-17

## ABSTRACT

The review "The Dipole-Quadrupole Theory of Surface Enhanced Raman Scattering" is devoted to explanation of SERS in terms of the strong dipole and especially quadrupole light-molecule interactions arising in surface fields strongly varying in space in the region of strongly irregular surface roughness. The influence of the quadrupole interaction is a matter of principle, since it permits to explain appearance of forbidden bands arising in a great number of experiments on SERS on symmetrical molecules. This essential detail in fact is absent in all other SERS theories, that does not permit to create a closed theory, which explain the majority of experimental facts accompanying SERS by a unitary approach. Moreover as it is demonstrated in the review the huge enhancement in the phenomenon of Single Molecule Detection by the SERS method can be explained exclusively by the strong quadrupole light-molecule interaction.

The first part of the review is devoted to a brief description of main SERS characteristics and critical analysis of theoretical approaches. Further the theory of electromagnetic field near some model kinds of rough surfaces and some other systems and the theory of the SER cross-section for arbitrary and symmetrical molecules are presented in detail. The obtained expressions permit to establish selection rules for contributions in the SER cross-section and analyze the SER spectra of symmetrical molecules. This analysis permits to corroborate main specific features of the SER spectra of symmetrical molecules and some anomalies, which exist in these spectra for some specific conditions. The existence of electrodynamic forbiddance of the quadrupole scattering mechanism for the methane molecule and molecules with cubic symmetry groups is established. The above theory permits to explain not only the huge enhancement in the phenomenon of Single –Molecule Detection but the blinking of the SERS signal too. The first layer effect, arising in the first layer of absorbed molecules also is considered in detail. It appears that the nature of this phenomenon is the electrodynamic enhancement, which is very large just in the first layer. The other phenomena, accompanying SERS are accounted for. It is demonstrated that the theory is in a good agreement with the SERS experiments and explains quite a number of characteristics related to the SERS phenomenon.

# TABLE OF CONTENTS

# 1. INTRODUCTION

Surface Enhanced Raman scattering (SERS) is a well known effect, discovered in 1974 by M. Fleishmann, P.J. Hendra and A.J. McMillan [1]. Later D.L. Jeanmaire and R.P. Van Duyne [2] established that the reason of SERS is the strong surface roughness arising after an oxygen-reduction cycle. Further investigations revealed that a large number of Surface Enhanced processes arise on rough surfaces of silver and other metals. They are: Surface Enhanced Infrared Absorption (SEIRA), Surface Enhanced Resonance Raman scattering (SERRS), Surface Enhanced Hyper Raman scattering (SEHRS), Surface Enhanced Hyper Resonance Raman scattering (SEHRRS), Surface Enhanced Second Harmonic Generation and enhancement of luminescence, photochemical and some other processes. In 1997 the strong SERS was observed on specially prepared rough surfaces with very large roughness and also on large metal particles. [3,4]. These experiments gave rise to a novel type of Surface Enhanced spectroscopy. This is the so-called spectroscopy of Single Molecule Detection. At present there are a large number of publications on all these issues. However different authors consider the mechanism of SERS, its theory and the theories of other surface enhanced processes from their own points of view. Here we restrict our consideration to only the theory of the SERS phenomenon because the other surface enhanced processes have a large number of specific features and require a more detailed analysis. Therefore the goal of this review is to present the theory which explains mechanism of SERS by the increase of the dipole and especially quadrupole light-molecule interactions in a strongly irregular region of the surface roughness. This opinion and the theory is not widely accepted however the theory can account for quite a number of SERS characteristics. For a more detailed exposition of the theory and of the supplementary material let us begin with consideration of the most general characteristic features of SERS.

## 2. MAIN SPECIFIC FEATURES OF THE SERS PHENOMENON

The main experimental condition for observation of SERS is existence of surface roughness with a very widely varying characteristic size, approximately from ~ 5 to (30-150) nm. Originally SERS was observed on silver surfaces. However it was found later that $Au, Cu, Pt, Pd, Ti, Ni, Co, Cd, Fe$ and some other metals exhibit this phenomenon. SERS is observed on various types of rough systems [5]. They are:

1. coldly evaporated silver films,
2. silver island films,
3. colloidal particles,
4. polycrystalline silver foils cleaned and probably roughened by $Ar$-ion bombardment in ultra high vacuum,
5. photo chemically roughened silver surfaces, tunnel junctions and others.

The enhancement factor for these systems depends on a number of parameters and methods used for surface preparation.

At present it appears, that the strongest enhancement is caused by molecules adsorbed on a roughness with a very large characteristic size~150nm and on very large colloidal particles of special type with a very low concentration of adsorbed molecules. In ordinary SERS, with a mono or multilayer coverage of the substrate the enhancement factor varies approximately in the interval $10^3 - 10^6$ for various systems and depends on methods of surface preparation, as it has been mentioned above. In addition, it depends on the type of the metal substrate in the case of scattering on the same molecule. In special experiments on surfaces with very large roughness and on large colloidal particles the enhancement factor as high as $10^{14}$ was achieved, which gave a basis to the so-called Single Molecule Spectroscopy by the SERS method K. Kneipp et al. [3], S.R. Emory and S. Nie [4]. Numerous experiments, carried out by A. Campion et al. ([6, 7] for example) revealed that, in fact, molecules adsorbed on smooth single crystal surfaces do not enhance Raman scattering (Fig. 1).

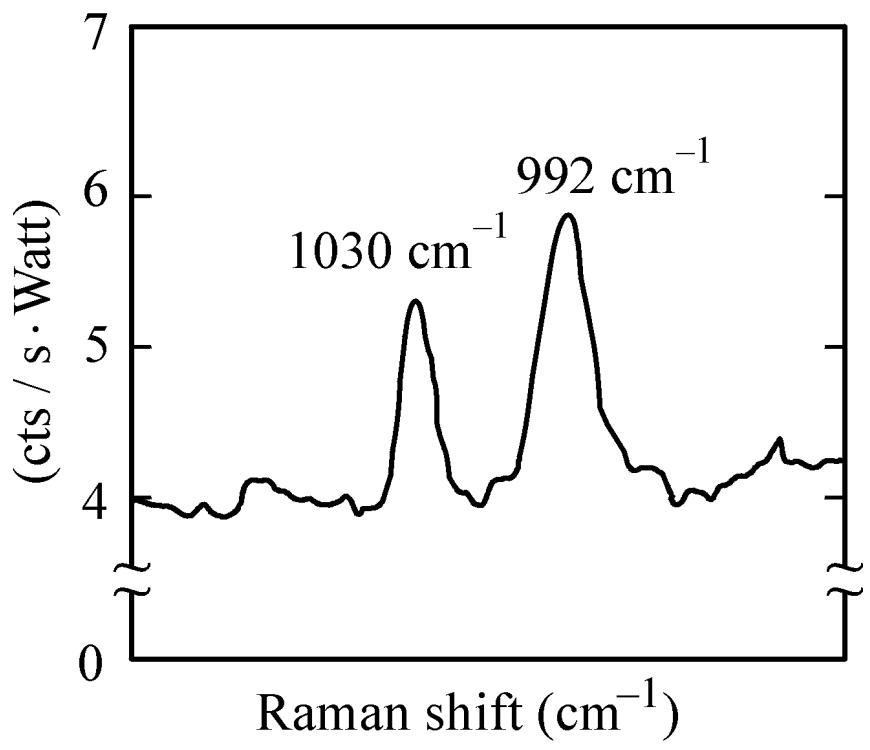

Fig. 1. Ordinary Raman spectra from pyridine on silver $\approx 1$ monolayer on $Ag(111)$, $T_S = 110K$; 1000mW of $514.5nm$ radiation, $10cm^{-1}$ bandpass. Symmetric $(992/996^{-1})$ and antisymmetric breathing mode $(1030/1037cm^{-1})$. The intensities of these bands demonstrate the absence of the enhancement on a single surface of silver.

The only study in which a certain enhancement was observed ($\sim 4\times10^2$) was performed by M. Udagawa et al. [8]. However there is an opinion that this enhancement may be due to a residual roughness as pointed out in [8].

In [9] A. Campion et al. found, that the Pyromellitic Dianhydride (PMDA) molecules adsorbed on $Cu\,(111)$ and $Cu\,(100)$ enhance the Raman scattering by approximately a factor of $30$-100 depending on a crystal face, and the polarization and frequency of the incident light. At present it is the only published system, which enhances Raman scattering on smooth single surfaces.

SERS is observed on virtually all molecules. However the extent of the enhancement occasionally depends on the type of a molecule in the case of adsorption on the same surfaces. For example it is sufficiently small for CO, $N_2$, methane, ethane and water, I. Pockrand [5].

At present it is well established, [5,10], that the enhancement in SERS in the first layer of adsorbed molecules is about $10^2$ times stronger, than that in the second and upper layers. This phenomenon named the short-range effect in SERS is frequently associated with distortion of the electron shell as a result of adsorption. Another name of this effect is the chemical effect. At present its mechanism is frequently identified with so-called charge transfer enhancement mechanism. However, as it is shown below, this effect is associated with a very strong change of the electromagnetic field near some prominent areas of the surface with a large curvature [11, 12]. In principle due to discovery of the charge transfer enhancement mechanism, mentioned above [9] for PMDA molecule, the short range effect may be a combination of both the chemical enhancement mechanism and very large difference in the enhancement in the first and the second layers, or purely electrodynamic effect (see below).

As it was demonstrated by P. N. Sanda et al. [13], molecules adsorbed in the second and upper layers enhance the scattering too. This phenomenon was named as a long-range effect (Fig. 2).

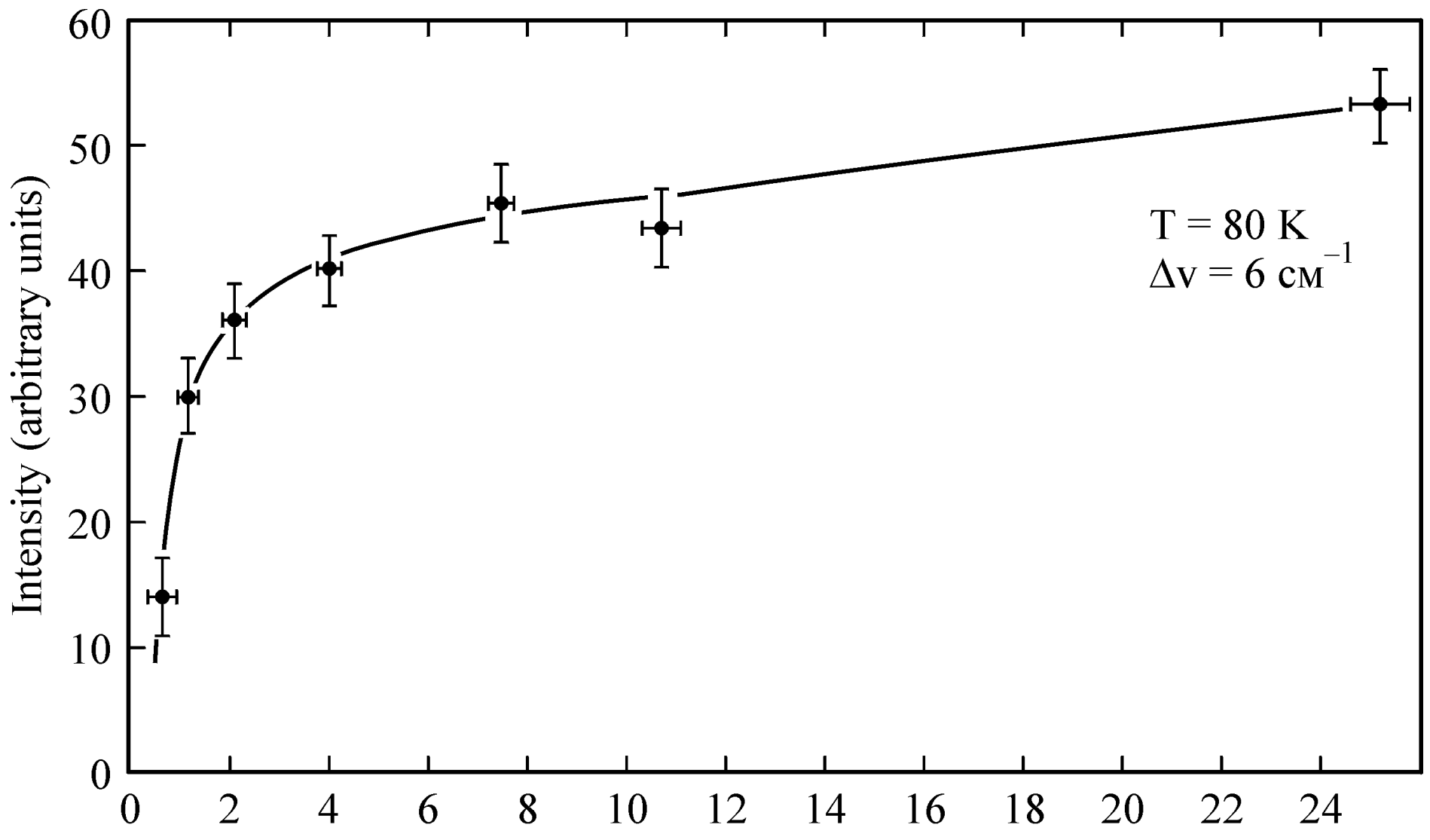

Fig. 2. The intensity of Raman scattering for the line $990cm^{-1}$ of pyridine molecule depending on coverage (in monolayer equivalents). The enhancement coefficient, which correspond to one monolayer is $10^4$. The increase of the intensity for the upper layers demonstrates the existence of the long-range effect in SERS

As it was pointed out in [14-16] the strongest enhancement occurs at some special places, named as active sites. These sites have various properties depending on the type of a roughness. In particular they disappear upon annealing for coldly evaporated thin films [5] (Fig. 3a, 3b).

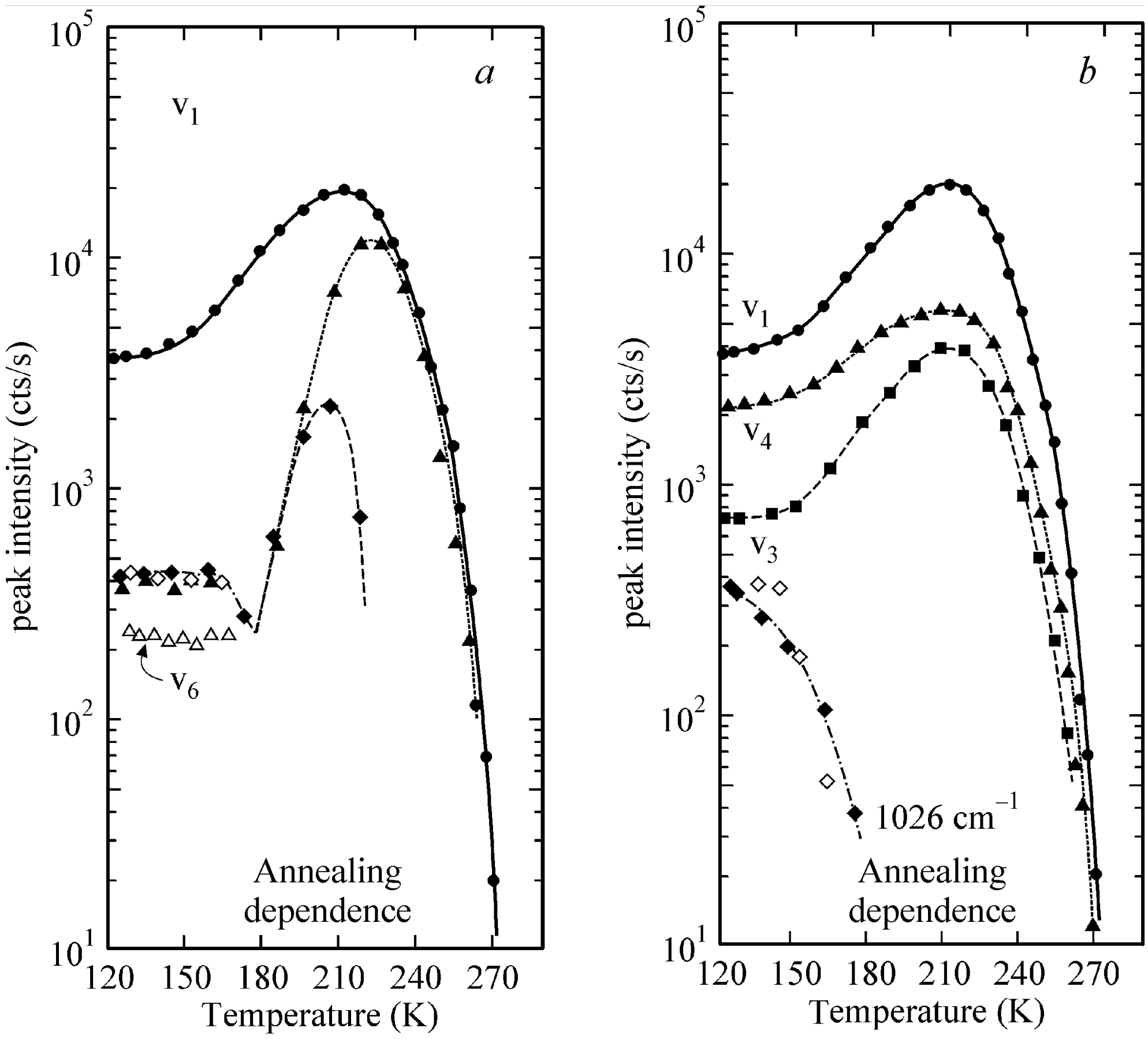

Fig. 3a Annealing of the surface and disappearance of the Raman peak intensities from the breathing mode $\nu_1$ of pyridine on $Ag$. Filled-in symbols: coldly evaporated film (dots 0.2 Langmuir, pyridine in the first layer; triangles: 200 Langmuir, pyridine in the first layer; rhombs: 200 Langmuir, pyridine in the volume). Open symbols: SERS inactive, room temperature deposited film exposed to 200 Langmuir (rhombs: $\nu_1$; triangles: $\nu_6$). Lines are guides to the eye. Annealing at room temperatures irreversibly destroys the enhancement properties of coldly evaporated silver films [5].

Fig. 3b Temperature variation of various SER peak intensities from coldly evaporated $Ag$ films exposed to 0.2 Langmuir of pyridine (filled-in symbols; lines are guides to the eye). Open rhombs: measured annealing of "impurity" line at $1050 cm^{-1}$. 200 mw of $514.5 nm$ radiation, $4 cm^{-1}$ bandpass, and $\approx 1K/\min$ temperature variation. Here annealing at room temperatures irreversibly destroys the enhancement properties of coldly evaporated silver films too [5].

As it was demonstrated in [17-20] the dependence of the SER signal on the wavelength of the incident light differs for some types of roughness, such as coldly evaporated silver films, from the usual frequency dependence $(\hbar\omega)^4$ characteristic of the common Raman scattering (Fig. 4).

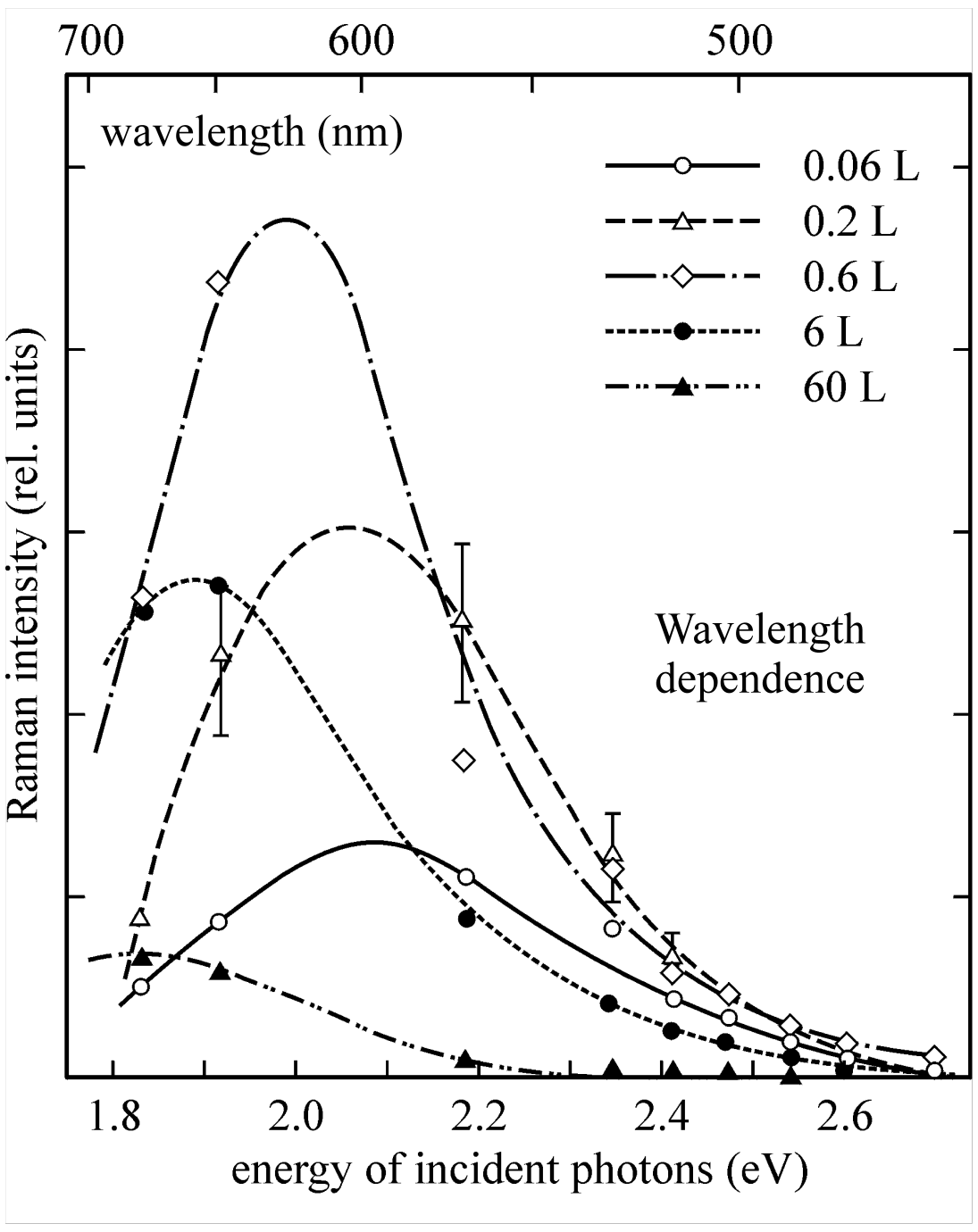

Fig. 4. SER excitation profiles from pyridine adsorbed on $Ag$ in the first layer for various exposures as indicated (symmetric breathing vibration). Lines are guides to the eye [5]. The Raman excitation profiles do not correspond to the $(\hbar\omega)^4$ law, characteristic for Raman scattering in a free space.

Analysis of polarization dependences of the SER signal demonstrates that they are depolarized both for $TE$ and $TH$ polarizations of the incident field.

The SERS signal is accompanied by a continuous background scattering [21-23] (Fig. 5). This background is observed without adsorbed molecules, and therefore may be attributed to an inherent property of a substrate. As it was pointed out by C. J. Chen et al. [23] the intensity of the background increases with the increase of the roughness degree.

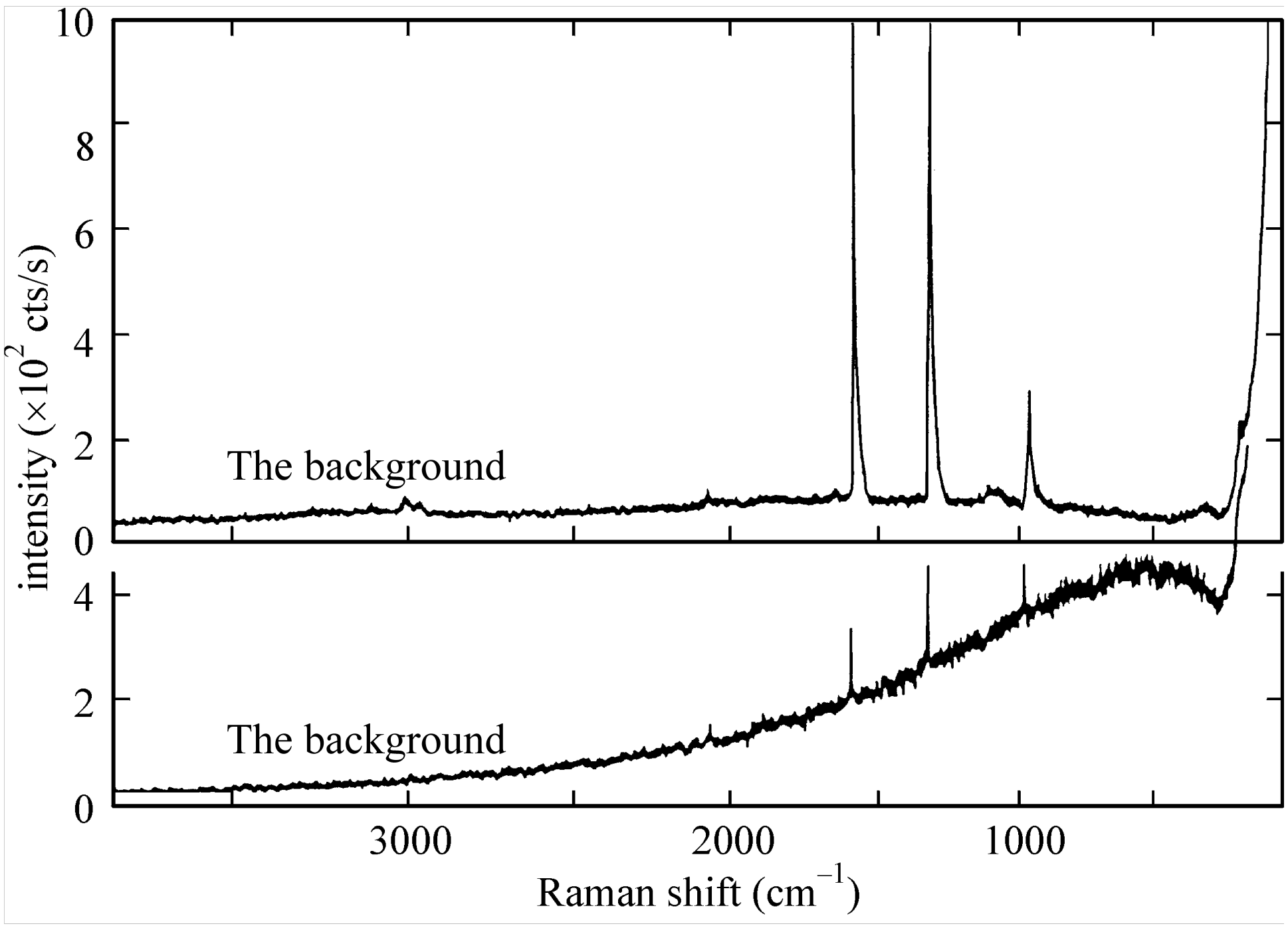

Fig. 5. SER spectra from coldly evaporated $Ag$ films exposed to 3 Langmuir of ethylene. Upper trace: sample exposed and measured at $120K$. Lower trace: sample annealed to $260K$ $(\approx 1K/\text{min})$, recooled to $120K$, exposed and measured. 200 mw of 514 nm radiation, $4.5cm^{-1}$ bandpass. The wide continuous background is observed for both spectra.

Analysis of the SER spectra is the most interesting issue. As it is pointed out in a large number of papers (see [5]) the Raman shifts for the adsorbed molecules are almost the same as those in a free space, especially in the case of $Ag$. In the SER spectra of symmetrical molecules, strong bands forbidden in the ordinary Raman scattering appear. Also there is a selective enhancement of separate Raman bands. As usual the band associated with the breathing mode, transforming under the unitary irreducible representation is most enhanced. The other bands, associated with the totally symmetric vibrations are enhanced too. However the enhancement factor changes in a wide range. Besides it depends on the type of the rough surface and on the kind of the metal substrate in the case when the same molecules are considered. In principle there are some exceptions from the above rules. Thus the situation is sufficiently complicated. Further we shall describe some reasons, when such exceptions occur.

## 3. SOME OF THE EXISTING THEORIES

At present there are numerous viewpoints concerning the SERS mechanism, reflected in a number of theories. A sufficiently comprehensive review of these theories can be found in [5]. However at present the most widely accepted theories are the plasmon-polariton theory, rod-effect and charge transfer or the chemical enhancement mechanism. In the author’s opinion, the most correct is the dipole-quadrupole theory which is considered here.

The plasmon-polarition theory implies that some surface plasmon modes can be excited in the substrate. These modes increase the electric field which affect the molecule and enhance the optical processes. Since plasmons can not be excited on flat surfaces at the frequencies used, their excitation is commonly attributed to the existence of the surface roughness. However the conception of surface plasmons as some special excitations is indefinite for arbitrary geometry. They must have different configuration, depending on the geometry of the surface and the dielectric constant of a metal. The field in the metal may consist of a number of different excitations and the total field is a sum of all these and other modes. Therefore the problem of reflectance and transmittance of the plane electromagnetic wave on the rough metal surface should be quantitatively solved in order to find the electric field and its characteristics in the metal and in the upper medium. In this situation the only methods to solve the problem are numerical ones. In particular one can point out the method based on the Kirchhoff formula [24]. Regretfully this method enables to calculate only the field in a simple two-dimensional configuration and was applied in [25] and some another papers. However further development of computers and their mathematical software will apparently permit to solve problems with a more complex configuration.

Another objection against plasmon theory used in its present form and its ability to account for the experimental results is that the appearance of forbidden lines can not be attributed only to the enhancement of the electric field as it is widely accepted in literature. In practice there are a large number of forbidden lines in the SER spectra of symmetrical molecules, which can not be explained only by the enhancement of the electric field and the dipole interaction ([5, 26, 27] for example). Therefore other ways to explain the enhancement should be found.

Another theory, which tries to explain SERS is the charge-transfer theory or the chemical effect suggested by A. Otto. A detailed exposition of this theory can be found in [28]. In accordance with this theory the enhancement must be observed on rough surfaces in the first layer of adsorbed molecules. Therefore this effect is frequently named as a first layer effect. The main objection against this effect is that the mechanism should be operative on single surfaces too. However no enhancement is observed on single surfaces of silver (see e.g. numerous experiments performed by A. Campion [6,7]). The only system in which the enhancement can be observed, known to the author is the PMDA molecule adsorbed on single copper surface [9]. However the molecule strongly changes its configuration as a result of adsorption and its Raman spectrum is markedly shifted with respect to that of the free molecule. In addition, the enhancement factor is very weak with a typical value~30-100 and does not correspond to the strong SERS. It has a resonance nature, whereas in ordinary SERS experiments on $Ag$ the frequency dependence of the enhancement is not resonant (Fig. 4). It was demonstrated in our investigations [11-12], that the first layer effect is associated with a large difference in the values of electric field strength and the field derivatives near the wedges or tips of the surface in the first and the second or other layers. Therefore it appears that the first layer effect and the whole SERS is of purely electrodynamic nature.

The dipole-quadrupole theory described here explains SERS by an increase of the electric field and its derivatives near wedges or tips of the surface, or near prominent areas with a large curvature. This mechanism includes both the rod effect and quantum-mechanical features of the quadrupole interaction of light with molecules. From our viewpoint this approach is the most fruitful for explanation of the SER spectra and other phenomena accompanying SERS. Further we shall consider this mechanism in detail and demonstrate that almost all main specific features of SERS characteristics can be explained in terms of this mechanism.

# 4. SOME PROPERTIES OF THE ELECTROMAGNETIC FIELD NEAR A ROUGH METAL SURFACE

Further we consider the model of a rough surface and justify our approach from the point of view of explanation of the SERS phenomena. As it was pointed out in [5] the enhancement coefficient essentially depends on the type of the rough surface. The strongest enhancement occurs on surfaces with a very sharp roughness. Because this review is devoted to explanation of the strong SERS we consider the model, which reflects the main features of the electromagnetic field near the surface and having the most enhancement of the fields and their derivatives. As it has been mentioned above the most fruitful way to calculate the electromagnetic field near a massive rough metal surface is to use numerical methods based on Kirchhoff's formulae [24, 25]. However it is a very difficult method, yielding random results for an arbitrary rough surface. Therefore we do not solve the diffraction problem, but try to obtain the main field characteristics, which can reflect the qualitative picture on the real rough surface. It is well known that monochromatic light in a free space is described by plane electromagnetic waves.

$$\begin{aligned} \overline{E} &= \overline{e} e^{i\overline{k}_0 \overline{r}} \\ \overline{H} &= \overline{h} e^{i\overline{k}_0 \overline{r}} \end{aligned} \tag{1}$$

where $\overline{e}$ and $\overline{h}$ are the polarization vectors of the electric and magnetic fields, $k_0 = 2\pi / \lambda$ is the wave vector and $\lambda$ is the wavelength of the incident radiation. The characteristic scale of the space variation of this field is the wavelength $\lambda$. In the presence of a flat surface the electromagnetic field can be represented as a superposition of the incident, reflected and transmitted waves. These solutions are necessary in order to satisfy the boundary conditions of the continuity of the tangential constituents of the field.

$$E_t^{\mathrm{I}} = E_t^{\mathrm{II}}$$
$$H_t^{\mathrm{I}} = H_t^{\mathrm{II}} \qquad (2)$$

Here I and II designate two media. However the wavelength $\lambda$ remains the characteristic scale of the spatial variation of this field. In the case of a rough surface the situation is absolutely different. The necessity for satisfying the boundary conditions (2) causes appearance of a strong localized surface field at the rough surface. It is obvious that, for a random profile, the mathematical description of this field is impossible because of its strong irregularity. However in order to receive a clearer qualitative notion about the behavior of the surface field, we can consider a strong jagged regular lattice of a triangular profile (Fig.6).

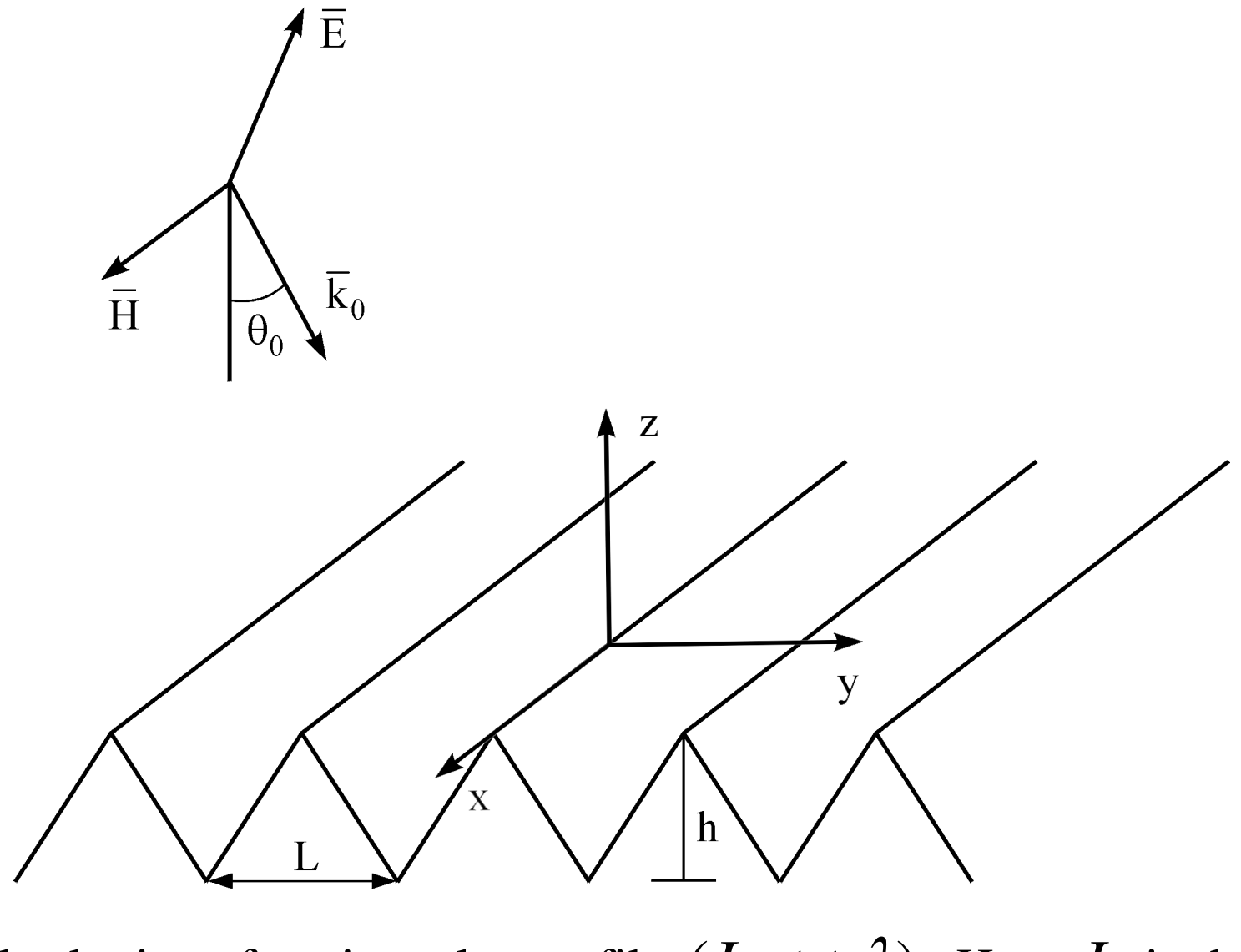

Fig. 6 Regular lattice of a triangular profile $(L << \lambda)$. Here $L$ is the period of the lattice and $\lambda$ is the wavelength of the incident light, $h$ is the height of the lattice.

The electromagnetic field above this lattice can be represented in the form

$$\overline{E} = \overline{E}_{inc} + \overline{E}_{surf.sc} \qquad (3)$$

Where,

$$\begin{aligned} \overline{E}_{inc} &= \overline{E}_{0,inc} e^{-ik_0 \cos\theta_0 z + ik_0 \sin\theta_0 y} \\ \left|\overline{E}_{0,inc}\right| &= 1 \end{aligned} \qquad (4)$$

$\theta_0$ - is the angle of incidence,

$$\overline{E}_{surf.sc.} = \sum_{n=-\infty}^{+\infty} \overline{g}_n e^{i\alpha_n y + i\gamma_n z} \qquad (5)$$

$$\alpha_n = \frac{2\pi n + k_0 \sin\theta_0 L}{L} \qquad (6)$$

$$\gamma_n = \sqrt{k_0^2 - \alpha_n^2} \qquad (7)$$

Here $\overline{g}_n$ are the amplitudes, $n$ is the number of a spatial harmonic. For the period of the lattice $L << \lambda$, the spatial harmonic with $n = 0$ is a direct reflected wave, while all the others are heterogeneous plane waves strongly localized near the surface. The minimum localization size has the harmonic with $n = 1$. All the others are localized considerably strongly. The exact solution of the diffraction problem on the lattice reduces to determination of the coefficients $\overline{g}_n$. The electric field can be obtained at any point by summation of series (5). However solution of such problems is very difficult in practice. Therefore it is necessary to develop ideas that would be sufficient to solve our tasks.

The main specific feature of the surface field is a steep or singular increase of the value of the electric field near wedges or the so-called rod effect. This type of behavior is independent of a particular surface profile. It is determined only by the existence of sharp wedges. Besides it is independent on the dielectric properties of the lattice and exists in lattices with any dielectric constants that differ from the dielectric constant of vacuum. Let us consider the solution of Maxwell equations in the region of the infinite wedge with the angle $\alpha$ near the top (Fig. 7).

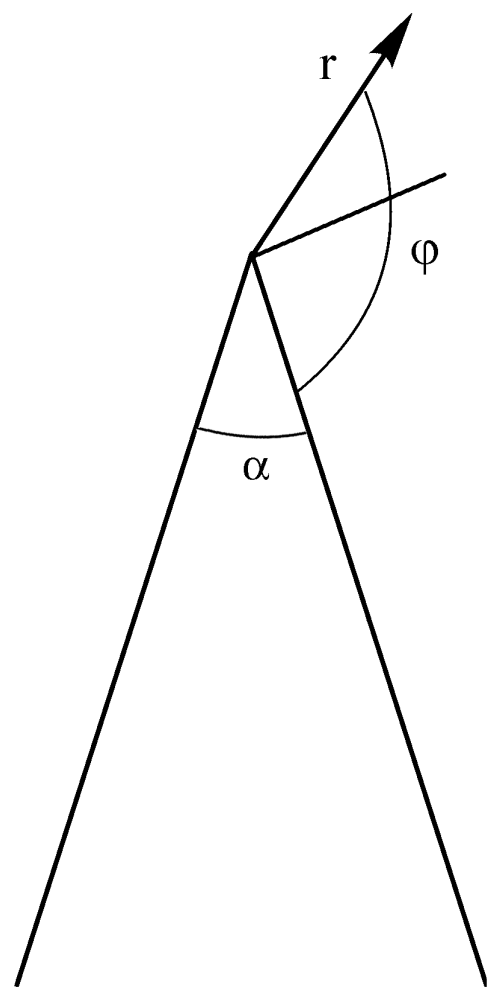

Fig.7 Infinite wedge

To solve the problem for the $TE$ and $TH$ polarizations in the general case we have to find the solutions of the wave equations

$$\begin{aligned} &\Delta U + k_0^2 U = 0 \qquad \textit{the first medium (vacuum)} \\ &\Delta U + k_0^2 \varepsilon U = 0 \qquad \textit{the second medium} \end{aligned} \tag{8}$$

with the boundary conditions (2). Here $U = E_x$ for the $TE$ polarization of the incident field and $U = H_x$ for the $TH$ polarization. For the $TE$ polarization the components $H_r$ and $H_\varphi$ can be determined as

$$H_r = \frac{1}{i\omega\mu}\frac{1}{r}\frac{\partial E_x}{\partial \varphi}$$
$$H_\varphi = -\frac{1}{i\omega\mu}\frac{\partial E_x}{\partial r} \tag{9}$$

In the case of the $TH$ polarization the components $E_r$ and $E_\varphi$ have the form

$$E_r = -\frac{1}{i\omega\varepsilon_0}\frac{1}{r}\frac{\partial H_x}{\partial \varphi}$$
$$E_\varphi = \frac{1}{i\omega\varepsilon_0}\frac{\partial H_x}{\partial r} \tag{10}$$

The exact solution of (8) can be found for a simpler case of an ideally conductive wedge with the boundary condition for the tangential component of the electric field

$$E_t = 0 \tag{11}$$

Then

$$U = \sum_{\lambda_n} \left[ b_{\lambda_n} J_{\lambda_n}(k_0 r) + d_{\lambda_n} J_{-\lambda_n}(k_0 r) \right] \ \Phi_{\lambda_n}(\varphi) \tag{12}$$

where

$$\Phi_{\lambda_n} = \sin \lambda_n \varphi \qquad \textit{for the} \quad TE \ \textit{polarization}$$
$$\Phi_{\lambda_n} = \cos \lambda_n \varphi \qquad \textit{for the} \quad TH \ \textit{polarization} \tag{13}$$

$$\lambda_n = \frac{\pi n}{2\pi - \alpha}, \quad n = 1,2,3\ldots \tag{14}$$

$J_{\lambda_n}$ and $J_{-\lambda_n}$ are Bessel functions, which are the solutions of Bessel equations

$$\frac{d^2 J}{dr^2} + \frac{1}{r}\frac{dJ}{dr} + \left( k_0^2 - \frac{\lambda_n^2}{r^2} \right) J = 0 \tag{15}$$

The value of $U$ must satisfy the condition of a finite quantity of energy in an arbitrary small volume $V$ around the wedge, that is equivalent to the condition

$$\int_V \left( \varepsilon \left| \overline{E} \right|^2 + \mu \left| \overline{H} \right|^2 \right) dV \underset{V \to 0}{\to} 0 \tag{16}$$

Taking into account that $J_{-\lambda_n}$ are singular functions it is necessary to put

$$d_{\lambda_n} = 0 \tag{17}$$

Then we have for the $TH$ polarization in a small region near the top of the wedge

$$\begin{aligned} E_r &= -i \sqrt{\frac{\mu}{\varepsilon_0}} b_0 \left( \frac{\lambda}{2\pi r} \right)^{\beta} \lambda_1 \sin \lambda_1 \varphi \\ E_\varphi &= -i \sqrt{\frac{\mu}{\varepsilon_0}} b_0 \left( \frac{\lambda}{2\pi r} \right)^{\beta} \lambda_1 \cos \lambda_1 \varphi \end{aligned} \tag{18}$$

Here $\sqrt{\frac{\mu}{\varepsilon_0}}$ is the impedance of a free space, $b_0$ is some value, which depends on configuration of the incident field,

$$\beta = 1 - \lambda_1 = \frac{\pi - \alpha}{2\pi - \alpha} \tag{19}$$

Here, it is meant that the size of the wedge is larger than $\lambda$ and we must solve the wave equations. For a finite wedge in the lattice with $L << \lambda$ we can write, in accordance with electrostatic approximation the following approximate formulas

$$E_r = -\left|\overline{E}_{inc}\right| C_0 \left(\frac{l_1}{r}\right)^{\beta} \sin(\lambda_1 \varphi)$$
$$E_{\varphi} = -\left|\overline{E}_{inc}\right| C_0 \left(\frac{l_1}{r}\right)^{\beta} \cos(\lambda_1 \varphi) \qquad (20)$$

Here $C_0$ is some numerical coefficient. Its value depends on the geometry of the lattice, the angle of incidence and some another factors, $l_1 \sim (L, h)$ is the characteristic size of the lattice. The specific feature of the field behavior (20) is appearance of the singularity $(l_1 / r)^{\beta}$, which describes geometrical nature of the field enhancement. For example it determines the following behavior of the projections $g_{n,z}$ in expression (5) [29].

$$g_{n,z} \sim |n|^{\beta - 1} \qquad (21)$$

Indeed, substitution of (21) into (5) for $E_{z,surf.sc.}$ gives the same singularity on the top of the wedge with coordinates $(x, y = 0, z = 0)$

$$E_{z,surf.sc.} \sim \sum_{\substack{n=-\infty \\ n \neq 0}}^{+\infty} |n|^{\beta-1} e^{2\pi |n| z / L} \sim 2\int_0^{\infty} t^{\beta-1} e^{-2\pi z t / L} dt \sim 2\left(\frac{L}{2\pi z}\right)^{\beta} \qquad (22)$$

For the wedge angles changing in the interval $0 < \alpha < \pi$ the $\beta$ value varies within the range $0 < \beta < 1/2$ and the coefficients $g_{n,z}$ slowly decrease as $n$ increases. Thus the singular behavior of the field arises because of specific summation of the surface waves at the top of the wedge. For an arbitrary rough surface the field does not have a mode character, however its singular behavior near the wedges preserves. We have considered the behavior of the electric field in the region of a two-dimensional roughness such as the wedge. In the region of a three-dimensional roughness of the cone type (Fig. 8)

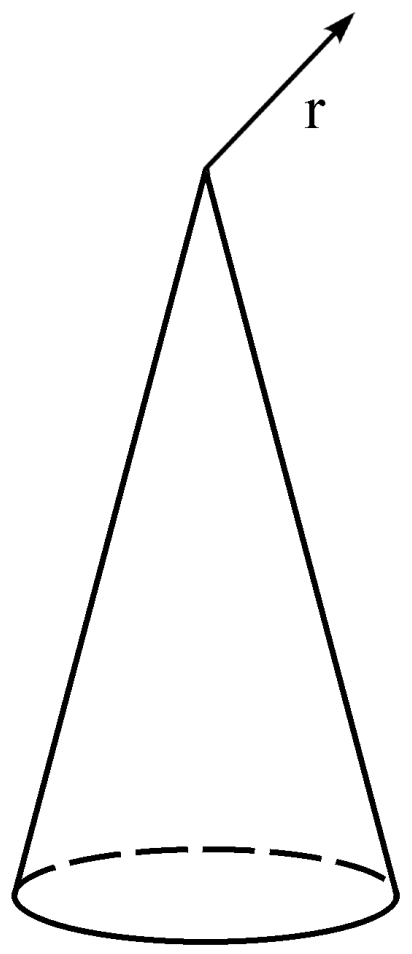

Fig. 8 The roughness of a cone type

the formula for estimation of the field has an approximate form, similar to (20)

$$E_r \sim \left|\overline{E}_{inc}\right| C_0 \left(\frac{l_1}{r}\right)^{\beta} \tag{23}$$

where $\beta$ depends on the cone angle and varies within the interval $0 < \beta < 1$. Here we designate the numerical coefficient and the characteristic size of the cone by the same letters $C_0$ and $l_1$. Using formulas (20) and (23) one can note a very important property: a strong spatial variation of the field. For example

$$\frac{1}{E_r}\frac{\partial E_r}{\partial r} \sim -\left(\frac{\beta}{r}\right) \tag{24}$$

is significantly larger then the value $2\pi / \lambda$ , which characterizes the variation of the electric field in a free space. If one considers more realistic models of the rough surface than the regular metal lattice, it is obvious, that there is a strong enhancement of the perpendicular component of the electric field at prominent places with a large curvature (Fig. 9) while the tangential components are comparable with the intensity with the incident field and tends to zero due to screening by conductive electrons.

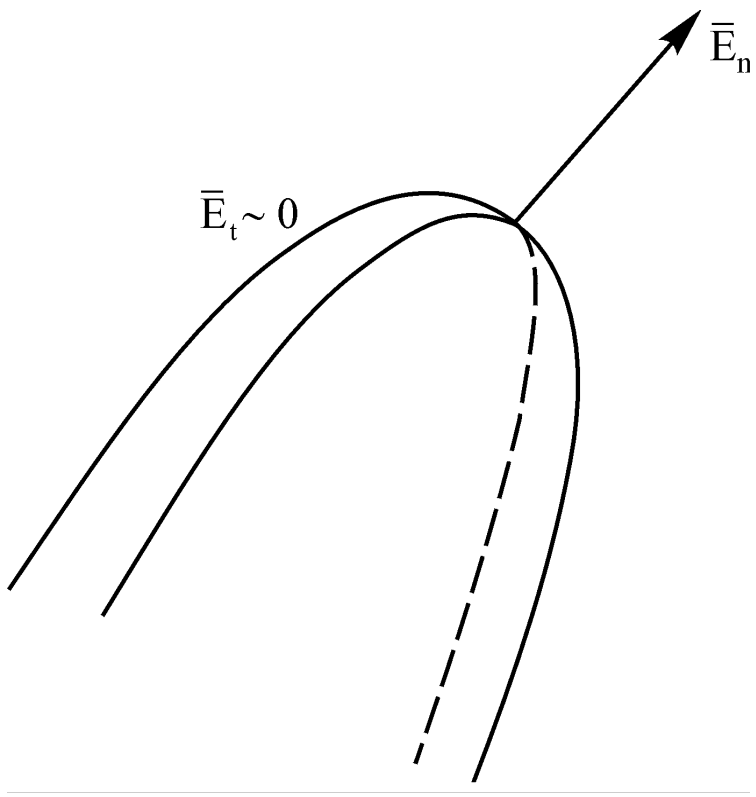

Fig. 9 More realistic model of the roughness. The normal component of the field and the derivatives $\partial E_\alpha / \partial x_\alpha, \alpha = (x, y, z)$ are enhanced near the top of the roughness.

Besides, the electromagnetic field strongly varies in space with a characteristic length $l_E$ equal to the characteristic roughness size. This type of behavior is not an exclusive property of the ideally conductive roughness and preserves near the surface with a finite dielectric constant. As it was demonstrated in [30] the field near the wedge or cone has a similar form as (23), however the $\beta$ value depends on the complex dielectric constant in a complicated manner. In [31, 32] model calculations of the electric field near metal nanowires with triangular, squared and some other cross-sections were performed. It should be noted that the authors used a finite value of the complex dielectric constant in this case. In spite of the authors observe several broad resonances in a high frequency region (approximately at $\lambda \leq 400nm$), the main enhancement occurs near the wedges of nanowires and reaches very large values. The appearance of the broad resonances is apparently associated with regular cross-sections of nanowires and arising due to the change in the relative values of the wavelength and the characteristic size of the cross-sections of nanowires. It should be noted that the obtained behavior corresponds to that in our model since the main enhancement was measured in the vicinity of the wedges and corresponds to the enhancement arising just in the rod effect. Therefore the formulas (20) apparently are valid in the vicinity of the wedges of nanowires. The influence exerted by a different environment, the angle of the wedge and by the finite value of the complex dielectric constant of the metal is reflected in possible various values of $C_0$ and $\beta$ in (20) on the features in both models. Thus the main property of the electric field (its singular behavior near the wedges) is preserved in the numerical results in [31, 32] and partially coincide with our above mentioned results that can be valid for more complex cases of SERS. It should be noted that in experimental investigations of the dependence of the enhancement on the wavelength on real rough surfaces [5] the resonance behavior is absent (Fig. 4). Anyway, the main property of the electromagnetic field and its derivatives is their steep increase near sharp points of the surface or at prominent places with a strong curvature.

## 5. EXPRESSIONS FOR THE SER CROSS-SECTIONS OF ARBITRARY AND SYMMETRICAL MOLECULES

The analytical expression for the SER cross-section is very important, because it permits us to interpret a large number of SERS regularities, especially for symmetrical molecules and to corroborate the SERS enhancement mechanism. In the set of our papers, which do not use some peculiarities we demonstrated that the increase of the cross-section is caused by the strong dipole and quadrupole light-molecule interactions near the prominent places with a large curvature or tips of metal surfaces [30]. It was also found that the quadrupole interaction become zero in molecules with cubic symmetry groups [12] and the SERS enhancement is determined only by the increase of the electric field. Besides some features of the SER spectra of symmetrical molecules were explained using these results. Here we obtain some modified expressions for the SER cross –sections of arbitrary and symmetrical molecules and verify the validity of the results obtained in our papers. The cross-section with the dipole and quadrupole interactions of light with molecules taken into account was obtained and used in [30, 33, 12, 35, and 36]. It should be remind that it can be calculated using wave-functions of the time dependent Schrödinger equation.

$$-i\hbar\frac{\partial\Psi}{\partial t}=\left[\hat{H}_{mol}+\hat{H}_{e-r}^{inc}+\hat{H}_{e-r}^{scat}\right]\Psi \tag{25}$$

where

$$\hat{H}_{mol}=\hat{H}_{e}+\hat{H}_{n}+\hat{H}_{e-n} \tag{26}$$

and

$$\hat{H}_{e}=-\hbar^{2}/2m\sum_{i}\Delta_{\bar{r}_i}+\frac{1}{2}\sum_{\substack{i,k\\ i\neq k}}\frac{e^{2}}{r_{ik}}-\sum_{iJ}\frac{e^{2}Z_{J}^{*}}{\left|\bar{R}_{iJ}^{0}\right|} \tag{27}$$

$$\widehat{H}_n = -\frac{\hbar^2}{2}\sum_J \frac{1}{M_J}\Delta_{\bar{R}_J} + \frac{1}{2}\sum_{\substack{J,K \\ J\neq K}} \frac{e^2 Z_J^* Z_K^*}{\left|\bar{R}_{JK}\right|} \qquad (28)$$

$$\widehat{H}_{e-n} = -\sum_{iJ} \frac{e^2 Z_J^*}{\left|\bar{R}_{iJ}\right|} + \sum_{iJ} \frac{e^2 Z_J^*}{\left|\bar{R}_{iJ}^0\right|} \qquad (29)$$

Here $\widehat{H}_{mol}, \widehat{H}_e, \widehat{H}_n, \widehat{H}_{e-n}$ are the Hamiltonians of the molecule, electrons (in the field of motionless nuclei), nuclei and electron-nuclei interactions. $\bar{r}_i$ is the radius vector of $i$ electron, $r_{ik}$ is the distance between $i$ and $k$ electrons, $\bar{R}_{iJ}^o$ is the radius vector between the motionless $J$ nucleus and $i$ electron, $\bar{R}_{iJ}$ is the radius vector between the $J$ nucleus and $i$ electron, $\bar{R}_J$ is the radius vector of the $J$ nucleus, $\bar{R}_{JK}$ is the radius vector between $J$ and $K$ nuclei, $M_J$ is the mass of the $J$ nucleus, $Z_J^*$ its atomic number. All other designations in (25-29) are conventional. The Hamiltonians of interaction of electrons with the incident and scattering fields $\widehat{H}_{e-r}^{\text{inc}}$ and $\widehat{H}_{e-r}^{\text{scat}}$ in (25) can be obtained by well known method of expansion of corresponding vector potentials $\bar{A}_i$ in the expression (30). Here the index $i$ numerates electrons and the value of $e$ is positive.

$$\widehat{H}_{e-r} = -\sum_i \frac{ie\hbar}{mc}\bar{A}_i \bar{\nabla}_i \qquad (30)$$

Here and further we put the scalar potential $\Phi = 0$ and $div\bar{A} = div\bar{E} = 0$. Then by neglecting of the magneto-dipole interaction, the light-electron interaction Hamiltonians for the incident and scattered fields are

$$\widehat{\mathrm{H}}_{e-r}^{\mathrm{inc}} = \left|\overline{E}_{inc}\right| \frac{(\overline{e}^{*}\overline{f}^{*}_{e})_{inc} e^{i\omega_{inc}t} + (\overline{e}\overline{f}_{e})_{inc} e^{-i\omega_{inc}t}}{2} \tag{31}$$

$$\widehat{\mathrm{H}}_{e-r}^{\mathrm{scat}} = \left|\overline{E}_{scat}\right| \frac{(\overline{e}^{*}\overline{f}^{*}_{e})_{scat} e^{i\omega_{scat}t} + (\overline{e}\overline{f}_{e})_{scat} e^{-i\omega_{scat}t}}{2} \tag{31a}$$

$\left|\overline{E}_{inc}\right|$ and $\left|\overline{E}_{scat}\right|$, $\omega_{inc}$ and $\omega_{scat}$ are the amplitudes and frequencies of the incident and scattered fields, $\overline{e}$ is a polarization vector,

$$f_{e\alpha} = d_{e\alpha} + \frac{1}{2E_{\alpha}} \sum_{\beta} \frac{\partial E_{\alpha}}{\partial x_{\beta}} Q_{e\alpha\beta} \tag{32}$$

is an $\alpha$ component of the generalized coefficient of light-electrons interaction [30, 34, 35], $d_{e\alpha}$ and $Q_{e\alpha\beta}$ are corresponding components of the dipole moment vector and the quadrupole moments tensor of electrons. (It should be noted, that here and further we use for designation of components of vectors and tensors the indices $i, j, k$ and $\alpha, \beta, \gamma$, which mean the values $x, y, z$. In addition the values $x_{\alpha}, x_{\beta}, x_{\gamma}$ designate the coordinates $x, y, z$). In the first stage it is necessary to obtain the molecular wave functions of the unperturbed Hamiltonian, which satisfy the equation

$$\widehat{\mathrm{H}}_{mol}\Psi = \mathrm{E}\Psi \tag{33}$$

This procedure is well described in [30, 33] and in Appendix 1 and we can write the following expressions for the full wave function of the ground state $\Psi_{n\overline{V}}$ and its vibrational part $\alpha_{\overline{V}}$. Taking into account the time dependence, we have

$$\Psi_{n\bar{V}}=\left[\Psi_n^{(0)}+\sum_{\substack{l\\ l\neq n}}\frac{\sum_s R_{nls}\sqrt{\frac{\omega_s}{\hbar}}\xi_s\Psi_l^{(0)}}{(\mathrm{E}_n^{(0)}-\mathrm{E}_l^{(0)})}\right]\alpha_{\bar{V}}\exp-(i\mathrm{E}_{n\bar{V}}t)/\hbar \qquad (34)$$

$$\alpha_{\bar{V}}=\prod_s N_s H_{V_s}\left(\sqrt{\frac{\omega_s}{\hbar}}\xi_s\right)\exp\left(-\frac{\omega_s\xi_s^2}{2\hbar}\right) \qquad (34a)$$

Here the full wave function is characterized by the electron wave number $n$ and by the set of the vibration wave numbers $\bar{V}=(V_1,V_2...V_s...)$, which consists of the wave numbers of separate vibrations $V_s$. $\Psi_n^{(0)}$ and $\Psi_l^{(0)}$ are the solutions for the ground and excited states of the equation

$$\hat{\mathrm{H}}_e\Psi^{(0)}=\mathrm{E}^{(0)}\Psi^{(0)} \qquad (35)$$

$\mathrm{E}_n^{(0)}$ and $\mathrm{E}_l^{(0)}$ are their energies. $\xi_s$ is the normal coordinate of the $s$ vibration mode.

$$R_{nls}=\sum_{J\alpha}\sqrt{\frac{\hbar}{\omega_s}}\bar{X}_{Js\alpha}\frac{\partial\langle l|\hat{\mathrm{H}}_{e-n}|n\rangle}{\partial X_{J\alpha}} \qquad (36)$$

is the coefficient, which characterizes excitation of the $\Psi_l^{(0)}$ wave function by the $s$ vibrational mode from the ground state $n$ to the excited states $l$, $\omega_s$ is a frequency of the $s$ vibrational mode, $\bar{X}_{Js\alpha}$ is an $\alpha$ component of the displacement vector of the $J$ nucleus in the $s$ vibrational mode, $X_{J\alpha}$ is an $\alpha$ component of the coordinate of the $J$ nucleus. Here and further we mean, that the derivatives of the matrix elements of $\hat{\mathrm{H}}_{e-n}$ are taken for $X_{J\alpha}^0$ values which correspond to the equilibrium positions of nuclei. $N_s$ and $H_{V_s}$ in (34a) are normalization constants and the Hermitian polynomials for separate vibrations.

$$\mathrm{E}_{n\bar{V}} = \mathrm{E}_n^{(0)} + \sum_s \hbar\omega_s (V_s + 1/2) \qquad (37)$$

is the expression for the full energy of the state. The specific form for the time dependent perturbed wave function, which takes into account the interaction of light with the molecule (31, 31a) can be obtained from the general expression for this function in accordance with [37].

$$\Psi_{n\bar{V}}^{(1)}(t) = \sum_{k\bar{V}_1} a_{(k\bar{V}_1),(n\bar{V})}(t)\Psi_{k\bar{V}_1} \qquad (38)$$

Here $\Psi_{n\bar{V}}^{(1)}(t)$ is the time dependent perturbated wave function, $\Psi_{k\bar{V}_1}$ are the eigenfunctions of the unperturbed Hamiltonian. $\bar{V}_1$ are sets of vibrational wave numbers of excited states. The cross-section of Raman scattering is expressed via the second terms of the expansions of the perturbation coefficients $a_{(n,\bar{V}\pm 1),(n,\bar{V})}^{(2)}(t)$, where the index $\bar{V} \pm 1$ designates the change of one of vibrational numbers $V_s$ on one unit and refer to the Stokes and AntiStokes scattering respectively. Since the coefficients $a_{(n,\bar{V}\pm 1),(n,\bar{V})}^{(2)}(t)$ are very small, the full electron wave function is determined mainly by the wave function of the ground state. We neglect by the difference between the vibrational functions and vibrational frequencies of the ground and virtual electron states that strongly simplified our calculations. The wave functions of the virtual states can be easily determined by generalization of the procedure, used to obtain $\Psi_{n\overline{V}}$ in [30, 33] and in Appendix 1 by changing of indices. If we use the above-mentioned concept concerning the vibrational states and frequencies the virtual wave functions $\Psi_{m,\overline{V}}$ have the form

$$\Psi_{m\overline{V}} = \left[ \Psi_m^{(0)} + \sum_{\substack{k \\ k \neq m}} \frac{\sum_s R_{mks} \sqrt{\frac{\omega_s}{\hbar}} \xi_s \Psi_k^{(0)}}{(\mathrm{E}_m^{(0)} - \mathrm{E}_k^{(0)})} \right] \alpha_{\overline{V}} \exp - (i\mathrm{E}_{m\overline{V}} t)/\hbar \tag{39}$$

Here

$$R_{mks} = \sum_{J\alpha} \sqrt{\frac{\hbar}{\omega_s}} \overline{X}_{Js\alpha} \frac{\partial \langle k | \hat{\mathrm{H}}_{e-n} | m \rangle}{\partial X_{J\alpha}} \tag{40}$$

similarly to $R_{nls}$ (36). The cross-section of Raman scattering is expressed in terms of the expression

$$a^{(2)}_{(n,\overline{V}\pm 1),(n,\overline{V})}(t) = \left(-\frac{i}{\hbar}\right)^2 \sum_{\substack{m \\ m \neq n}} \left[ \int_0^t \langle n,\overline{V}\pm 1 | \hat{\mathrm{H}}^{inc}_{e-r} + \hat{\mathrm{H}}^{scat}_{e-r} | m,\overline{V}\pm 1 \rangle dt_1 \times \right.$$

$$\times \int_0^{t_1} \langle m,\overline{V}\pm 1 | \hat{\mathrm{H}}^{inc}_{e-r} + \hat{\mathrm{H}}^{scat}_{e-r} | n,\overline{V} \rangle dt_2 + \tag{41}$$

$$\left. + \int_0^t \langle n,\overline{V}\pm 1 | \hat{\mathrm{H}}^{inc}_{e-r} + \hat{\mathrm{H}}^{scat}_{e-r} | m,\overline{V} \rangle dt_1 \int_0^{t_1} \langle m,\overline{V} | \hat{\mathrm{H}}^{inc}_{e-r} + \hat{\mathrm{H}}^{scat}_{e-r} | n,\overline{V} \rangle dt_2 \right]$$

Taking into account the expressions for the Hamiltonians of interaction of light with electrons (31, 31a) and expressions (34, 39) for the wave functions of the ground and excited states one can obtain the following expression for the Raman cross-section in a free space

$$d\sigma_{vol} = \frac{\omega_{inc}\omega_{scat}^{3}}{16\hbar^{2}\varepsilon_{0}^{2}\pi^{2}c^{4}} \times \begin{pmatrix} (V_s+1)/2 \\ V_s/2 \end{pmatrix} \left| A_{V_s}\begin{pmatrix} St \\ anSt \end{pmatrix} \right|_{vol}^{2} dO \tag{42}$$

where

$$A_{V_s}\begin{pmatrix} St \\ \overline{\phantom{x}} \\ anSt \end{pmatrix} = \sum_{\substack{m,l \\ m,l\neq n}} \frac{R_{nls}\langle n|(\overline{e}^{*}\overline{f}^{*}_{e})_{scat}|m\rangle\langle m|(\overline{e}\overline{f}_{e})_{inc}|l\rangle}{(\mathrm{E}_{n}^{(0)}-\mathrm{E}_{l}^{(0)})(\omega_{mn}\pm\omega_{s}-\omega_{inc})} +$$

$$\sum_{\substack{m,l \\ m,l\neq n}} \frac{R^{*}_{nls}\langle l|(\overline{e}^{*}\overline{f}^{*}_{e})_{scat}|m\rangle\langle m|(\overline{e}\overline{f}_{e})_{inc}|n\rangle}{(\mathrm{E}_{n}^{(0)}-\mathrm{E}_{l}^{(0)})(\omega_{mn}-\omega_{inc})} +$$

$$\sum_{\substack{m,l \\ m,l\neq n}} \frac{R_{nls}\langle n|(\overline{e}\overline{f}_{e})_{inc}|m\rangle\langle m|(\overline{e}^{*}\overline{f}^{*}_{e})_{scat}|l\rangle}{(\mathrm{E}_{n}^{(0)}-\mathrm{E}_{l}^{(0)})(\omega_{mn}\pm\omega_{s}+\omega_{scat})} +$$

$$\sum_{\substack{m,l \\ m,l\neq n}} \frac{R^{*}_{nls}\langle l|(\overline{e}\overline{f}_{e})_{inc}|m\rangle\langle m|(\overline{e}^{*}\overline{f}^{*}_{e})_{scat}|n\rangle}{(\mathrm{E}_{n}^{(0)}-\mathrm{E}_{l}^{(0)})(\omega_{mn}+\omega_{scat})} +$$

$$\sum_{\substack{m,k \\ m\neq n,k}} \frac{\langle n|(\overline{e}^{*}\overline{f}^{*}_{e})_{scat}|m\rangle R^{*}_{mks}\langle k|(\overline{e}\overline{f}_{e})_{inc}|n\rangle}{(\mathrm{E}_{m}^{(0)}-\mathrm{E}_{k}^{(0)})(\omega_{mn}\pm\omega_{s}-\omega_{inc})} +$$

$$\sum_{\substack{m,k \\ m\neq n,k}} \frac{\langle n|(\overline{e}^{*}\overline{f}^{*}_{e})_{scat}|k\rangle R_{mks}\langle m|(\overline{e}\overline{f}_{e})_{inc}|n\rangle}{(\mathrm{E}_{m}^{(0)}-\mathrm{E}_{k}^{(0)})(\omega_{mn}-\omega_{inc})} +$$

$$\sum_{\substack{m,k \\ m\neq n,k}} \frac{\langle n|(\overline{e}\overline{f}_{e})_{inc}|m\rangle R^{*}_{mks}\langle k|(\overline{e}^{*}\overline{f}^{*}_{e})_{scat}|n\rangle}{(\mathrm{E}_{m}^{(0)}-\mathrm{E}_{k}^{(0)})(\omega_{mn}\pm\omega_{s}+\omega_{scat})} +$$

$$\sum_{\substack{m,k \\ m \neq n,k}} \frac{\langle n|(\overline{e}\overline{f}_e)_{inc}|k\rangle R_{mks} \langle m|(\overline{e}^*\overline{f}_e^*)_{scat}|n\rangle}{(\mathrm{E}_m^{(0)} - \mathrm{E}_k^{(0)})(\omega_{mn} + \omega_{scat})} + \qquad (43)$$

Here

$$\omega_{mn} = \frac{\mathrm{E}_m^{(0)} - \mathrm{E}_n^{(0)}}{\hbar} \qquad (44)$$

The SER cross –section can be obtained from expression (42); with taking into account of the fact that the incident field affecting the molecule is the surface field. The same refers to the emitted field. Therefore one should multiply (42) by the expression

$$\frac{|\overline{E}_{inc}|^2_{surf}}{|\overline{E}_{inc}|^2_{vol}} \frac{|\overline{E}_{scat}|^2_{surf}}{|\overline{E}_{scat}|^2_{vol}} \qquad (45)$$

and take the surface fields as the incident and scattered fields under the sign of moduli in (42). Finally the expression for the SER cross-section has the form

$$d\sigma_{surf} = \frac{\omega_{inc}\omega_{scat}^3}{16\hbar^2\varepsilon_0^2\pi^2c^4} \frac{|\overline{E}_{inc}|^2_{surf}}{|\overline{E}_{inc}|^2_{vol}} \frac{|\overline{E}_{scat}|^2_{surf}}{|\overline{E}_{scat}|^2_{vol}} \times$$
$$\times \begin{pmatrix} (V_s+1)/2 \\ V_s/2 \end{pmatrix} \left| A_{V_s}\left(\frac{St}{anSt}\right) \right|^2_{surf} dO \qquad (46)$$

Here the signs $surf$ and $vol$ designate the values of the cross-sections and electromagnetic fields at the surface and in the volume. It should be noted, that $(\overline{E}_{inc})_{surf}$ is the surface field generated by the field $(\overline{E}_{inc})_{vol}$ incident on the surface, while $(\overline{E}_{scat})_{surf}$ is the surface field generated by the field $(\overline{E}_{scat})_{vol}$ incident on the surface from the direction for which the direction of the wave reflected from the surface coincides with the direction of the scattering.

In order to estimate the enhancement coefficient one should note that not all the terms of the quadrupole interaction are essential for SERS. It is necessary to take into account that the quadrupole moments $Q_{\alpha\alpha}$ are the values with a constant sign, while the moments $Q_{\alpha\beta}$ for $\alpha \neq \beta$ and $d_\alpha$ are of the changeable sign. This fact causes some additional "oscillation" of integrands in the matrix elements $\langle l|Q_{\alpha\beta}|n\rangle$ ($\alpha \neq \beta$) and in the matrix elements $\langle l|d_\alpha|n\rangle$, which results in significant relative increase of the matrix elements $\langle l|Q_{\alpha\alpha}|n\rangle$ compared with $\langle l|Q_{\alpha\beta}|n\rangle$ and $\langle l|d_\alpha|n\rangle$. Here we omit the index $e$ near the components of the dipole moments vector and the quadrupole moments tensor. Then the relation of some mean values of matrix elements of the quadrupole and the dipole moments can be estimated as

$$\frac{\overline{\langle n|Q_{\alpha\beta}|l\rangle}}{\overline{\langle n|d_\alpha|l\rangle}} \approx Ba \qquad (47)$$

where $B \sim 1$ for $(\alpha \neq \beta)$ as it is usually accepted in quantum mechanics, while $B >> 1$ for $\alpha = \beta$, $a$ is the molecule size [30, 34,35]. This fact can be considered as a quantum mechanical reason of the enhancement of the quadrupole interaction complementary to the enhancement of the electric field derivatives $\frac{\partial E_\alpha}{\partial x_\alpha}$. The estimation of some mean enhancement coefficient in SERS formally coincides with estimation, made in [30, 34, 35]. Here the ratio $\frac{\left|A_{V_s}\right|^2_{surf}}{\left|A_{V_s}\right|^2_{vol}}$ is substituted with a ratio of some mean values of matrix elements of the quadrupole and dipole transitions (47) and with the ratios of squared moduli of some values of derivatives of the incident and scattered surface fields to the squared moduli of the values of these fields in the volume. Taking into account the previous results and the results of paragraph 4 one can obtain the following expression for the estimation of the enhancement coefficient of the cross-section near the model roughness of the wedge or the cone type [30, 35].

$$G_0 = \frac{d\sigma_{surf}}{d\sigma_{vol}} = \frac{\left|\overline{E}_{inc}\right|^2_{surf}}{\left|\overline{E}_{inc}\right|^2_{vol}} \frac{\left|\overline{E}_{scat}\right|^2_{surf}}{\left|\overline{E}_{scat}\right|^2_{vol}} \frac{\left|A_{V\,s}\right|^2_{surf}}{\left|A_{V\,s}\right|^2_{vol}}$$
$$\sim \frac{(Ba)^4}{16} \frac{\left|\overline{\partial E^{inc}_\alpha / \partial x_\alpha}\right|^2_{surf}}{\left|\overline{E}_{inc}\right|^2_{vol}} \frac{\left|\overline{\partial E^{scat}_\alpha / \partial x_\alpha}\right|^2_{surf}}{\left|\overline{E}_{scat}\right|^2_{vol}} \sim \qquad (48)$$
$$\sim C_0^4 \beta^4 \left(\frac{B}{2}\right)^4 \left(\frac{l_1}{r}\right)^{4\beta} \left(\frac{a}{r}\right)^4$$

As it is mentioned above $\left|\overline{\partial E^{inc}_\alpha / \partial x_\alpha}\right|_{surf}$ and $\left|\overline{\partial E^{scat}_\alpha / \partial x_\alpha}\right|_{surf}$ are some values of the derivatives of the incident and scattered fields near the surface. For estimation of the $Ba$ value it is necessary to make the following simplifications. Because the inner electron shell of molecules for excited states remains almost intact, then for estimations we usually take the value $\overline{\langle n | Q_{\alpha\alpha} | n \rangle}$ instead of

$\overline{\langle n|Q_{\alpha\alpha}|l\rangle}$ and the value $\sqrt{e^2\hbar/2m\omega_{nl}}\times\sqrt{\overline{f}_{nl}}$ for $\overline{\langle n|d_\alpha|l\rangle}$, which is expressed in terms of some mean value of the oscillator strength $\overline{f}_{nl}=0.1$ while $\omega_{nl}$ correspond to the edge of absorption ($\lambda=2500A$). [30, 35]. The first approximation is forced because of the absence of data on quadrupole transitions. Since the configuration of the electron shell is close to the configuration of nuclei the value $\overline{\langle n|Q_{\alpha\alpha}|n\rangle}$ was calculated as a $Q_{n,\alpha\alpha}$ component of the quadrupole moments of nuclei. For reasonable values of the parameters $C_0\sim 1, \beta\sim\frac{1}{2}$, $l_1\sim 10nm$ and $Ba$ value $\sim 18nm$ corresponding to the pyridine molecule, the electrodynamic enhancement $\sim 10^6$ can be obtained for the distances $r\sim 1.5nm$. It is evident that for such distances from the cone the values of the field derivatives can correspond to the real values near the rough surface. A more real situation is the following. We can see from (48) that the strongest enhancement is observed for molecules placed at the top of the cone. The enhancement factor for molecules situated at some distance from the surface decreases very strong. The estimation taken for the pyridine molecule placed at the top of the cone results in the enhancement factor ~ $4\times 10^{14}$. Therefore the overall enhancement $10^6$ for the layer of pyridine molecules apparently arise from the small amount of molecules situated at the prominent places with a very large curvature, which are responsible for the very strong enhancement factor. The averaging of the enhancement factor over all molecules results in decrease of the overall enhancement till the value $10^6$. It should be noted that the enhancement coefficient in SERS strongly depends on the type of the roughness. The above estimates show that the quadrupole interaction become stronger than the dipole interaction for a very sharp roughness. It can be demonstrated that depending on numerous parameters of substrate the quadrupole interaction can be either stronger or weaker than the dipole interaction. Also there may be some cases, in which the quadrupole interaction may not manifest oneself, since the substrate surface is smooth. In actual practice comparison of experimental values with the theory is very complicated. The real amplitude of the scattering signal is a sum of signals from molecules adsorbed in different conditions. Therefore our estimation is very rough approximation, which demonstrates that in principal large enhancement coefficients are possible and the quadrupole enhancement mechanism can be operative. For a more precise comparison of the theory and results of experiments, carried out under various conditions it is better to consider the SER spectra of symmetrical molecules. The existence of forbidden and allowed lines in these spectra can give more precise prove about validity of the dipole-quadrupole SERS mechanism.

The general expression for the cross-section of symmetrical molecules can be obtained from expression (46) by taking into account the effects of degeneration of the molecule states and designating the vibrational states by two indices$(s, p)$ instead of $s$. Here $s$ designates the group of degenerate vibrations, whereas $p$ enumerates the states within the group. Then the expression for the cross-section of symmetrical molecules takes the form

$$d\sigma_{surf}=\frac{\omega_{inc}\omega_{scat}^{3}}{16\hbar^{2}\varepsilon_{0}^{2}\pi^{2}c^{4}}\frac{\left|\overline{E}_{inc}\right|_{surf}^{2}}{\left|\overline{E}_{inc}\right|_{vol}^{2}}\frac{\left|\overline{E}_{scat}\right|_{surf}^{2}}{\left|\overline{E}_{scat}\right|_{vol}^{2}}\times$$

$$\times\begin{pmatrix}(V_{(s,p)}+1)/2\\ V_{(s,p)}/2\end{pmatrix}\sum_{p}\left|A_{V_{(s,p)}}\begin{pmatrix}St\\ anSt\end{pmatrix}\right|_{surf}^{2}dO \tag{49}$$

where

$$A_{V_{(s,p)}}\left(\frac{St}{anSt}\right)=\sum_{\substack{m,l\\ m,l\neq n}}\frac{R_{nl(s,p)}\langle n|(\overline{e}^{*}\overline{f}_{e}^{*})_{scat}|m\rangle\langle m|(\overline{e}\overline{f}_{e})_{inc}|l\rangle}{(\mathrm{E}_{n}^{(0)}-\mathrm{E}_{l}^{(0)})(\omega_{mn}\pm\omega_{(s,p)}-\omega_{inc})}+$$

$$\sum_{\substack{m,l\\ m,l\neq n}}\frac{R_{nl(s,p)}^{*}\langle l|(\overline{e}^{*}\overline{f}_{e}^{*})_{scat}|m\rangle\langle m|(\overline{e}\overline{f}_{e})_{inc}|n\rangle}{(\mathrm{E}_{n}^{(0)}-\mathrm{E}_{l}^{(0)})(\omega_{mn}-\omega_{inc})}+$$

$$\sum_{\substack{m,l\\ m,l\neq n}}\frac{R_{nl(s,p)}\langle n|(\overline{e}\overline{f}_{e})_{inc}|m\rangle\langle m|(\overline{e}^{*}\overline{f}_{e}^{*})_{scat}|l\rangle}{(\mathrm{E}_{n}^{(0)}-\mathrm{E}_{l}^{(0)})(\omega_{mn}\pm\omega_{(s,p)}+\omega_{scat})}+$$

$$\sum_{\substack{m,l\\ m,l\neq n}}\frac{R_{nl(s,p)}^{*}\langle l|(\overline{e}\overline{f}_{e})_{inc}|m\rangle\langle m|(\overline{e}^{*}\overline{f}_{e}^{*})_{scat}|n\rangle}{(\mathrm{E}_{n}^{(0)}-\mathrm{E}_{l}^{(0)})(\omega_{mn}+\omega_{scat})}+$$

$$\sum_{\substack{m,k\\ m\neq n,k}}\frac{\langle n|(\overline{e}^{*}\overline{f}_{e}^{*})_{scat}|m\rangle R_{mk(s,p)}^{*}\langle k|(\overline{e}\overline{f}_{e})_{inc}|n\rangle}{(\mathrm{E}_{m}^{(0)}-\mathrm{E}_{k}^{(0)})(\omega_{mn}\pm\omega_{(s,p)}-\omega_{inc})}+$$

$$\sum_{\substack{m,k\\ m\neq n,k}}\frac{\langle n|(\overline{e}^{*}\overline{f}_{e}^{*})_{scat}|k\rangle R_{mk(s,p)}\langle m|(\overline{e}\overline{f}_{e})_{inc}|n\rangle}{(\mathrm{E}_{m}^{(0)}-\mathrm{E}_{k}^{(0)})(\omega_{mn}-\omega_{inc})}+$$

$$\sum_{\substack{m,k \\ m \neq n,k}} \frac{\langle n|(\overline{e}\overline{f}_e)_{inc}|m\rangle R^{*}_{mk(s,p)}\langle k|(\overline{e}^{*}\overline{f}^{*}_e)_{scat}|n\rangle}{(E^{(0)}_m - E^{(0)}_k)(\omega_{mn} \pm \omega_{(s,p)} + \omega_{scat})} +$$

$$\sum_{\substack{m,k \\ m \neq n,k}} \frac{\langle n|(\overline{e}\overline{f}_e)_{inc}|k\rangle R_{mk(s,p)}\langle m|(\overline{e}^{*}\overline{f}^{*}_e)_{scat}|n\rangle}{(E^{(0)}_m - E^{(0)}_k)(\omega_{mn} + \omega_{scat})} + \qquad (50)$$

In spite of we use the double designation for the vibrational states we preserve the continuous numbering for the electronic states, which is not important for our goals. In addition we mean that the values $V_{(s,p)}$ are equal for the same $s$ and various $p$, since the molecule usually is in an equilibrium state. The difference between the cross-sections, obtained in [35] and the modified expressions for the cross-sections for arbitrary and symmetrical molecules obtained here is appearance of additional terms in (43) and (50), with $R_{mks}$ and $R_{mk(s,p)}$ and also with conjugated coefficients, which describe some additional processes with another sequence of absorption and emission of photons and excitation of virtual electron states by vibrations in various phases of the Raman process. Now let us introduce the scattering tensor $C_{V_{(s,p)}}[f_i, f_j]$. Its evident form is

$$C_{V_{(s,p)}}\lfloor f_i, f_j \rfloor = \sum_{\substack{m,l \\ m,l \neq n}} \frac{R_{nl(s,p)}\langle n|f_i|m\rangle\langle m|f_j|l\rangle}{(E^{(0)}_n - E^{(0)}_l)(\omega_{mn} \pm \omega_{(s,p)} - \omega_{inc})} +$$

$$\sum_{\substack{m,l \\ m,l \neq n}} \frac{R^{*}_{nl(s,p)}\langle l|f_i|m\rangle\langle m|f_j|n\rangle}{(E^{(0)}_n - E^{(0)}_l)(\omega_{mn} - \omega_{inc})} +$$

$$\sum_{\substack{m,l \\ m,l\neq n}} \frac{R_{nl(s,p)}\langle n|f_j|m\rangle\langle m|f_i|l\rangle}{(\mathrm{E}_n^{(0)}-\mathrm{E}_l^{(0)})(\omega_{mn}\pm\omega_{(s,p)}+\omega_{scat})}+$$

$$\sum_{\substack{m,l \\ m,l\neq n}} \frac{R^*_{nl(s,p)}\langle l|f_j|m\rangle\langle m|f_i|n\rangle}{(\mathrm{E}_n^{(0)}-\mathrm{E}_l^{(0)})(\omega_{mn}+\omega_{scat})}+$$

$$\sum_{\substack{m,k \\ m\neq n,k}} \frac{\langle n|f_i|m\rangle R^*_{mk(s,p)}\langle k|f_j|n\rangle}{(\mathrm{E}_m^{(0)}-\mathrm{E}_k^{(0)})(\omega_{mn}\pm\omega_{(s,p)}-\omega_{inc})}+$$

$$\sum_{\substack{m,k \\ m\neq n,k}} \frac{\langle n|f_i|k\rangle R_{mk(s,p)}\langle m|f_j|n\rangle}{(\mathrm{E}_m^{(0)}-\mathrm{E}_k^{(0)})(\omega_{mn}-\omega_{inc})}+$$

$$\sum_{\substack{m,k \\ m\neq n,k}} \frac{\langle n|f_j|m\rangle R^*_{mk(s,p)}\langle k|f_i|n\rangle}{(\mathrm{E}_m^{(0)}-\mathrm{E}_k^{(0)})(\omega_{mn}\pm\omega_{(s,p)}+\omega_{scat})}+$$

$$\sum_{\substack{m,k \\ m\neq n,k}} \frac{\langle n|f_j|k\rangle R_{mk(s,p)}\langle m|f_i|n\rangle}{(\mathrm{E}_m^{(0)}-\mathrm{E}_k^{(0)})(\omega_{mn}+\omega_{scat})}+ \quad (51)$$

Here $f_i$ and $f_j$ are some formal parameters, that designate different dipole and quadrupole moments. Taking into account the expression (32) for the generalized coefficient of interaction of light with electrons one can obtain similar general expression for $A_{V_{(s,p)}}$ for symmetrical molecules as in [12, 35]. This expression can be separated into four contributions

$$A_{V_{(s,p)}} = S_{d-d} + S_{d-Q} + S_{Q-d} + S_{Q-Q} \tag{52}$$

where

$$S_{d-d} = \sum_{i,k} C_{V_{(s,p)}} \left[d_i, d_k\right] (e^*_{scat,i} e_{inc,k})_{surf} \tag{53}$$

is the contribution of the dipole-dipole $(d - d)$ scattering,

$$S_{d-Q} = \sum_{i,\gamma\delta} C_{V_{(s,p)}} \left[d_i, Q_{\gamma\delta}\right] \left( e^*_{scat,i} \frac{1}{2\left|\overline{E}_{inc}\right|} \frac{\partial E^{inc}_{\gamma}}{\partial x_{\delta}} \right)_{surf} \tag{54}$$

is the contribution of the dipole-quadrupole $(d - Q)$ scattering,

$$S_{Q-d} = \sum_{\alpha\beta,k} C_{V_{(s,p)}} [Q_{\alpha\beta}, d_k] \left( \frac{1}{2\left|\overline{E}_{scat}\right|} \frac{\partial E^{scat^*}_{\alpha}}{\partial x_{\beta}} e_{inc,k} \right) \tag{55}$$

is the contribution of the quadrupole-dipole $(Q - d)$ scattering and

$$S_{Q-Q} = \sum_{\alpha\beta,\gamma\delta} C_{V_{(s,p)}} \left[Q_{\alpha\beta}, Q_{\gamma\delta}\right] \left( \frac{1}{2\left|\overline{E}_{scat}\right|} \frac{\partial E^{scat^*}_{\alpha}}{\partial x_{\beta}} \frac{1}{2\left|\overline{E}_{inc}\right|} \frac{\partial E^{inc}_{\gamma}}{\partial x_{\delta}} \right)_{surf} \tag{56}$$

is the contribution of the quadrupole-quadrupole $(Q - Q)$ scattering.

## 6. SELECTION RULES FOR SYMMETRICAL MOLECULES

As it follows from (49) the SER cross-section of symmetrical molecules is expressed in terms of the sum of $\left|A_{V_{(s,p)}}\right|^2$ for all $p$. Since the vibrational states with different $p$ are degenerate and transform one through another under symmetry operations it is sufficient to analyze $A_{V_{(s,p)}}$ only for one $p$. The results for other values will be the same. The value $A_{V_{(s,p)}}$ is expressed via

contributions (53-56), which contain the scattering tensor (51), and depends on dipole and quadrupole moments. Therefore the intensity of the vibrational line is determined by definite combinations of contributions in all $A_{V_{(s,p)}}$, which depend on combinations of moments. When all the contributions are equal to zero, the appearance of the vibrational band is forbidden. If however, there are nonzero contributions, the corresponding band may be allowed. The last fact depends on the type of the moments which are contained in the scattering tensor. For a theoretical-group analysis of the contributions it is more convenient to pass to those containing the moments that transform under irreducible representations of the symmetry group. Because not all the moments are important in the scattering process it is useful to divide them into two groups main and minor ones. The first group includes the moments that are most important for the scattering process, while the second ones are the moments that are inessential for SERS. This classification strongly depends on a number of parameters, such as the material of substrate, its complex dielectric constant, the type of the roughness, the number of molecular layers covering the substrate and possibly some others. All these factors are of a qualitative character and the experimental conditions should be clearly understood for distinguishing between the main and minor moments. Our classification is apparently valid primarily for sufficiently thick layers of $Ag$, with sufficiently strong roughness degree and monolayer coverage of the substrate. Consideration for $Cu$, $Au$ and transition metals under usual observation conditions requires a more careful analysis. In accordance with our previous consideration the main moments are those, associated with the enhanced electric field (the $d_z$ moment for the monolayer coverage, which is perpendicular to the surface) and the quadrupole moments $Q_{xx}, Q_{yy}, Q_{zz}$ having a constant sign. The minor moments may depend on the coverage of substrate and for the monolayer coverage they are $d_x, d_y, Q_{xy}, Q_{xz}, Q_{yz}$ [12, 35] for all point symmetry groups. It should be noted, that in some experimental situations (e.g., thick layers of adsorbed molecules for example) the moments $d_x$ and $d_y$ may also belong to the main moments due to arbitrary orientation of the molecules and the moments to the enhanced electric field. Therefore our classification should be carefully considered for each particular experimental situation, as it has been mentioned above. In order to obtain selection rules for the contributions it is necessary to pass to the moments that transform under irreducible representations of the symmetry group and to the contributions that are expressed in terms of these moments. Analysis of the tables of irreducible representations of all point symmetry groups shows, that all the dipole and minor quadrupole moments transform under irreducible representations of the point groups, whereas $Q_{xx}, Q_{yy}, Q_{zz}$ may not transform in such a manner. Therefore we should pass to the moments $Q_1, Q_2, Q_3$, which transform under irreducible representations. Then the moments with a constant sign will be the main moments, while the moments with a variable sign the minor ones. The $Q_{xx}, Q_{yy}, Q_{zz}$ moments are expressed in terms of $Q_1, Q_2, Q_3$ in the following manner

$$Q_{xx} = a_{11}Q_1 + a_{12}Q_2 + a_{13}Q_3$$

$$Q_{yy} = a_{21}Q_1 + a_{22}Q_2 + a_{23}Q_3$$

$$Q_{zz} = a_{31}Q_1 + a_{32}Q_2 + a_{33}Q_3 \qquad (57)$$

Now using (57) and the following property of the tensor

$$C_{V_{(s,p)}}[f_i,(a_1 f_j + a_2 f_k)] = a_1 C_{V_{(s,p)}}\lfloor f_i, f_j \rfloor + a_2 C_{V_{(s,p)}}[f_i, f_k]$$

$$C_{V_{(s,p)}}[(a_1 f_i + a_2 f_j), f_k] = a_1 C_{V_{(s,p)}}[f_i, f_k] + a_2 C_{V_{(s,p)}}\lfloor f_j, f_k \rfloor \qquad (58)$$

we can obtain the following expression for the SER cross-section in terms of the moments that transform under irreducible representations

$$d\sigma_{surf} = \frac{\omega_{inc}\omega_{scat}^3}{16\hbar^2\varepsilon_0^2\pi^2 c^4} \frac{\left|\bar{E}_{inc}\right|_{surf}^2}{\left|\bar{E}_{inc}\right|_{vol}^2} \frac{\left|\bar{E}_{scat}\right|_{surf}^2}{\left|\bar{E}_{scat}\right|_{vol}^2}$$

$$\times \begin{pmatrix} (V_{(s,p)}+1)/2 \\ V_{(s,p)}/2 \end{pmatrix} \sum_p \left| T_{d-d} + T_{d-Q} + T_{Q-d} + T_{Q-Q} \right|_{surf}^2 dO \qquad (59)$$

Here the precise expressions under the sign of modulus are the following

$$T_{d-d} = S_{d-d} \qquad (60)$$

$$T_{d-Q} = \sum_{\substack{i,\gamma\delta \\ \gamma\neq\delta}} C_{V_{(s,p)}}[d_i, Q_{\gamma\delta}] \left( e_{scat,i}^* \frac{1}{2\left|\bar{E}_{inc}\right|} \frac{\partial E_\gamma^{inc}}{\partial x_\delta} \right)_{surf} +$$

$$\sum_{i,\gamma k} a_{\gamma k} C_{V_{(s,p)}}[d_i, Q_k] \left( e_{scat,i}^* \frac{1}{2\left|\bar{E}_{inc}\right|} \frac{\partial E_\gamma^{inc}}{\partial x_\gamma} \right)_{surf} \qquad (61)$$

$$T_{Q-d} = \sum_{\substack{\alpha\beta,i \\ \alpha\neq\beta}} C_{V_{(s,p)}}[Q_{\alpha\beta}, d_i] \left( \frac{1}{2\left|\bar{E}_{scat}\right|} \frac{\partial E_\alpha^{scat^*}}{\partial x_\beta} e_{inc,i} \right)_{surf} +$$

$$+\sum_{\alpha k,i} a_{\alpha,k} C_{V_{(s,p)}}[Q_k, d_i]\left( \frac{1}{2\left|\overline{E}_{scat}\right|} \frac{\partial E_{\alpha}^{scat^*}}{\partial x_{\alpha}} e_{inc,i} \right)_{surf} \quad (62)$$

$$T_{Q-Q} = \sum_{\substack{\alpha\beta,\gamma\delta \\ \alpha\neq\beta,\gamma\neq\delta}} C_{V_{(s,p)}}[Q_{\alpha\beta}, Q_{\gamma\delta}]\left( \frac{1}{2\left|\overline{E}_{scat}\right|} \frac{\partial E_{\alpha}^{scat^*}}{\partial x_{\beta}} \frac{1}{2\left|\overline{E}_{inc}\right|} \frac{\partial E_{\gamma}^{inc}}{\partial x_{\delta}} \right)_{surf} +$$

$$\sum_{\substack{\alpha,i,\gamma\delta \\ \gamma\neq\delta}} a_{\alpha,i} C_{V_{(s,p)}}[Q_i, Q_{\gamma\delta}]\left( \frac{1}{2\left|\overline{E}_{scat}\right|} \frac{\partial E_{\alpha}^{scat^*}}{\partial x_{\alpha}} \frac{1}{2\left|\overline{E}_{inc}\right|} \frac{\partial E_{\gamma}^{inc}}{\partial x_{\delta}} \right)_{surf} +$$

$$\sum_{\substack{\alpha\beta,\gamma,k \\ \alpha\neq\beta}} a_{\gamma,k} C_{V_{(s,p)}}[Q_{\alpha\beta}, Q_k]\left( \frac{1}{2\left|\overline{E}_{scat}\right|} \frac{\partial E_{\alpha}^{scat^*}}{\partial x_{\beta}} \frac{1}{2\left|\overline{E}_{inc}\right|} \frac{\partial E_{\gamma}^{inc}}{\partial x_{\gamma}} \right)_{surf} +$$

$$\sum_{\alpha,i,\gamma,k} a_{\alpha,i} a_{\gamma_{,k}} C_{V_{(s,p)}}[Q_i, Q_k]\left( \frac{1}{2\left|\overline{E}_{scat}\right|} \frac{\partial E_{\alpha}^{scat^*}}{\partial x_{\alpha}} \frac{1}{2\left|\overline{E}_{inc}\right|} \frac{\partial E_{\gamma}^{inc}}{\partial x_{\gamma}} \right)_{surf} \quad (63)$$

The selection rules for the contributions are determined by the combinations of moments in the scattering tensor (51) as it follows from (60-63). One can obtain conditions to which the vibration and the moments symmetry types in tensor (51) should satisfy, when

$$C_{V_{(s,p)}}\left[f_i, f_j\right] \neq 0 \quad (64)$$

First let us obtain the conditions when one of the lines in (51) for example the first one

$$R_{n,l,(s,p)}\langle n|f_i|m\rangle\langle m|f_j|l\rangle \neq 0 \quad (65)$$

Let us limit our consideration by the case, when the Jahn-Teller theorem is valid for the molecule symmetry group. This case refers to the nondegenerate ground state, which transforms after a single irreducible representation. The condition (65) reduces to the following ones, which have to be satisfied simultaneously

$$R_{n,l,(s,p)} \neq 0 \tag{66a}$$

$$\langle n | f_i | m \rangle \neq 0 \tag{66b}$$

$$\langle m | f_j | l \rangle \neq 0 \tag{66c}$$

The expression for $R_{nl(s,p)}$ can be expressed via $\langle l | \mathrm{H}_{e-n} | n \rangle$. Indeed, taking into account the relations (A1-9) (A1-14) and (A1-23) in Appendix 1, $\langle l | \mathrm{H}_{e-n} | n \rangle$ can be generalized and written in normal coordinates for symmetrical molecules as

$$\langle l | \widehat{\mathrm{H}}_{e-n} | n \rangle = \sum_{(s,p)} \sqrt{\frac{\omega_{(s,p)}}{\hbar}} \xi_{(s,p)} \times \sum_{J\alpha} \sqrt{\frac{\hbar}{\omega_{(s,p)}}} \overline{X}_{J,(s,p),\alpha} \times \frac{\partial \langle l | \widehat{\mathrm{H}}_{e-n} | n \rangle}{\partial X_{J\alpha}} \tag{67}$$

Considering excitation only one vibrational mode $(s, p)$ in (67) and using (36), one can obtain

$$\sqrt{\frac{\omega_{(s,p)}}{\hbar}} \xi_{(s,p)} R_{n,l,(s,p)} = \langle l | \widehat{\mathrm{H}}_{e-n} | n \rangle . \tag{68}$$

Thus the $R_{nl(s,p)}$ values are not equal to zero, when

$$\langle l | \widehat{\mathrm{H}}_{e-n} | n \rangle \neq 0 \tag{69}$$

Since the vibrational states transform after irreducible representations of the symmetry group of the molecule the symmetry of the Hamiltonian $\widehat{\mathrm{H}}_{e-n}$ coincides with the symmetry of the $(s, p)$ vibration. Then from (66a), taking into account (68, 69) and the fact that the ground state $n$ transforms under single irreducible representation of the molecule symmetry group, using the general selection rules for matrix elements one can obtain

$$\Gamma_l \in \Gamma_{(s,p)}\Gamma_n \tag{70}$$

Here the corresponding signs $\Gamma$ designate irreducible representations which determine transformational properties of the $n, l$ and vibrational $(s, p)$ functions. From (66c), knowing the irreducible representation of the $f_j$ moment one can determine the irreducible representations of the intermediate states $m$.

$$\Gamma_m \in \Gamma_{f_j}\Gamma_l \in \Gamma_{f_j}\Gamma_{(s,p)}\Gamma_n \tag{71}$$

The condition (66b) determines the irreducible representation of the $f_i$ moment, when (65) is valid

$$\Gamma_{f_i} \in \Gamma_n\Gamma_m \in \Gamma_n\Gamma_{f_j}\Gamma_{(s,p)}\Gamma_n \in \Gamma_{f_j}\Gamma_{(s,p)} \tag{72}$$

The expression (72) is satisfied when

$$\Gamma_{(s,p)} \in \Gamma_{f_i}\Gamma_{f_j} \tag{73}$$

and determines the irreducible representations of the vibrational mode referred to the Raman band, caused by the scattering via the $f_i$ and $f_j$ moments. It should be noted that the expression (73) is obtained for the first line of the scattering tensor (51). Considering all another lines in the same manner one can convince, that the condition (73) is also the same for them and hence it is valid for the whole scattering tensor. It should be noted, that the relation (73), was obtained in [35] for some reduced form of the scattering tensor. Here it appears that this relation is valid for the modified scattering tensor and presents the selection rules for contributions in (60-63). In addition the selection rules (73) are valid for all contributions in $A_{V_{(s,p)}}$ for all $p$ as it has been mentioned above.

## 7. CLASSIFICATION OF THE CONTRIBUTIONS BY THE DEGREE OF ENHANCEMENT

In accordance with classification of the moments all the contributions (60-63) or, more precisely their separate constituents, associated with certain combinations of the moments, can be classified by the degree of enhancement. It should be noted that this classification is valid for molecules adsorbed on silver and may be also for rough surfaces of gold and copper, with a strong roughness and monolayer coverage, for which our classification of moments is correct. The sequence below reflects the decrease in the degree of enhancement.

1. the strongest, essential $(Q_{main} - Q_{main})$ scattering type;

2. essential $(d_z - Q_{main})$ and $(Q_{main} - d_z)$ scattering types;

3. essential $(d_z - d_z)$ scattering type.

The contributions of the first type must be strongly enhanced, while the contributions of the second and the third types are also enhanced, but apparently to a lesser extent.

4. $(d_{\min or} - Q_{main})$ and $(Q_{main} - d_{\min or})$ scattering types;

5. $(Q_{\min or} - Q_{main})$ and $(Q_{main} - Q_{\min or})$ scattering types;

6. $(d_{\min or} - d_z)$ and $(d_z - d_{\min or})$ scattering types;

7. $(Q_{\min or} - d_z)$ and $(d_z - Q_{\min or})$ scattering types.

These types of the processes may also be enhanced but apparently to a less extent than the first three. The other $(d-d),(d-Q),(Q-d)$ and $(Q-Q)$ contributions, in which both the dipole and quadrupole moments are of the minor type, are apparently small. Therefore we name them as inessential parts in SERS.

## 8. MOST ENHANCED BANDS IN SERS

In accordance with the classification of the moments and classification of the contributions by the degree of enhancement the most enhanced contributions are those, which contain two main quadrupole moments. This statement is valid for the conditions, mentioned above. Then using the transformed contributions (60-63), we can state that the sum of the most enhanced contributions is that caused by the two main quadrupole moments. Then its evident form is given by the expression

$$\sum_{\substack{\alpha,i,\gamma,k \\ i,k-main \\ moments}} a_{\alpha,i} a_{\gamma,k} C_{V_{(s,p)}} [Q_i, Q_k] \left( \frac{1}{2\left|\overline{E}_{scat}\right|} \frac{\partial E_{\alpha}^{scat^*}}{\partial x_{\alpha}} \frac{1}{2\left|\overline{E}_{inc}\right|} \frac{\partial E_{\gamma}^{inc}}{\partial x_{\gamma}} \right)_{surf} \quad (74)$$

Taking into account the selection rules for the contributions (73), and the fact that the main quadrupole moments $Q_i$ transform under the unitary irreducible representation we obtain, that the corresponding $\Gamma_{(s,p)}$ is also unitary. Thus usually for $Ag$ substrates and for the strong roughness the most enhanced bands in SERS for symmetrical molecules are those due to vibrations transforming after the unitary irreducible representation.

# 9. $(d_z - Q_{main}), (Q_{main} - d_z)$ AND $(d_z - d_z)$ CONTRIBUTIONS AND APPEARANCE OF FORBIDDEN BANDS IN THE SER SPECTRA OF SYMMETRICAL MOLECULES

As it has been pointed out the strongly enhanced constituents of the $(d_z - Q_{main}), (Q_{main} - d_z)$ and $(d_z - d_z)$ types should also exist in addition to the strongly enhanced contributions of the $(Q_{main} - Q_{main})$ scattering type. If the molecule belongs to the symmetry group, for which the $d_z$ moment transforms under the unitary irreducible representation, the $(d_z - Q_{main})$ and $(Q_{main} - d_z)$ constituents will cause an additional change in the intensities of the bands associated with vibrations transformed under the unitary irreducible representation. If the $d_z$ moment transforms under another irreducible representation, the constituents of these type will enhance the bands associated with vibrations related to the same representations as the $d_z$ moment. If the contributions of the $(d - d)$ scattering type in these lines are forbidden, the existence of the constituents of the $(d_z - Q_{main})$ type will give rise to strong bands forbidden in the ordinary Raman scattering. In practice such bands are observed in the SER spectra of a great number of symmetrical molecules.

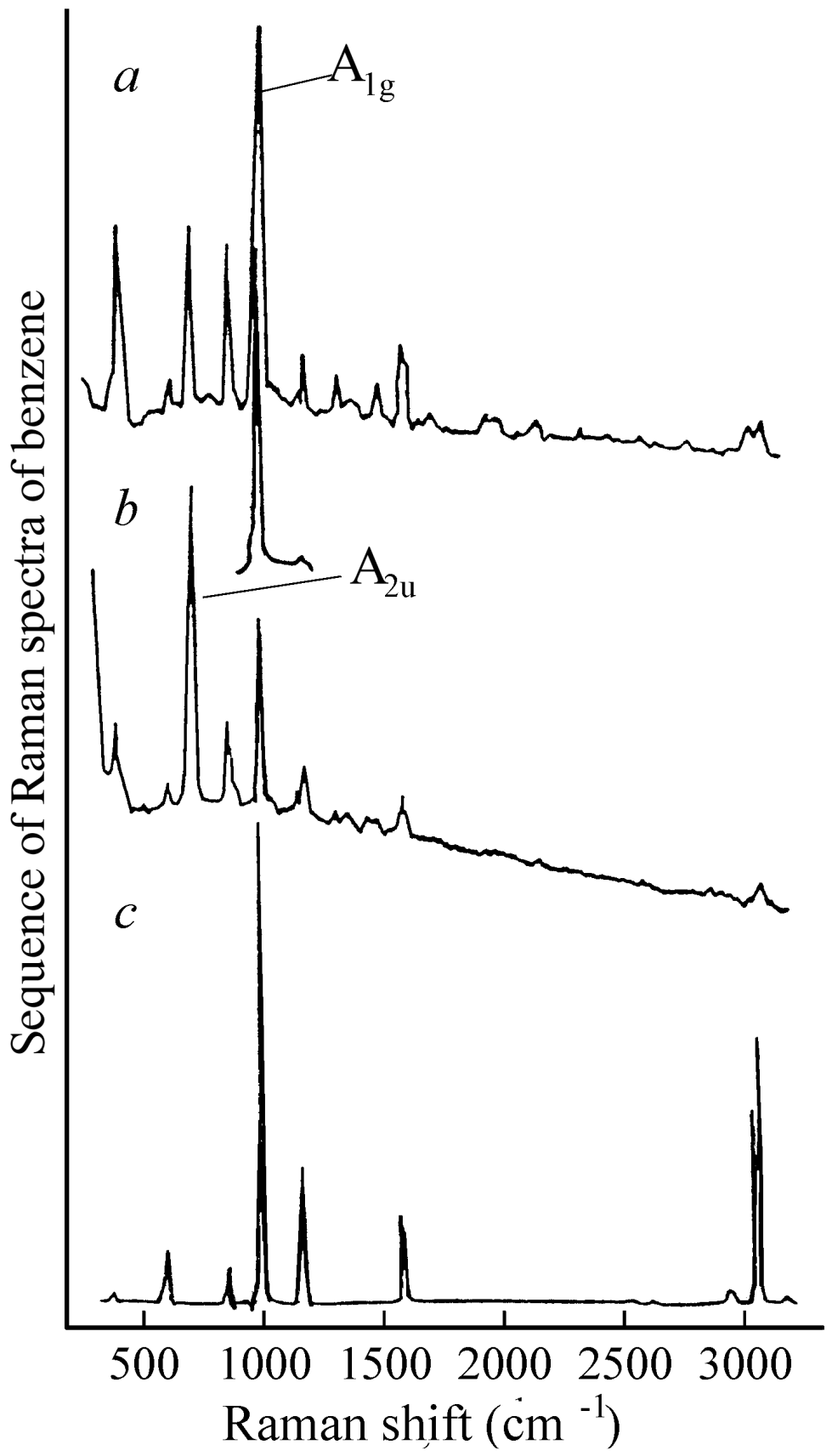

Fig. 10 a .The SER spectrum of benzene on $Ag$, b. The same on $Li$, c. The spectrum of polycrystalline benzene. One can see a large band of a $A_{2u}$ type for benzene adsorbed on lithium, which is much stronger then the breathing mode of $A_{1g}$ type.

It should be noted that as a rule these lines, have a lower intensity than those caused by the $(Q_{main} - Q_{main})$ contributions. However some exceptions exist. For example it is the SER spectra of benzene on lithium [38] (Fig. 10). This fact points out a great number of unknown factors, which form the intensities of spectral lines. Thus the bands caused by the contributions of the $(d_z - Q_{main})$ and $(Q_{main} - d_z)$ scattering types can also be considered as the most enhanced in SERS. Since the $d_z$ moments in the $(d_z - d_z)$ contribution are identical and belong to the same irreducible representation, we can demonstrate that it enhances the bands associated with vibrations transformed under the unitary irreducible representation. It is the reason, that the maximum enhancement of these spectral lines is frequently attributed to the electrical field and to increase of the dipole interaction but not to enhancement of the quadrupole interaction.

## 10. CONTRIBUTIONS CAUSED BY ONE MINOR AND ONE MAIN MOMENTS

The contributions of this type are $(d_{\min or} - Q_{main})$, $(Q_{main} - d_{\min or})$, $(Q_{\min or} - Q_{main})$, $(Q_{main} - Q_{\min or})$, $(d_{\min or} - d_z)$, $(d_z - d_{\min or})$, $(Q_{\min or} - d_z)$ and $(d_z - Q_{\min or})$. As it has been noted above, since one of the optical transitions occurs via the main moment and another via the minor one, these constituents will be enhanced to a significantly lesser extent than those considered earlier. In general case they can contribute to the bands caused by the $(Q_{main} - Q_{main})$ and $(d_z - d_z)$, as well as by the $(d_z - Q_{main})$ and $(Q_{main} - d_z)$ scatterings. It must change to a certain extent the intensities of these lines. In the case when the $(Q_{main} - Q_{main})$, $(d_z - Q_{main})$ $(Q_{main} - d_z)$ and $(d_z - d_z)$ contribute to another lines the contributions discussed may lead to appearance of additional enhanced spectral lines in the SER spectra. If the contributions of the $(d_{\min or} - d_{\min or})$,$(d_{main} - d_{\min or})$ and $(d_{\min or} - d_{main})$ types to these bands are also forbidden, then the remaining constituents discussed must give rise to new enhanced spectral lines, forbidden in the ordinary Raman spectrum. The intensities of these bands related to the most enhanced bands can give a notion about the role of the minor moments in SERS.

## 11. DISTURBANCE OF THE SYMMETRY OF MOLECULES UPON ADSORPTION AND TWO POSSIBLE MECHANISMS OF THE APPEARANCE OF FORBIDDEN BANDS IN THE SER SPECTRA

It should be noted, that the appearance of forbidden bands in the SER spectra is attributed both with the possible disturbance of symmetry upon adsorption and to the quadrupole light-molecule interaction [38]. In our opinion these two mechanisms can be distinguished in some cases and the main mechanism can be identified. For this to be done it is necessary to compare the Raman spectra of molecules interacting with surfaces with different degrees of roughness. The symmetry disturbance is a consequence of the surface-molecule interaction, which must change the power constants and thus the vibrational frequencies of the molecule. However it is well known, that a change in the degree of the roughness does not cause sufficiently strong shifts of the Raman bands. Therefore we can conclude that the surface-molecule interaction is virtually invariable. For example, this conclusion can be verified by analyzing the temperature dependence of the SER spectra of benzene on lithium [38] (Fig. 11).

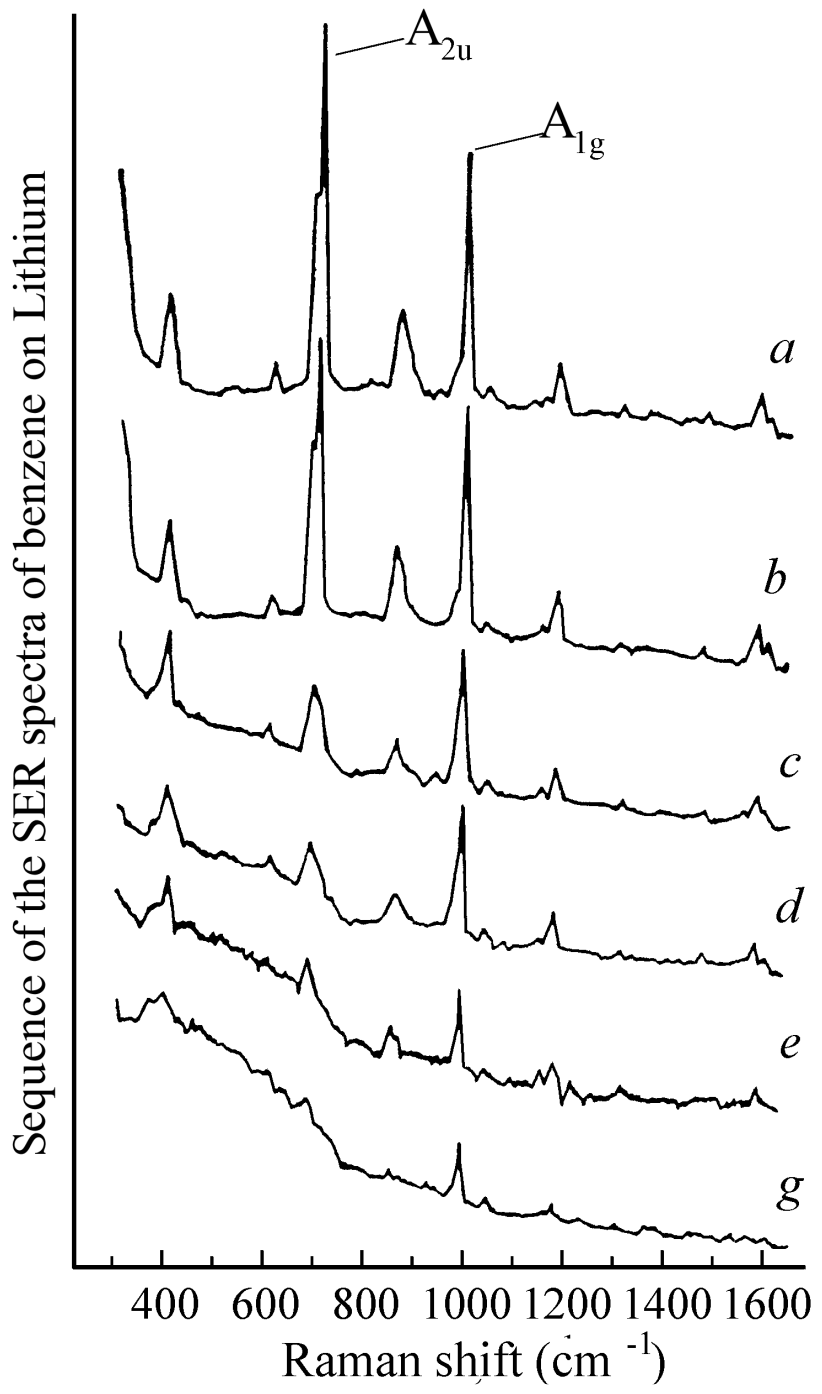

Fig. 11 The sequence of the SER spectra of benzene on lithium at various temperatures; a-12$K$, b-28$K$, c-40$K$, d-72$K$, e-114$K$, g-137$K$ .Annealing causes the decrease in intensity, however the positions of the lines are invariable. Thus the reason of SERS is the quadruple interaction but not the symmetry disturbance.

The increase of the temperature leads to surface annealing and thus changes the roughness degree. However the spectral position of the Raman lines is virtually invariable. This occurs because the condition $a < l_E < \lambda$ is apparently valid in the SERS experiments [38]. Here as usual $a$ is the molecule size, and $l_E$ is the roughness characteristic size as it is designated above. The local environment of most molecules under this condition is like the local environment on the plane surface (in some mean sense). Therefore the large relative change of the intensities of the forbidden bands in [38] can not be attributed to the change of the disturbance of the molecule symmetry, and there remains the second mechanism, which connects the changes in the roughness degree with the increase of the electrical field and its derivatives and thus with the increase of the dipole and quadrupole interactions. This leads us to conclusion, that when the surface-molecule interaction is weak and the Raman bands frequencies are remain almost unchanged upon adsorption, the main reason of the forbidden bands appearance is the quadrupole interaction, but not the symmetry disturbance.

## 12. ANALYSIS OF THE SER SPECTRA

For corroboration of the theory presented it is of great interest to analyze the spectra of symmetrical molecules, when we can neglect the symmetry disturbance, caused by the molecule-surface interaction. Usually this situation occurs for physisorbed molecules with the spectral position of the Raman lines being almost invariable with respect to that for molecules in a free state. It can arise in chemisorbed molecules when the molecule symmetry remains the same as that in a free state (e.g. in vertical chemisorbed pyridine). Of some interest is the case of symmetry preservation in chemisorbed phthalimide and potassium phthalimide (PIMH and PIMK), when upon loss H or K respectively, interacting with substrate via nitrogen, the corresponding anion preserves the $C_{2v}$ symmetry. In order to exclude the influence of the intermolecular interaction on the SER spectra, we try to analyze only those situations, in which the substrate coverage by molecules is sufficiently small and equal to several Langmuir units.

Ethylene - $C_2H_4$. The molecule has a plane geometry and belongs to the $D_{2h}$ group. It is physisorbed on $Ag$ and $Cu$ at low temperatures without any significant symmetry disturbance [5] with its plane parallel to the substrate, forming weak $p$ bonds. The corresponding spectrum is shown in (Fig. 12) [5].

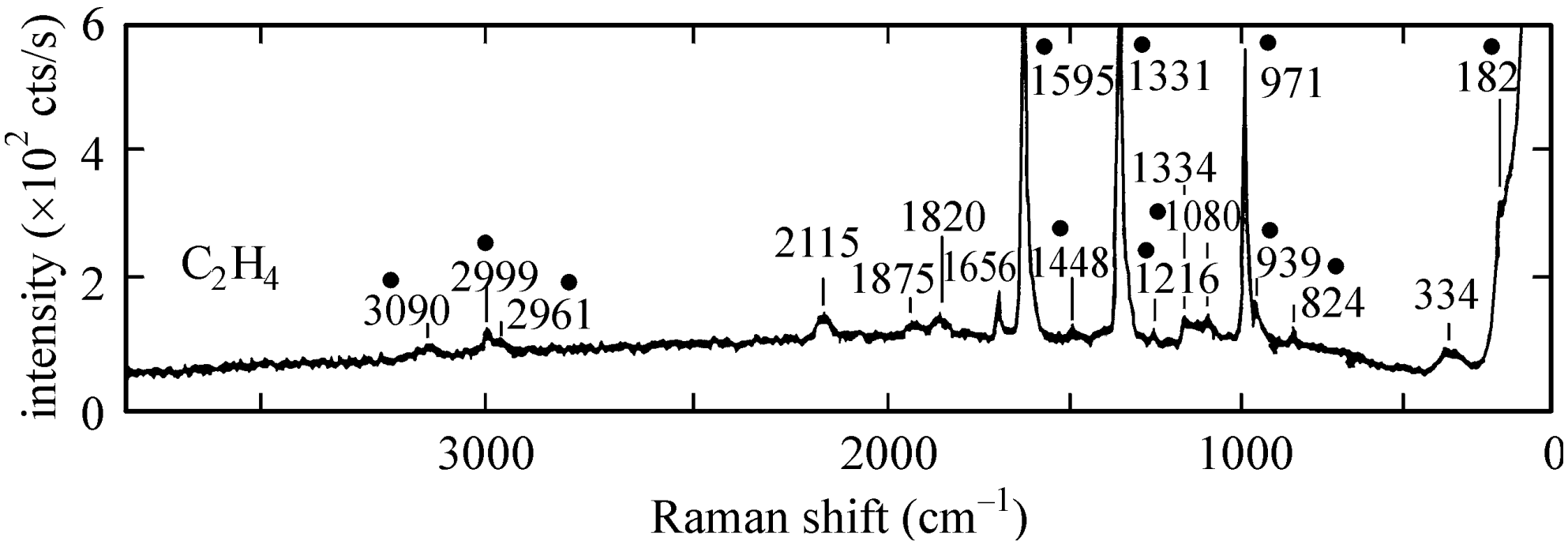

Fig. 12 SER spectra from coldly evaporated silver films exposed to 36 Langmuir of ethylene. Dots mark ethylene skeletal vibrations. 200 mw of $514.5 nm$ radiation, 4.5 cm bandpass [5].

The small deviation of the vibrational frequencies from their values for a free molecule can serve as a criterion of the weakness of the symmetry disturbance. The contributions of various scattering types in the SER bands of $C_2H_4$ are listed in Table I.

Table 1. Part 1 Contributions of various scattering types in the SER bands of ethylene $C_2H_4$ adsorbed onto $Ag$. The sequence of contributions approximately corresponds to that of the degrees of enhancement.

| Mode Symmetry | Mode number | Wavenumber $(cm^{-1})$ | Inessential part in SERS | | | $d_{\min or} - d_z$ $d_z - d_{\min or}$ | $Q_{\min or} - d_z$ $d_z - Q_{\min or}$ |
|---|---|---|---|---|---|---|---|
| | | | $(Q-Q)$ | $(d-Q)$ $(Q-d)$ | $(d-d)$ | | |
| $A_g$ | 3 | 1331 | $Q_{xy}Q_{xy}$ $Q_{xz}Q_{xz}$ $Q_{yz}Q_{yz}$ | | $d_xd_x$ $d_yd_y$ | | |
| $A_g$ | 2 | 1595 | | | | | |
| $A_g$ | 1 | 2999 | | | | | |
| $A_u$ | 4 | 1080 | | $d_xQ_{yz}$ $d_yQ_{xz}$ | | | $d_zQ_{xy}$ |
| $B_{1g}$ | 6 | 1216 | $Q_{xz}Q_{yz}$ | | $d_xd_y$ | | |
| $B_{1g}$ | 5 | 3090 | | | | | |
| $B_{1u}$ | 7 | 971 | | $d_xQ_{xz}$ $d_yQ_{yz}$ | | | |
| $B_{2g}$ | 8 | 939 | $Q_{xy}Q_{yz}$ | | | $d_xd_z$ | |
| $B_{2u}$ | 10 | 824 | | $d_xQ_{xy}$ | | | $d_zQ_{yz}$ |
| $B_{2u}$ | 9 | 3090 | | | | | |
| $B_{3u}$ | 12 | 1448 | | $d_yQ_{xy}$ | | | $d_zQ_{xz}$ |
| $B_{3u}$ | 11 | 2961 | | | | | |

Table 1. Part 2 Contributions of various scattering types in the SER bands of ethylene $C_2H_4$ adsorbed onto $Ag$ (This part is continuation of the Part 1). The sequence of contributions approximately corresponds to that of the degrees of enhancement.

| $Q_{\min or} - Q_{main}$ <br> $Q_{main} - Q_{\min or}$ | $d_{\min or} - Q_{main}$ <br> $Q_{main} - d_{\min or}$ | Essential part in SERS | | | Relative Intensity (%) |
|---|---|---|---|---|---|
| | | $d_z - d_z$ | $d_z - Q_{main}$ <br> $Q_{main} - d_z$ | $Q_{main} - Q_{main}$ | |
| | | $d_z d_z$ | | $(Q_{xx}, Q_{yy}, Q_{zz})$ | 100 |
| | | | | $(Q_{xx}, Q_{yy}, Q_{zz})$ | 69 |
| | | | | | 2 |
| | | | | | 2 |
| $Q_{xy}(Q_{xx}, Q_{yy}, Q_{zz})$ | | | | | 2 |
| | | | | | 2 |
| | | | $d_z(Q_{xx}, Q_{yy}, Q_{zz})$ | | 25 |
| $Q_{xz}(Q_{xx}, Q_{yy}, Q_{zz})$ | | | | | 2 |
| | $d_y(Q_{xx}, Q_{yy}, Q_{zz})$ | | | | 2 |
| | | | | | 2 |
| | $d_x(Q_{xx}, Q_{yy}, Q_{zz})$ | | | | 2 |
| | | | | | 2 |

It can be seen, that two bands at $\nu_3 = 1331 cm^{-1}$ and $\nu_2 = 1595 cm^{-1}$ associated with the vibrations transforming under the unitary irreducible representation $A_g$ and with the strong $(Q_{main} - Q_{main})$ scattering via the main moments $Q_1 = Q_{xx}, Q_2 = Q_{yy}$ and $Q_3 = Q_{zz}$ of the $D_{2h}$ group are enhanced to the strongest extent. The sufficiently weak enhancement of the band $\nu_1 = 2992 cm^{-1}$ ($A_g$ symmetry) is due to the following: within the limits of the bands group, determined by vibrations of the same symmetry, the band caused by the breathing mode will be most enhanced. This apparently results from the fact of the largest deformation of the electron shell of the molecule for this band and largest values of the set of $R_{nl(s,p)}$ and $R_{mk(s,p)}$ coefficients and their conjugated values. For other totally symmetric vibrations, the deformation is weaker, that cause the lesser values of the above coefficients and the lesser enhancement of these bands. However the deformation and the absolute values of the coefficients $R_{nl(s,p)}$, $R_{mk(s,p)}$ and their conjugated values for the breathing mode are determined by numerous factors dependent on atoms, which constitute the molecule. Therefore this rule can be violated in some particular cases. A correct explanation of the above fact can be only based on numerical methods. However the complexity of the problem gives no way of performing such calculations at present. It should be noted that there exists a rather strong band $\nu_7 = 971 cm^{-1}$ associated with the vibration with the $B_{1u}$ symmetry. It is forbidden in the ordinary Raman scattering in the dipolar approximation. Its appearance in the SER spectrum of ethylene can be explained by the essential $(d_z - Q_{main})$ and $(Q_{main} - d_z)$ scattering types. In our opinion its significantly stronger intensity with respect to the intensities of the $B_{2u}$ and $B_{3u}$ bands associated with the $(d_{\min or} - Q_{main})$, $(Q_{main} - d_{\min or})$, $(d_z - Q_{\min or})$ and $(Q_{\min or} - d_z)$ scattering types is due to the scattering just via the two main moments. Moreover the small frequency shift of the band under discussion from its value for the molecules in the gaseous phase and its relatively large intensity with respect to the most enhanced lines, indicate in our opinion, the strong light-molecule dipole and quadrupole interactions, but not the symmetry disturbance. It should be noted, that the small intensity ~2% of the $B_{2u}$ and $B_{3u}$ spectral lines associated with the above mentioned constituents and the intensity of other bands, determined exclusively by the scattering types with the minor moments points out a negligible role of the latter in the SER spectrum of ethylene. Let us explain some designations in Table 1 which are also used in further similar tables. In the rows the moments combinations that contribute to the scattering band are listed. The combination $f_i f_k$ designates two possible combinations $f_i f_k$ and $f_k f_i$. Similarly, simplified, a note of type $f_i(f_k, f_l)$ means $f_i f_k$, $f_i f_l$ $f_k f_i$ $f_l f_i$ and that of type $(f_i f_k)(f_m f_l)$ means: $f_i f_m$ $f_i f_l$, $f_k f_m$, $f_k f_l$, $f_m f_i$, $f_l f_i$, $f_m f_k$, $f_l f_k$.

Pyridine $C_5H_5N$, adsorbed on $Cu, Ag$ and $Au$ [5] has a plain geometry and belongs to the $C_{2v}$ symmetry group. The corresponding type of a spectrum is shown in (Fig. 13) [5]. It can be adsorbed in two ways:

1. physisorbed with its plane parallel to the substrate;
2. chemisorbed by the edge of the molecule to the surface, being bound to metal atoms of the substrate via the nitrogen lone-pair. As in the case of ethylene a small change in the vibrational frequencies can serve as a criterion of weak symmetry disturbance. The contributions of the different scattering types to the SER bands of pyridine adsorbed on $Ag$ and $Cu$ are listed in Table II. In accordance with observation conditions and the values $\nu_1$ and $\nu_3$ these data refer to the vertical chemisorbed pyridine [5]. It can be seen, that the bands assigned to the modes, which transform under the unitary irreducible representation $A_1$ are enhanced to the strongest extent. In particular the strongest is the band at $\nu_1 = 1005cm^{-1}$ caused by the mode "close" to the breathing one (as in ethylene it can be associated mainly with the largest values $R_{nl(s,p)}, R_{mk(s,p)}$ and their conjugated values. The bands assigned to the high frequency vibrations of the same symmetry $\nu_{10}, \nu_2, \nu_7$ are virtually absent. Here as in ethylene the values $R_{nl(s,p)}$, $R_{mk(s,p)}$ and their conjugated values are apparently small for these lines. It should be noted that the $(d_z - Q_{main})$ $(Q_{main} - d_z)$ and $(d_z - d_z)$ scatterings contribute to the bands due to the vibrations with the $A_1$ symmetry, and thus additionally enhance them.

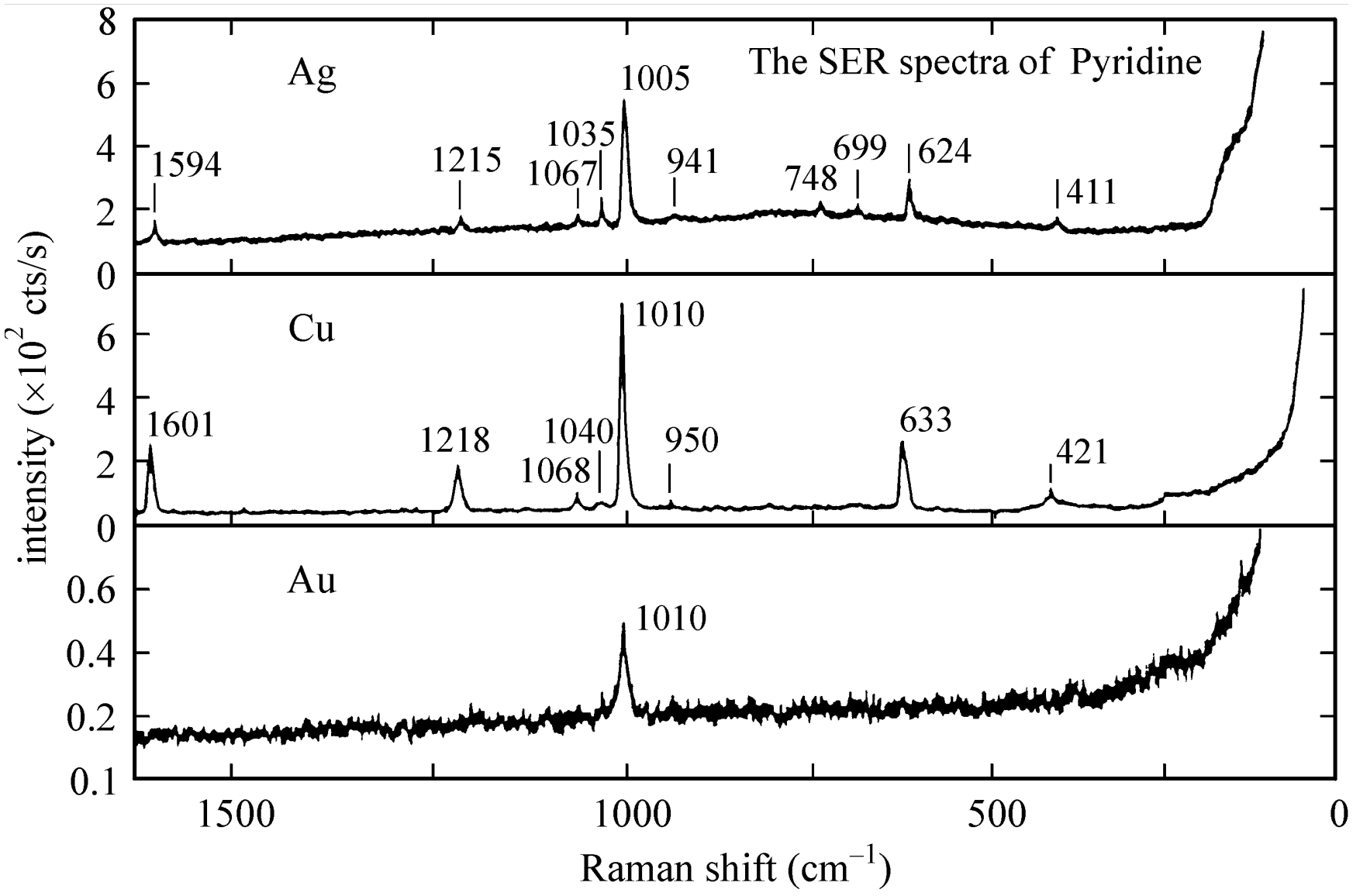

Fig. 13 SER spectra from pyridine on coldly evaporated $Ag$ (0.2 Langmuir), $Cu$ (2 Langmuir) and $Au$ (2 Langmuir) films. 60 mw of $674.6nm$ radiation, $4cm^{-1}$ bandpass. $Au$ film has been warmed to $210K$ and recooled to $120K$ before measurement. The data demonstrate the strongest enhancement of the breathing mode of pyridine

Table 2 Part 1. Contributions of various scattering types in the SER bands of pyridine, adsorbed onto $Ag$ and $Cu$. The sequence of contributions approximately corresponds to that of the degrees of enhancement.

| Mode Symmetry | Mode Number | $(Q-Q)$ | $(d-Q)$<br>$(Q-d)$ | $(d-d)$ | $d_{\min or} - d_z$<br>$d_z - d_{\min or}$ | $Q_{\min or} - Q_z$<br>$d_z - Q_{\min or}$ | $Q_{\min or} - Q_{main}$<br>$Q_{main} - Q_{\min or}$ |
|---|---|---|---|---|---|---|---|
| $A_1$ | 3<br>1<br>6<br>8<br>5<br>9<br>4<br>10<br>2<br>7 | $Q_{xy}Q_{xy}$<br>$Q_{xz}Q_{xz}$<br>$Q_{yz}Q_{yz}$ | $d_xQ_{xz}$<br>$d_yQ_{yz}$ | $d_xd_x$<br>$d_yd_y$ | | | |
| $A_2$ | 21<br>20<br>22 | $Q_{xz}Q_{yz}$ | | $d_xd_y$ | | $Q_{xy}d_z$ | $Q_{xy}(Q_{xx},Q_{yy},Q_{zz})$ |
| $B_1$ | 12<br>17<br>16<br>11<br>15<br>18<br>14<br>13<br>19 | $Q_{xy}Q_{yz}$ | $d_yQ_{xy}$ | | $d_xd_z$ | | $Q_{xz}(Q_{xx},Q_{yy},Q_{zz})$ |
| $B_2$ | 27<br>26<br>23<br>25<br>24 | $Q_{xy}Q_{xz}$ | | | $d_yd_z$ | $Q_{yz}d_z$ | $Q_{yz}(Q_{xx},Q_{yy},Q_{zz})$ |

Table 2 Part 2 Contributions of various scattering types in the SER bands of pyridine, adsorbed onto $Ag$ and $Cu$ (This part is continuation of the part 1). The sequence of contributions approximately corresponds to that of the degrees of enhancement.

| $d_{\min or} - Q_{main}$ $Q_{main} - d_{\min or}$ | $d_z - d_z$ | $d_z - Q_{main}$ $Q_{main} - d_z$ | $Q_{main} - Q_{main}$ | Wavenumber $(cm^{-1})$ $Ag$ 0,2$L$ | Relative Intencity (%) | Wavenumber $(cm^{-1})$ $Cu$ 2$L$ | Relative Intencity (%) |
|---|---|---|---|---|---|---|---|
| | $d_z d_z$ | $d_z(Q_{xx},Q_{yy},Q_{zz})$ | $(Q_{xx},Q_{yy},Q_{zz})$ $(Q_{xx},Q_{yy},Q_{zz})$ | 623 | 20 | 633 | 32 |
| | | | | 1006 | 100 | 1010 | 100 |
| | | | | 1037 | 20 | 1040 | 4 |
| | | | | 1069 | 6 | 1068 | 8 |
| | | | | 1215 | 59 | 1218 | 21 |
| | | | | 1480 | 2 | 1485 | 2 |
| | | | | 1593 | 59 | 1601 | 33 |
| | | | | 3033 | 4 | | |
| | | | | 3061 | 1 | | |
| | | | | | | | |
| | | | | 380 | 2 | | |
| | | | | | | | |
| | | | | 972 | 2 | | |
| $d_x(Q_{xx},Q_{yy},Q_{zz})$ | | | | 652 | 2 | | |
| | | | | | | | |
| | | | | 1150 | 3 | | |
| | | | | | | | |
| | | | | 1355 | 1 | | |
| | | | | 1444 | 1 | | |
| | | | | 1572 | 4 | 1573 | 2 |
| | | | | | | | |
| | | | | | | | |
| $d_y(Q_{xx},Q_{yy},Q_{zz})$ | | | | 413 | 3 | 421 | 9 |
| | | | | 696 | 3 | 701 | 2 |
| | | | | 749 | 4 | | |
| | | | | 880 | 1 | | |
| | | | | 942 | 4 | 950 | 2 |

Apparently the strong SERS in chemisorbed pyridine is due to this fact. It should be noted that a part of pyridine molecules can be physisorbed with their plane parallel to the substrate. Because the condition $divE = 0$ takes place, all the $\partial E_\alpha / \partial x_\alpha$ derivatives will be of the same order of magnitude. Therefore the bands transforming under the unitary irreducible representation slightly change their intensity due to existence of such molecules. However, because the $d_y$ moment becomes perpendicular to the surface, the bands with the $B_2$ symmetry will be additionally enhanced. As can be seen from Table II, the spectral line $\nu_{27}$ associated with this symmetry type has such a large intensity ~9% with respect to $\nu_1$. Its large intensity can be explained by the existence of such physisorbed molecules. The intensities of the bands, caused by the vibrations of the remaining symmetry types are weak. This points out the sufficiently inessential role of the minor moments in the SER spectrum of pyridine.

Phtalimide and potassium phthalimide. In [39] Aroca et al. published the results of detailed studies of the SER spectra of phthalimide (PIMH) and potassium phthalimide (PIMK) belonging to the $C_{2v}$ symmetry group (Fig. 14,15).

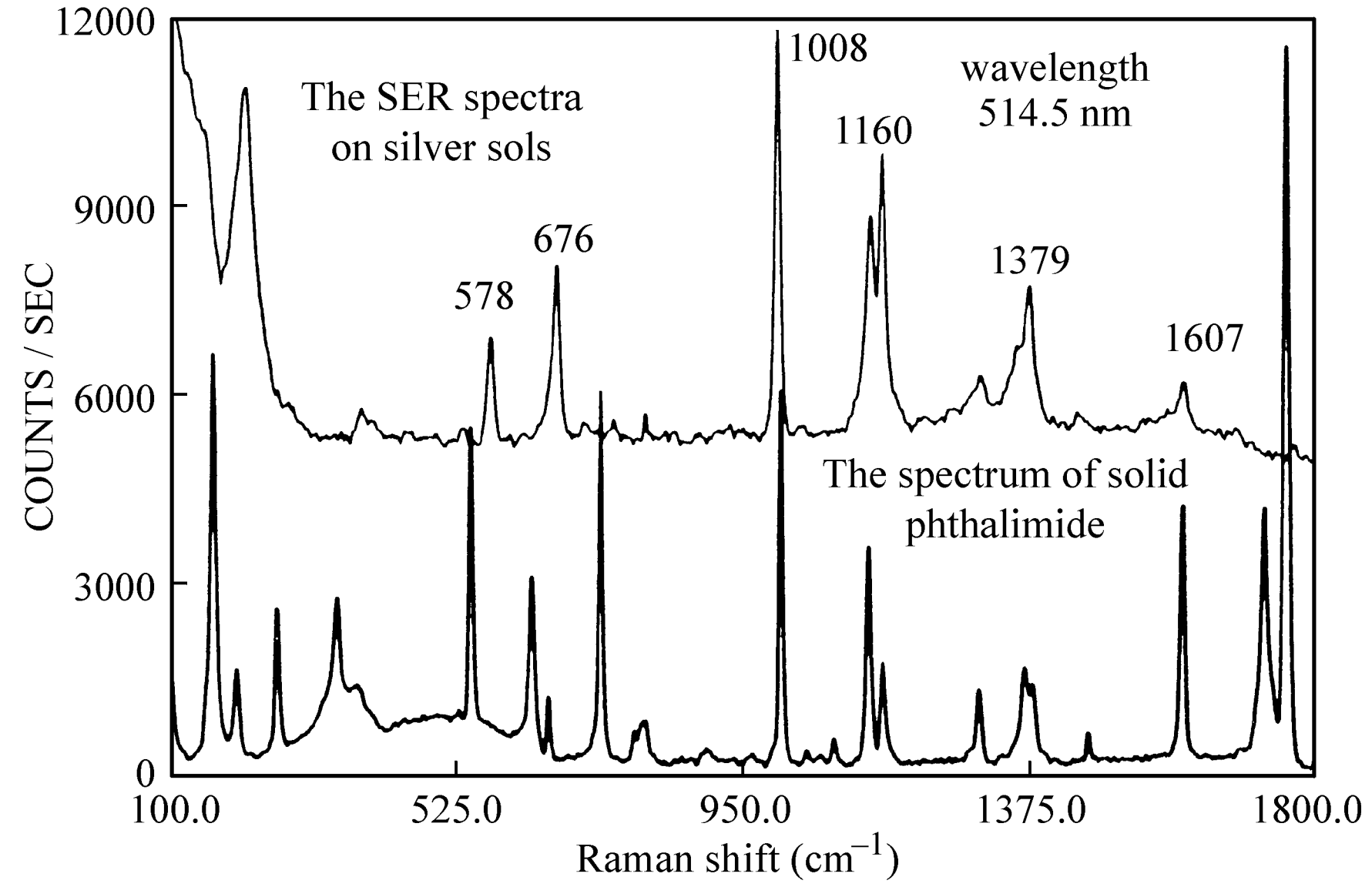

Fig. 14 Raman scattering of solid phthalimide (lower trace) and SERS (upper trace) on Carey-Lea sols. 50 mw of 514.5$nm$ laser line and spectral bandpass of 4 $cm^{-1}$

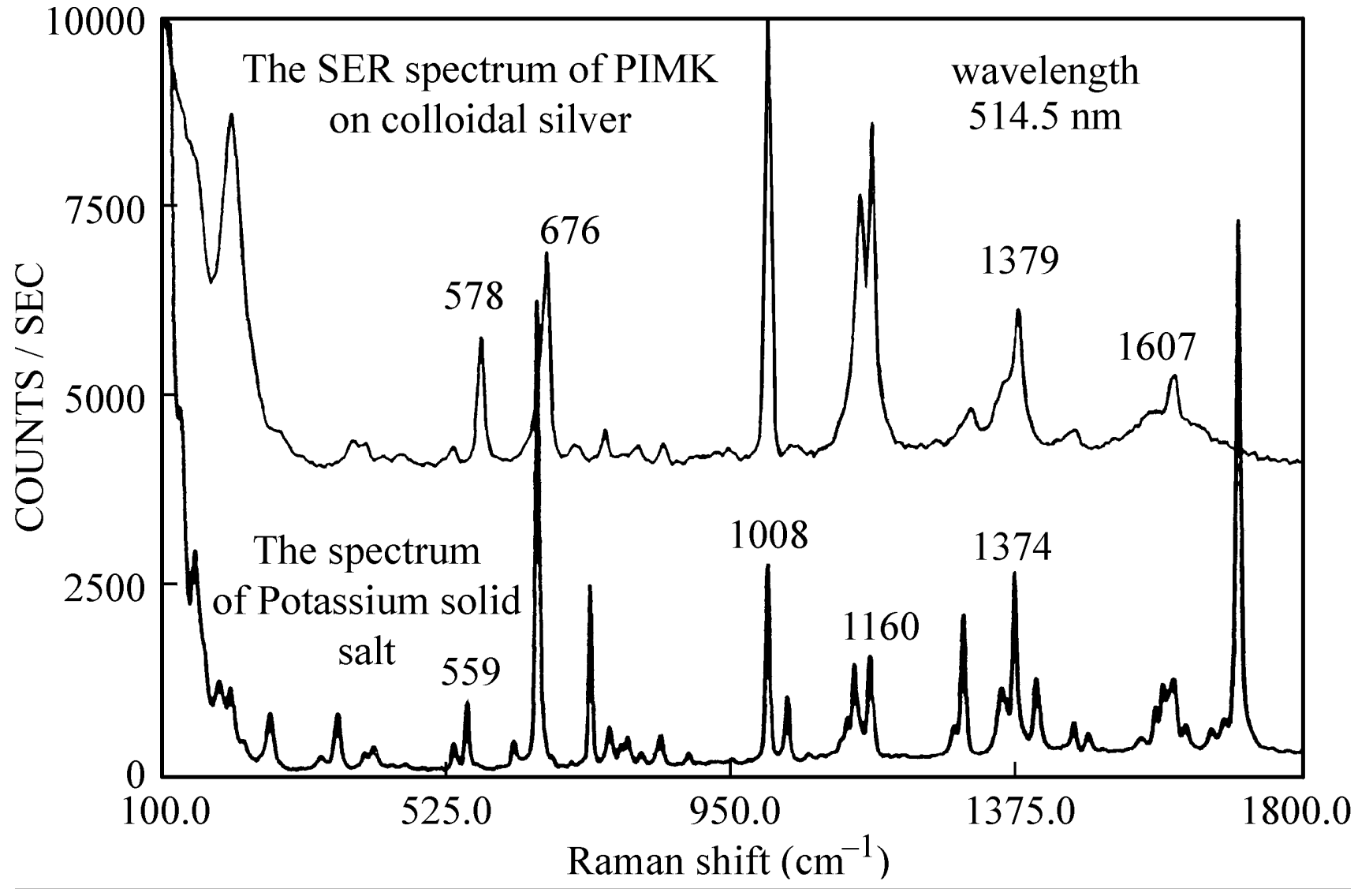

Fig. 15 Raman scattering of solid PIMK (lower trace) and SERS (upper trace) of $10^{-5}$ M solution on Carey-Lea silver sols.

Both molecules have a large number of vibrations. Unfortunately, the data of [39] are insufficient for estimating the number of adsorbed layers and thus revealing all the possible orientations of these molecules with respect to the substrate. However the main property of the SER spectra is the strongest enhancement of the bands, belonging to the vibrations transforming under the unitary irreducible representation, in accordance with Table 3. Furthermore the strong band at $\nu = 1607 cm^{-1}$ which can be assigned to vibrations with the $B_2$ symmetry was observed. If we consider that the coverage is of monolayer character and there are two possible orientations of the molecule with respect to the surface, as in the case of pyridine, then we can attribute its appearance to the essential $(d_z - Q_{main})$ and $(Q_{main} - d_z)$ scattering types from undissociated molecules physisorbed with their plane parallel to the substrate. The function of the $d_z$ moment in the coordinate system associated with the surface will be executed by the $d_y$ component of the dipole moment of the molecule. The existence of this band in the ordinary Raman spectra of solid PIMH and PIMK confirms that this line belongs to the undissociated molecules.

Table 3. Raman shifts of PIMH, PIMK and PIM on $Ag$. One can see the strongest enhancement of the totally symmetric vibrations

| PIMH (sol) Wavenumber $(cm^{-1})$ | PIMK (sol) Wavenumber $(cm^{-1})$ | SERS (PIM $Ag$) Wavenumber $(cm^{-1})$ | INTERPRETATION OF VIBRATIONS |
|---|---|---|---|
| | | 214 | Ag-N stretch |
| | 405 | 392 | $A_1$ |
| 550 | 558 | 578 | $A_1$ ring def |
| | 668 | 676 | $A_1$ ring def |
| 1012 | 1008 | 1008 | $A_1$ |
| 1143 | 1140 | 1144 | $A_1$ ring stretch |
| 1165 | 1161 | 1160 | $A_1$ C-H bend |
| 1373 | | 1379 | $A_1$ ring stretch |
| | 1581 | 1572 | $A_1$ |
| 1607 | 1607 | 1607 | $B_2$ ring stretch |
| 1756 | 1706 | | $A_1$ C=O stretch |
| 3057 | | 3057 | $A_1$ C-H stretch |
| 3074 | 3066 | 3078 | $A_1$ C-H stretch |

## 13. ANALYSIS OF THE SER SPECTRA OF MOLECULES WITH $D_{3h}$ AND $D_{6h}$ SYMMETRY GROUPS

In the preceding paragraph we considered theoretically the main regularities of the SER spectra of symmetrical molecules and corroborated the theoretical results by analysis of the SER spectra of symmetrical molecules with $D_{2h}$ and $C_{2v}$ symmetry groups. It is of great interest to confirm our theoretical conclusions by experimental data obtained for molecules with a higher symmetry [40]. Here we analyze the work of M. Moskovits et. al. [27] who considered the SER spectra of cyclic aromatic molecules adsorbed on silver. We consider the SER spectra of benzene 1,3,5 trideutereobenzene, , hexafluorobenzene, 1,3,5 trifluorobenzene, which belong to the $D_{3h}$ and $D_{6h}$ symmetry groups. The analysis of Raman spectra of the same molecules published in [27] is not performed because of lack of data about experimental conditions. The corresponding material is presented in Figs. 16-18 and Tables 4,5.

The main quadrupole moments for these groups are $Q_1 = Q_{xx} + Q_{yy}$, $Q_2 = Q_{zz}$. Due to a large coverage of substrate in these experiments (200 Langmuir) and to the multilayer character of adsorbed molecules and their possible arbitrary orientation in higher layers with respect to the electric field, all the $d$ moments may be essential for SERS. As it follows from experimental data for all these molecules, the bands belonging to the totally symmetric vibrations $A_{1g}$ (in $D_{6h}$ group) and $A_1^{'}$ (in $D_{3h}$ group) experience a strong enhancement. Besides the most enhanced bands in these molecules are the ones associated with the breathing vibration. This is apparently due to the largest values of the coefficients $R_{nl(s,p)}$ and $R_{mk(s,p)}$ for these bands and as a consequence the strongest deformation of the electron shell by the breathing mode. The only exception is hexafluorobenzene (Fig. 17), for which the most enhanced band is associated with $C-F$ symmetric stretching vibration. This may be due to a large electron density on fluorine atoms. Therefore the deformation of the electron shell and the $R_{nl(s,p)}$ and $R_{mk(s,p)}$ coefficients for this mode may exceed those for the breathing mode. In addition “interference” of the scattering contributions may be another reason. The total intensity of spectral lines is formed by the constituents (60-63), which contribute to the same line in accordance with the selection rules. These contributions contain various values of the components of the electromagnetic field derivatives, which are, in fact, random coefficients. Therefore, after summation, the total intensity of the spectral line may vary with the experimental conditions. This effect is a certain "interference" of the contributions. It may be the most pronounced for the lines, caused by the totally symmetric vibrations, or the lines that include the scattering via the $Q_{main}$ moment and arises mainly from the fact, that $div\overline{E} = 0$ and the derivatives $\partial E_{\alpha} / \partial x_{\alpha}$ with different $\alpha$ can have different signs (+ or - ). This can lead to a competition of the intensities of different lines. The limiting case of such "interference" effects is the electrodynamic forbiddance of the quadrupole SERS mechanism in molecules with cubic symmetry groups [12], for which the sums of contributions, containing the main quadrupole moments are exactly zero. This phenomenon will be described for SERS later. Another reason for such an anomaly will be discussed later too. Thus the hierarchy of the enhancement of various lines can change, depending on the values of the fields and their derivatives and experimental conditions, and may not correspond to the classification of the contributions presented above. In particular, the "interference" can apparently account for the strongest enhancement of the $A_{2u}$ band of benzene on lithium [38] (Fig. 10). In the case of hexafluorobenzene the "interference" may lead to the main enhancement of the $A_{1g}$ band associated with C-F symmetric stretching vibration. By the analysis of the SER spectra one can make a conclusion that the bands, due to the vibrations, transforming as the $d_z$ moment (of $A_2^{''}$ and $A_{2u}$ symmetry types) and caused by $(Q_{main} - d_z)$, $(d_z - Q_{main})$ scattering types experience enhancement too. In spite of this fact is not seen in Table 5 for hexafluorobenzene, it can be seen in Fig. 17. Because of overlapping with a pronounced background it is difficult to estimate the amplitude of this band. Let us consider why other bands appear in the SER spectra of the considered molecules. Because there are a number of factors, which determine the intensities of the Raman lines, such as $R_{nl(s,p)}$ and $R_{mk(s,p)}$ coefficients and their conjugated values, the main type of the scattering, the values of the fields and their derivatives and also influence of the interference effects, we consider only the fact of appearance of the bands allowed in the dipole and quadrupole approximation. It should be noted that a similar analysis was made in [27]. The authors analyzed the SER spectra under assumption of the influence of the "field gradient" mechanism or the quadrupole interaction. However their opinion about a strong value of this interaction differs from our point of view. They attribute the enhancement to the plasmon excitation in spherical particles and to the increase of the field derivatives only. As it has been demonstrated above, from our opinion the large value of the quadrupole interaction is both due to the large values of the matrix elements of the main quadrupole moments $Q_i$ and to the increase of the field derivatives. Therefore we attribute the enhancement of the totally symmetric bands to the strong $(Q_{main} - Q_{main})$ scattering type, whereas in [27] it is accounted for by the strong $(d_z - d_z)$ scattering type. Analysis of the SER spectra of 1,3,5 trideuteriobenzene and 1,3,5 trifluorobenzene (Tables 4,5) shows that, in accordance with our theory, the bands, having the enhanced contributions with $Q_{main}$ and $d_{main}$ moments exist in the SER spectra. They are of the $A_1^{'}, A_2^{'}, A_2^{''}, E^{'}$ and $E^{''}$ symmetry types. The bands with $A_2^{'}$ and $A_2^{''}$ symmetry are forbidden in the ordinary Raman scattering and appear under the influence of the quadrupole interaction. In addition the band $A_1^{''}$, which is forbidden in all types of the scattering is not observed in the spectra at all, in agreement with our theory. Similarly, the following bands are enhanced in the SER spectra of benzene (Table 4): $A_{1g}, E_{1g}, E_{2g}, A_{2u}, B_{1u}, B_{2u}, E_{1u}, E_{2u}$. Analysis of the contributions, caused at least by one of the main moments: $Q_{xx} + Q_{yy}$, $Q_{zz}$ and $d_z$ shows that the following bands must be enhanced in accordance with the theory: $A_{1g}, E_{1g}, E_{2g}, A_{2u}, E_{1u}, E_{2u}$.. The enhancement of the $B_{1u}$ and $B_{2u}$ bands may be explained by $(d_x, d_y)$ $(Q_{xx} - Q_{yy}, Q_{xy})$ types of scattering. It should be remind that the $d_x$ and $d_y$ moments can be considered as the main moments for large coverage of the substrate. Therefore the appearance of $B_{1u}$ and $B_{2u}$ bands can be understood. It should be noted however, that such type of the scattering as $(d_x, d_y)$ $(Q_{xz}, Q_{yz})$ which contributes to the $E_{2u}, A_{2u}$ and $A_{1u}$ bands is not strongly enhanced because of the absence of the $A_{1u}$ line in the spectrum. This anomaly is apparently associated with the negligible role of the $Q_{xz}$ and $Q_{yz}$ moments in optical transitions in benzene. The bands $E_{1u}, E_{2u}, A_{2u}, B_{1u}$ and $B_{2u}$ are forbidden in the ordinary Raman scattering. Thus the SER spectrum of benzene demonstrates a number of forbidden lines. It should be noted that all the observed lines correspond to the dipole and quadrupole electromagnetic mechanism of the enhancement. For the hexafluorobenzene molecule (Fig. 17, Table 5) the situation is poorer. Only the bands of the $A_{1g}, E_{1g}, A_{2u}, B_{1u}, B_{2u}, E_{1u}$, and $E_{2u}$ symmetry are observed. This fact is apparently associated with concrete experimental conditions. In this case the analysis does not differ fundamentally from that for benzene. A number of lines forbidden in the ordinary Raman scattering can be seen, which is due to the dipole and quadrupole electrodynamic mechanisms of the enhancement. All the bands observed can be accounted for by these mechanisms.

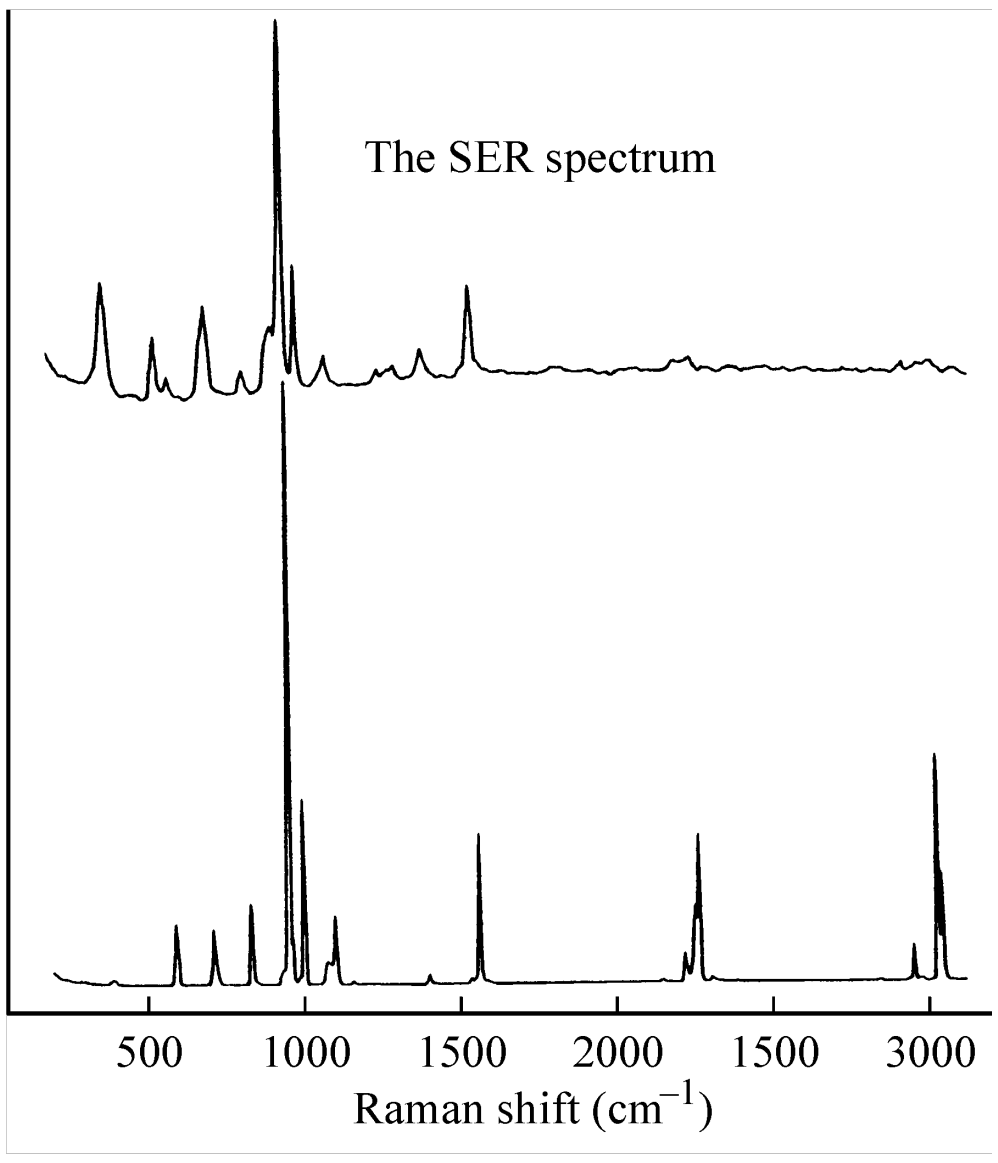

Fig. 16 SERS and Raman spectra of 1,3,5-trideuteriobenzene

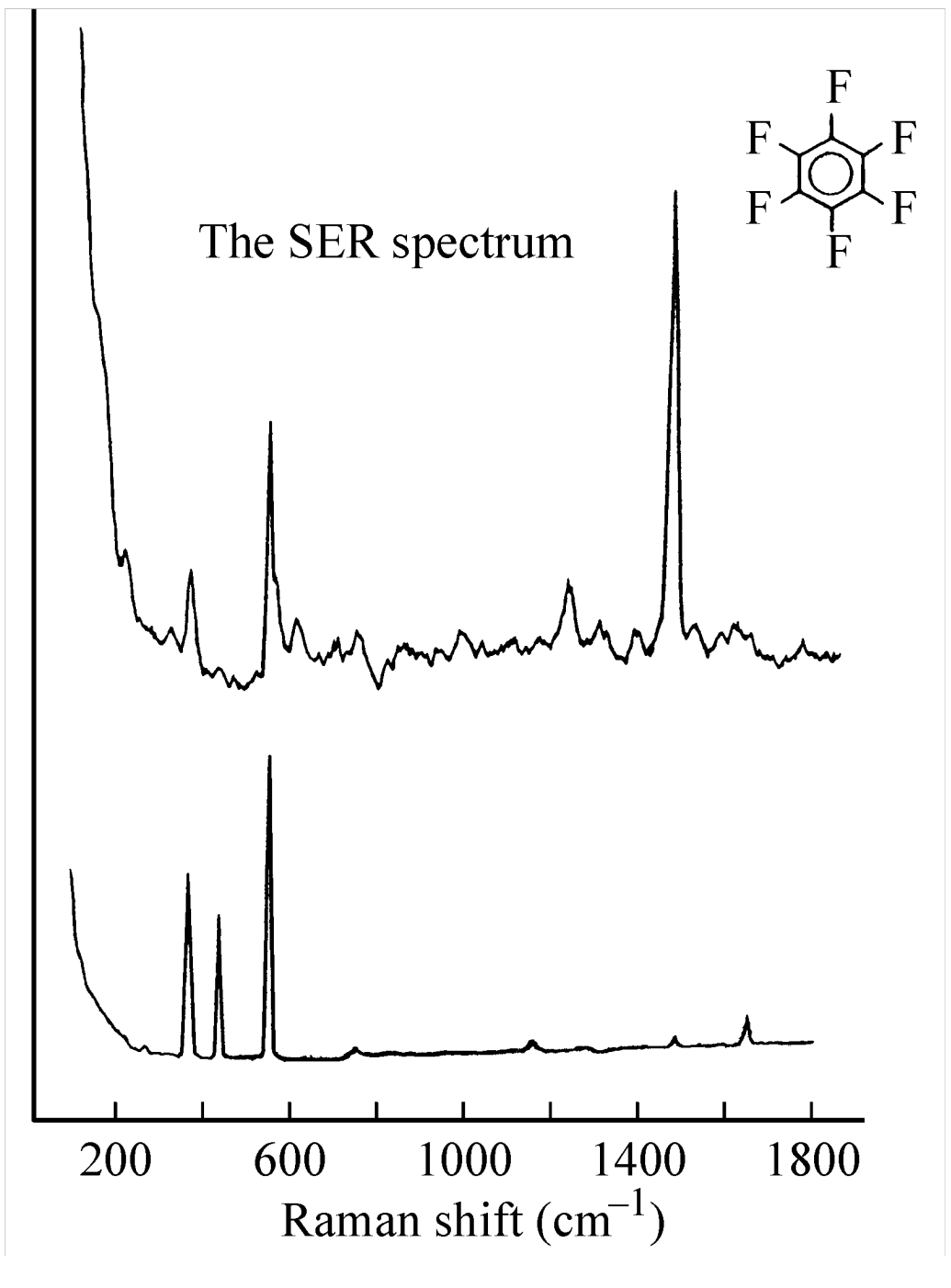

Fig. 17 SERS spectrum of~200 Langmuir of hexafluorobenzene and Raman spectrum of a thick polycrystalline sample of the same molecule, both deposited on silver at 15K.

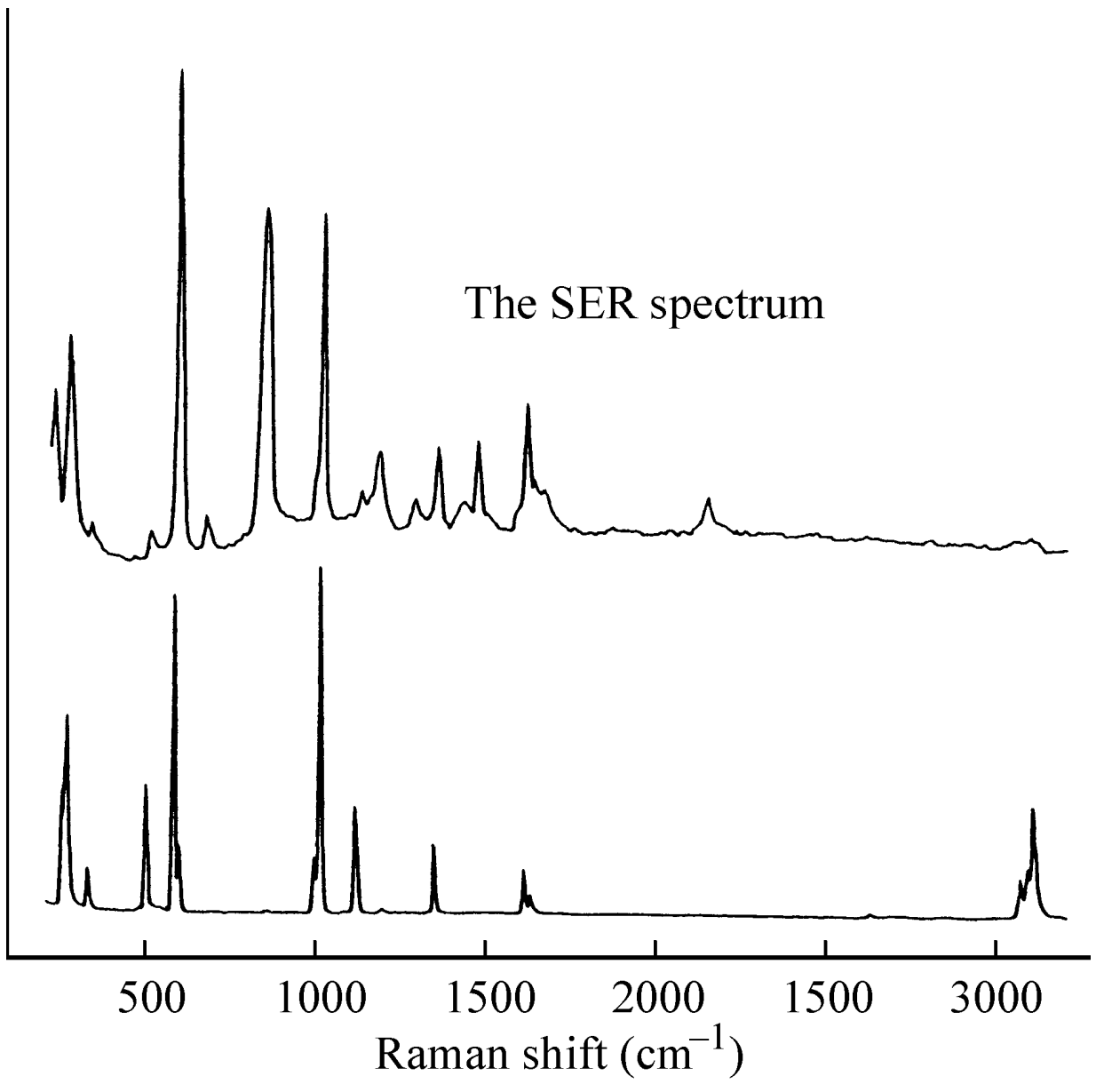

Fig. 18 SERS and Raman spectra of 1,3,5-trifluorobenzene

Table 4. Relative intensities of various Raman lines of benzene and benzene 1,3,5 - $d_3$ (a - Absent in the SER spectrum, b - possible overlapping with other modes prevents determination of the presence of this mode. Frequencies in parentheses are IR or calculated frequencies. vs-very strong, s-strong, m-medium, w-weak, vw-very weak)

| **BENZENE** | | | BENZENE-1,3,5-$d_3$ | | |
|---|---|---|---|---|---|
| sym | bulk | SERS | sym | bulk | SERS |
| 1 $A_{1g}$ | 990 vs | 982 vs | $A_1'$ | 956 vs | 952 s |
| 2 $A_{1g}$ | 3059 *s* | 3080 w | $A_1'$ | 3044 s | 3046 vw |
| 3 $A_{2g}$ | (1346) | a | $A_2'$ | 1265 vw | 1267 vw |
| 4 $B_{2g}$ | (703) | a | $A_2''$ | (697) | 696 w sh |
| 5 $B_{2g}$ | (989) | b | $A_2''$ | 911 vw | b |
| 6 $E_{2g}$ | 606 m | 606 w | $E'$ | 594 m | 594 w |
| 7 $E_{2g}$ | 3046 a | b | $E'$ | 2275 m | a |
| 8 $E_{2g}$ | 1596 m | 1587 m | $E'$ | 1574 s | 1569 m |
| 9 $E_{2g}$ | 1178 m | 1174 m | $E'$ | 1105 m | 1103 w |
| 10 $E_{1g}$ | 849 m | 864 m | $E''$ | 717 m | 711 m |
| 11 $A_{2u}$ | (670) | 697 m | $A_2''$ | 545 vw | 544 m |
| 12 $B_{1u}$ | (1008) | a | $A_1'$ | 1004 s | 1002 s |
| 13 $B_{1u}$ | (3062) | b | $A_1'$ | 2284 s | 2285 vw |
| 14 $B_{2u}$ | (1309) | 1311 w | $A_2'$ | (1322) | 1324 vw |
| 15 $B_{2u}$ | (1149) | 1149 w | $A_2'$ | (911) | b |
| 16 $E_{2u}$ | (404) | 397 m | $E''$ | 378 w | 374 m |
| 17 $E_{2u}$ | (966) | 970 vw | $E''$ | 932 w | 921 w |
| 18 $E_{1u}$ | (1036) | 1032 vw | $E'$ | 834 m | 835 w |
| 19 $E_{1u}$ | (1479) | 1473 w | $E'$ | 1413 w | 1410 w |
| 20 $E_{1u}$ | 3073 | b | $E'$ | 3061 m | a |

Table 5. Relative intensities of various Raman lines of hexafluorobenzene and 1,3,5 trifluorobenzene (a-Absent in the spectrum. b-Possible overlapping with other modes prevents unambiguous determination of the presence of this mode. Frequencies in parentheses are IR or calculated frequencies. Brackets indicate the assignment in dubious. vs-very strong, s-strong, m-medium, w-weak, vw-very weak)

| HEXAFLUOROBENZENE | | | **1,2,3-TRIFLUOROBENZENE** | | |
|---|---|---|---|---|---|
| sym | bulk | SERS | sym | bulk | SERS |
| 1 $A_{1g}$ | 558 s | 557 s | $A_1'$ | 576 s | 574 s |
| 2 $A_{1g}$ | 1493 w | 1484 s | $A_1'$ | 3080 | |
| 3 $A_{2g}$ | (691) | | $A_2'$ | (1165) | a |
| 4 $B_{2g}$ | 715 vw | | $A_2''$ | 662 vw | 665 w |
| 5 $B_{2g}$ | (247) | | $A_2''$ | 220 w | 221 m |
| 6 $E_{2g}$ | 443 s | | $E'$ | 550 ms | 498 w |
| 7 $E_{2g}$ | 1154 w | | $E'$ | 992 m | a |
| 8 $E_{2g}$ | 1659 m | | $E'$ | 1612 m | 1606 m |
| | | | | 1632 w | 1630 w |
| 9 $E_{2g}$ | 269 w | | $E'$ | 323 m | 325 w |
| | | | | 326 m | |
| 10 $E_{1g}$ | 373 s | 374 m | $E''$ | 850 vw | b |
| 11 $A_{2u}$ | 220 vw | | $A_2''$ | (843) | 832 s |
| 12 $B_{1u}$ | 641 vw | | $A_1'$ | 1009 s | 1003 s |
| 13 $B_{1u}$ | (1315) | 1315 w | $A_1'$ | 1342 m | 1345 m |
| 14 $B_{2u}$ | (1252) | 1242 m | $A_2'$ | (1294) | 1280 w |
| 15 $B_{2u}$ | (208) | | $A_2'$ | 553 vw | a |
| 16 $E_{2u}$ | (595) | 620 w | $E''$ | 595 m | 592 |
| 17 $E_{2u}$ | (175) | | $E''$ | 251 s | 255 |
| | | | | 263 s | |
| 18 $E_{1u}$ | (317) vw | | $E'$ | 1114 m | 1121 |
| 19 $E_{1u}$ | (1530) | | $E'$ | 1474 vw | 1462 |
| 20 $E_{1u}$ | (1000) | 1000 w | $E'$ | 3080 | |

## 14. ANOMALIES IN THE SER SPECTRA OF BENZENE ADSORBED ON LITHIUM AND HEXAFLUOROBENZENE ADSORBED ON SILVER

As it was pointed out in paragraphs 9, 13 depending on the experimental conditions the kind of a molecule and its symmetry group there may be a competition among the bands, associated with the $(Q_{main} - Q_{main})$ scatterings and also by the $(Q_{main} - Q_{main})$ and $(d_z - Q_{main})$ scatterings [27, 38]. In the previous paragraphs we pointed out that these anomalies may be due to some "interference" effects associated with different values of the electric field derivatives on the random surface. Here we suggest an additional explanation of these anomalies, which does not contradict to the previous one. If we assume that this competition is between quadrupole and dipole light-molecule interactions, than this supposition can not be valid. In any case when the dipole interaction with the $d_z$ moment is stronger, the breathing mode may be enhanced to the greatest extent because the $(d_z - Q_{main})$ scattering can not be stronger than the $(d_z - d_z)$ type of scattering, contributing to the breathing mode. Therefore a more correct assumption is that these anomalies are caused by the set of different $R_{nl(s,p)}$ and $R_{mk(s,p)}$ coefficients and their conjugated values in (51), which may result in different intensities for various types of vibrations. For benzene on lithium [38] (Fig. 10, Fig. 11) it may be due to decrease of electromagnetic enhancement factor, because lithium is not as favorable metal for SERS as silver. The latter has the lowest imaginary part of the complex dielectric constant and the value and heterogeneity of electromagnetic field on silver is stronger than that on lithium for the same roughness. Therefore the most important fact in this case may be enhancement due to the stronger influence of the set of the $R_{nl(s,p)}$, $R_{mk(s,p)}$ and conjugated coefficients for the mode, associated with the scattering via the $d_z$ moment or with $(d_z - Q_{main})$ scattering type. Similarly, the most enhanced band in hexafluorobenzene is that related to the totally symmetric stretching vibration, rather than to the breathing one [38] (fig. 17). This fact may be qualitatively accounted for by the strong influence of the set of the $R_{nl(s,p)}$ and $R_{mk(s,p)}$ coefficients for this vibration, compared to the coefficients, associated with the breathing mode. Apparently this is a consequence of a large electron density on fluorine atoms. In conclusion it should be noted, that our explanations are only qualitative, because the precise quantitative calculations of the intensities of spectral lines in SERS are impossible. Therefore the qualitative approach is the most fruitful.

## 15. ELECTRODYNAMIC FORBIDDANCE OF THE QUADRUPOLE ENHANCEMENT MECHANISM IN MOLECULES WITH CUBIC SYMMETRY GROUPS

It was noted in [12] that the strong quadrupole light-molecule interaction do not contribute to the SER spectra of methane and other SER spectra of molecules with cubic symmetry groups. This phenomenon was named the electrodynamic forbiddance of the quadrupole scattering mechanism. Because expression (49) for the SER cross-section with a modified scattering tensor coincides with a similar expression (1c) in [12] and expression for $A_{V_{(s,p)}}$ (52) and for $S_{d-Q}, S_{Q-d}$, and $S_{Q-Q}$ (54-56) coincide with similar expressions (6, 8-10) in [12], the reasoning made in [12] is still correct. Results, similar to those of [12] can be obtained by considering the transformed contributions (61-63). Taking into account the specific expressions for the $Q_{xx}, Q_{yy}, Q_{zz}$ moments in the methane molecule

$$
\begin{aligned}
Q_{xx} &= (Q_1 + \frac{2}{3}Q_3 + Q_2) \\
Q_{yy} &= (Q_1 + \frac{2}{3}Q_3 - Q_2) \\
Q_{zz} &= (Q_1 + \frac{2}{3}Q_3 - 2Q_2)
\end{aligned}
\tag{75}
$$

and expressions for $Q_1, Q_2, Q_3$

$$
\begin{aligned}
Q_1 &= 1/3(Q_{xx} + Q_{yy} + Q_{zz}) \\
Q_2 &= \frac{1}{2}(Q_{xx} - Q_{yy}) \\
Q_3 &= \frac{1}{4}(Q_{xx} + Q_{yy} - 2Q_{zz})
\end{aligned}
\tag{76}
$$

using the coefficient $a_{\gamma,k}$ from expressions (75) , we can make summations in various transformed contributions (61-63) over the indices $\alpha$ and $\gamma$. Then taking into account, that $divE = 0$ the transformed contributions, which contain the quadrupole moments have the form

$$T_{d-Q} = \sum_{\substack{i,\gamma\delta \\ \gamma\neq\delta}} C_{V_{(s,p)}}[d_i, Q_{\gamma\delta}] \left( e^*_{scat,i} \frac{1}{2\left|\overline{E}_{inc}\right|} \frac{\partial E^{inc}_{\gamma}}{\partial x_{\delta}} \right)_{surf} +$$

$$\sum_{i,\gamma} a_{\gamma 2} C_{V_{(s,p)}}[d_i, Q_2] \left( e^*_{scat,i} \frac{1}{2\left|\overline{E}_{inc}\right|} \frac{\partial E^{inc}_{\gamma}}{\partial x_{\gamma}} \right)_{surf} \tag{77}$$

$$T_{Q-d} = \sum_{\substack{\alpha\beta,i \\ \alpha\neq\beta}} C_{V_{(s,p)}}[Q_{\alpha\beta}, d_i] \left( \frac{1}{2\left|\overline{E}_{scat}\right|} \frac{\partial E^{scat^*}_{\alpha}}{\partial x_{\beta}} e_{inc,i} \right)_{surf} +$$

$$+ \sum_{\alpha,i} a_{\alpha,2} C_{V_{(s,p)}}[Q_2, d_i] \left( \frac{1}{2\left|\overline{E}_{scat}\right|} \frac{\partial E^{scat^*}_{\alpha}}{\partial x_{\alpha}} e_{inc,i} \right)_{surf} \tag{78}$$

$$T_{Q-Q} = \sum_{\substack{\alpha\beta,\gamma\delta \\ \alpha\neq\beta,\gamma\neq\delta}} C_{V_{(s,p)}}[Q_{\alpha\beta}, Q_{\gamma\delta}] \left( \frac{1}{2\left|\overline{E}_{scat}\right|} \frac{\partial E^{scat^*}_{\alpha}}{\partial x_{\beta}} \frac{1}{2\left|\overline{E}_{inc}\right|} \frac{\partial E^{inc}_{\gamma}}{\partial x_{\delta}} \right)_{surf} +$$

$$\sum_{\substack{\alpha,\gamma\delta \\ \gamma\neq\delta}} a_{\alpha,2} C_{V_{(s,p)}}[Q_2, Q_{\gamma\delta}] \left( \frac{1}{2\left|\overline{E}_{scat}\right|} \frac{\partial E^{scat^*}_{\alpha}}{\partial x_{\alpha}} \frac{1}{2\left|\overline{E}_{inc}\right|} \frac{\partial E^{inc}_{\gamma}}{\partial x_{\delta}} \right)_{surf} +$$

$$\sum_{\substack{\alpha\beta,\gamma \\ \alpha\neq\beta}} a_{\gamma,2} C_{V_{(s,p)}}[Q_{\alpha\beta}, Q_2] \left( \frac{1}{2\left|\overline{E}_{scat}\right|} \frac{\partial E^{scat^*}_{\alpha}}{\partial x_{\beta}} \frac{1}{2\left|\overline{E}_{inc}\right|} \frac{\partial E^{inc}_{\gamma}}{\partial x_{\gamma}} \right)_{surf} +$$

$$\sum_{\alpha,\gamma} a_{\alpha,2} a_{\gamma 2} C_{V_{(s,p)}}[Q_2, Q_2] \left( \frac{1}{2\left|\overline{E}_{scat}\right|} \frac{\partial E^{scat^*}_{\alpha}}{\partial x_{\alpha}} \frac{1}{2\left|\overline{E}_{inc}\right|} \frac{\partial E^{inc}_{\gamma}}{\partial x_{\gamma}} \right)_{surf} \tag{79}$$

It can be seen (77-79), that the transformed contributions contain only the dipole and minor quadrupole moment $Q_2$. Thus the whole spectrum of methane is determined by the scattering via the dipole and minor quadrupole moments, and, in accordance with these results it must be enhanced to a lesser extent, compared with the SER spectra of other symmetrical molecules.

## 16. DISCUSSION OF EXPERIMENTAL RESULTS

Indeed a study of SERS on methane [41] revealed a low intensity of the whole spectrum, relative to that typical for SERS. The observed bands are $\nu_1, \nu_2, \nu_3$ and $\nu_4$ with $A_1, E, F_2$ and $F_2$ symmetry respectively (Table 6).

Table 6. Vibrational wave numbers of the methane molecule for gaseous phase and for SERS

| VIB-RA-TION | SYMMETRY | WAVENUMBER IN A GAZEOUS PHASE $cm^{-1}$ | WAVENUMBER IN SERS FOR THE FIRST LAYER $(cm^{-1})$ | WAVENUMBER IN SERS FOR THE SECOND AND UPPER LAYERS |
|---|---|---|---|---|
| $\nu_1$ | $A_1$ | 2916.5 | 2884 | 2903 |
| $\nu_2$ | $E$ | 1533.6 | 1517 | |
| $\nu_3$ | $F_2$ | 3018.9 | 2996, 3027 | |
| $\nu_4$ | $F_2$ | 1305.9 | 1295 | |

The most enhanced band is $\nu_1$ which is splitted at large substrate coverage (Fig. 19). The band at $\nu_1 \approx 2884 cm^{-1}$ can be attributed to molecules, adsorbed in the first layer, whereas the band at $\nu_1 \approx 2903 cm^{-1}$ corresponds to the molecules physisorbed in the second and higher layers. The large frequency shift of this band for the molecules adsorbed in the first layer indicates that it belongs to the chemisorbed molecules. Therefore its enhancement can be explained by the following reasons:

1. existence of the strong scattering of the $(d_z - d_z)$ type caused by the main moments, which contributes to the bands with $A_1$ symmetry,

2. appearance of the quadrupole enhancement mechanism of the $(Q_1 - Q_1)$ scattering type because of violation of the electrodynamic forbiddance due to symmetry disturbance.

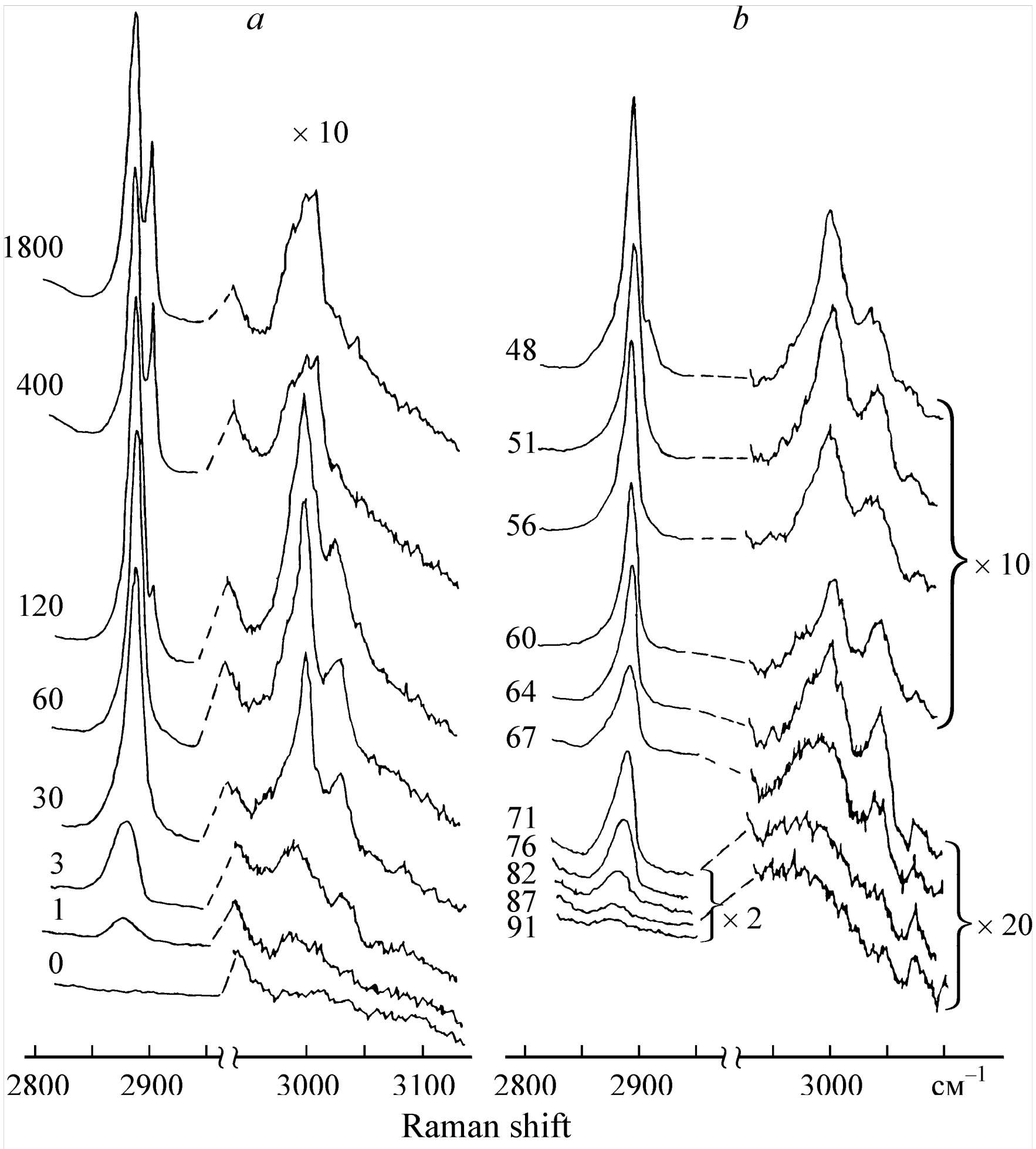

Fig. 19 Changes of the SER spectra of methane in the region of $\nu_1$ and $\nu_3$ vibrations: a - in the case of changes of filling of the surface by adsorbate $(T \approx 25K)$ (exposure in methane media); b - in the case of changes in temperature for exposure ~120 Langmuir. The figures near the curves: a - exposure (Langmuir), b - temperature $(K)$

The last factor must be absent for physisorbed molecules in the second and higher layers. Observation of the band at $\nu_1 = 2903cm^{-1}$ with a substantially lower intensity confirms this conclusion. Taking into account that the band at $\nu_1 = 2903cm^{-1}$ is associated with the breathing vibration, which must be most enhanced (as it follows from a large number of experimental investigations [5]) the results presented indicate that the spectrum enhancement of physisorbed methane is weak, compared with that usually observed in SERS. The enhancement is apparently caused by the strong dipole interaction and by the $(d_z - d_z)$ type of the scattering. Thus the significantly weaker enhancement of the SER spectra of methane relative to that for molecules that do not belong to the cubic symmetry groups may be experimental manifestation of the electrodynamic forbiddance of the quadrupole SERS mechanism in the methane molecule. Then the main mechanism of the enhancement for physisorbed molecules is an increase of the electric field component, which is perpendicular to the surface.

# 17. ABOUT THE ELECTRODYNAMIC FORBIDDANCE OF THE QUADRUPOLE SERS MECHANISM IN OTHER MOLECULES

The specific feature of the preceding analysis is that it is based only on the possibility of transformation of the quadrupole moments one into another under symmetry operations and their representation via symmetric combinations $Q_1, Q_2, Q_3$ in accordance with (75). The coincidence of the later moments to any particular irreducible representation was not used. Therefore we can conclude, that for the "cubic" molecules belonging to the $T, T_h, T_d, O$ and $O_h$ symmetry groups in which representation (75) takes place, the main contributions to SERS from the main quadrupole moments will be also electrodynamically forbidden.

# 18. ANOMALIES OF THE SER SPECTRA OF SYMMETRICAL MOLECULES ADSORBED ON TRANSITION METAL SUBSTRATES

In accordance with preceding consideration the main regularities of symmetrical molecules are the following:

1. The most enhanced bands are those related to the totally symmetric vibrations, and the breathing mode is usually the most enhanced among them.
2. Appearance of a large number of forbidden bands in molecules with high symmetry and also some other specific features of the SER spectra.

These facts and the anomalies mentioned above have been interpreted in terms of the dipole-quadrupole SERS mechanism. Originally, the above theory was created in order to explain the enhancement value of the spectra and some another features of the phenomena accompanying SERS for molecules adsorbed on $Ag$ , $Cu$, and $Au$. However an additional consideration is necessary for some specific systems. For example, the existence of SERS for pyridine and pyrazine adsorbed on a number of transition metal surfaces, such as $Fe$, $Co$, $Ni$, $Ru$, $Rh$, $Pd$, $Pt$ with a considerably weaker surface enhancement (by 2 to 3 orders of magnitude) as compared with those on $Ag$ , $Cu$ and $Au$ was reported in [42-47]. On some of these metal surfaces (e.g. $Ni$, $Rb$), it was found that the most enhanced band of pyridine is not associated with the breathing mode as in the preceding cases of benzene adsorbed on lithium and hexafluorobenzene adsorbed on silver. This fact also can be explained in terms of the above theory, by taking into account the weak quadrupole light-molecule (electrons) interaction for molecules, adsorbed on these particular metals. The most enhanced band in SERS of pyridine on noble metals is the breathing mode at $\nu_1 = 1005/1008 cm^{-1}$. However the most enhanced bands in SERS on $Ni$ are those at $\nu_5 \sim 1210\ cm^{-1}$ and $\nu_4 \sim 1595 cm^{-1}$, which are associated with totally symmetric vibrations of vertically adsorbed pyridine (Fig. 20) [42]. The variation of the relative intensity of these bands with the electrode potential is not so significant. Therefore it demonstrates a certain change in the enhancement mechanism compared with $Ag$ , $Cu$ and $Au$. The assignment of different scattering contributions of the adsorbed pyridine to the vibrational types is listed in Table 7. Besides their sequence qualitatively corresponds to the degree of enhancement. (It coincides with those for vertically adsorbed pyrazine. Table 7 additionally lists the wavenumbers and description of the intensities of pyrazine). In our opinion, the most enhanced contributions are now caused by the $(d_z - d_z)$ type of the scattering. However, in accordance with the selection rules, the $(d_z - Q_{\alpha\alpha})$ and $(Q_{\alpha\alpha} - Q_{\beta\beta})$ scattering types should also contribute.

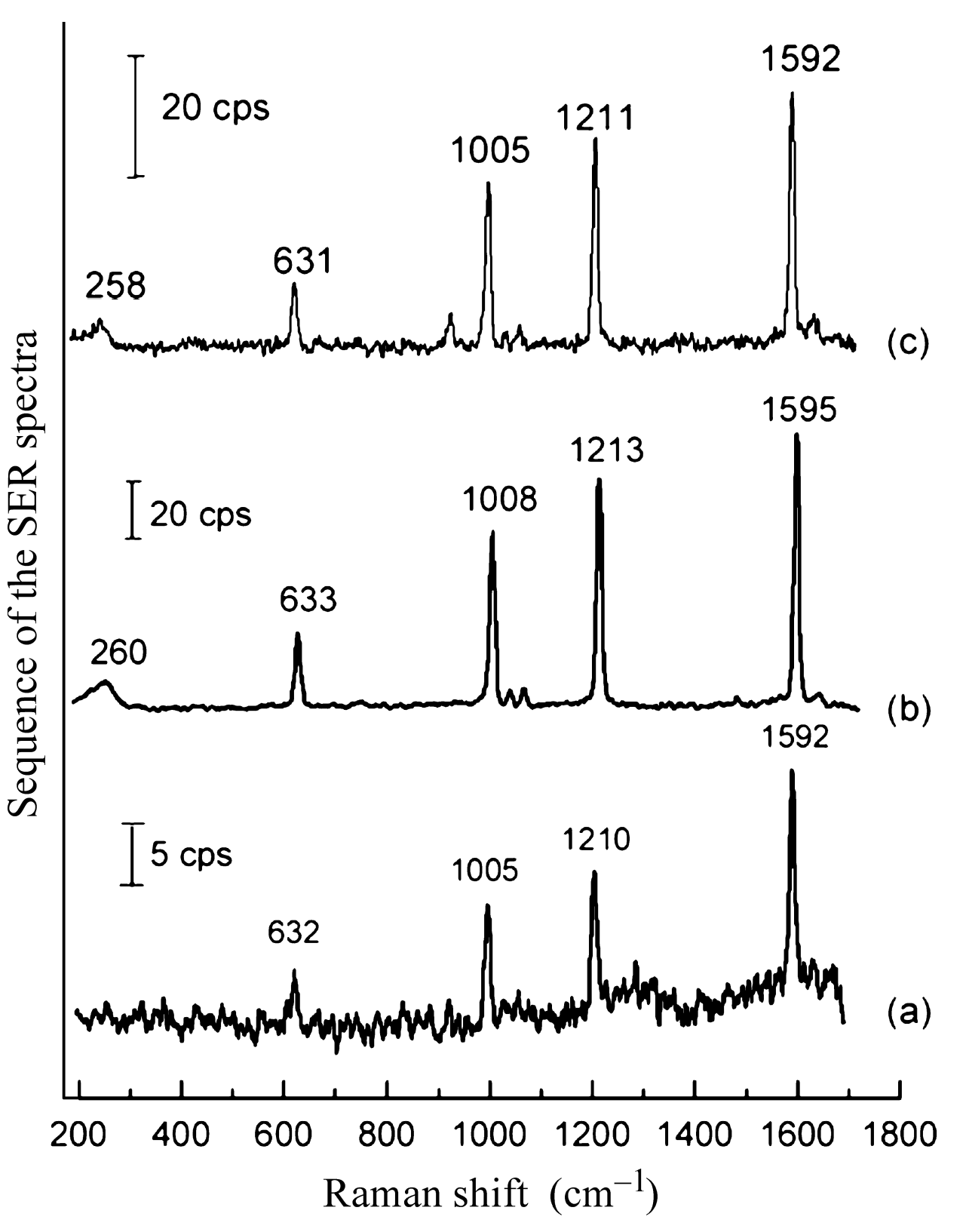

Fig. 20 Surface Raman spectra of pyridine, adsorbed on *Ni* electrodes, roughened by: (a) chemical etching; (b) chemical etching, then in situ (ORC) in a solution of 0.01M Py + 0.1 M KCl; (c) the same as (b) but in 0.01M Py + $0.1NaClO_4$. The potential was -1.2 V and the excitation line was 514.5 nm.

Table 7. The contribution of various scattering types for pyridine and pyrazine, vertically adsorbed on $Ni$ and some other electrodes. Here we specially reveal the scattering via $Q_{\alpha\alpha}$ and $d_z$ and other $Q$ and $d$ moments. The sequence of contributions approximately corresponds to that of the degrees of enhancement

| Mode symmetry | Wavenumbers of Pz (cm $^{-1}$) | $(Q_{\min or} - Q_{\min or})$ | $(d_{\min or} - Q_{\min or})$ | $(Q_{\min or} - Q_{\alpha\alpha})$ $(Q_{\alpha\alpha} - Q_{\min or})$ | $Q_{\min or} - d_z$ $d_z - Q_{\min or}$ | $(d_{\min or} - Q_{\alpha\alpha})$ $(Q_{\alpha\alpha} - d_{\min or})$ | $d_{\min or} - d_z$ $d_z - d_{\min or}$ | $(Q_{\alpha\alpha} - Q_{\beta\beta})$ | $(d_z - Q_{\alpha\alpha})$ $(Q_{\alpha\alpha} - d_z)$ | $(d_z - d_z)$ | Intensity |
|---|---|---|---|---|---|---|---|---|---|---|---|
| $A_1$ | 1576 | $Q_{xy}Q_{xy}$ $Q_{xz}Q_{xz}$ $Q_{yz}Q_{yz}$ | $d_x Q_{xz}$ $d_y Q_{yz}$ | | | | | $(Q_{xx}, Q_{yy}, Q_{zz})$ $(Q_{xx}, Q_{yy}, Q_{zz})$ | $d_z (Q_{xx}, Q_{yy}, Q_{zz})$ | $d_z d_z$ | S |
| | 1216 | | | | | | | | | | S |
| | 1045 | | | | | | | | | | m |
| | 1010 | | | | | | | | | | m |
| | 632 | | | | | | | | | | S |
| $A_2$ | | $Q_{xz}Q_{yz}$ | | $Q_{xy}(Q_{xx}, Q_{yy}, Q_{zz})$ | $Q_{xy} d_z$ | | | | | | Absent |
| $B_1$ | 1473 | $Q_{xy}Q_{yz}$ | $d_y Q_{xy}$ | $Q_{xz}(Q_{xx}, Q_{yy}, Q_{zz})$ | | $d_x (Q_{xx}, Q_{yy}, Q_{zz})$ | $d_x d_z$ | | | | Wv |
| | 1115 | | | | | | | | | | Wv |
| $B_2$ | 795 | $Q_{xy}Q_{xz}$ | | $Q_{yz}(Q_{xx}, Q_{yy}, Q_{zz})$ | $Q_{yz} d_z$ | $d_y (Q_{xx}, Q_{yy}, Q_{zz})$ | $d_y d_z$ | | | | Wv |
| | 450 | | | | | | | | | | Wv |

The problem of determination of the relative values for different contributions is very difficult. One of the possible reasons of the anomalies arising from the deviation from $(\hbar\omega)^4$ law [5], which depends on the type of the chemical preparation of the substrate, on its complex dielectric constant and some another factors in a complex manner. This fact is confirmed by the result that for the longer wavelengths of illumination - $\lambda = 632.8 nm$, instead of those used in our experiment $(\lambda = 514.5 nm)$ the mode $(\nu_4 \sim 1592-1595 cm^{-1})$ is weaker than the breathing mode $\nu_1 = 1005-1008 cm^{-1}$. However the most important fact, that may be responsible for the anomalies in the SERS spectra of pyridine on $Ni$ and another transition metals is the weak influence of the quadrupole light-molecule interaction. A more clear analysis should be based on experiments dealing with molecules of higher symmetry, for which the lines forbidden in the normal Raman spectra can be observed. As an example, we analyze the SER spectra of pyrazine adsorbed on $Ni$ surface [43]. The most difficult problem with this system is that the molecule can be adsorbed on the surface in two ways: vertically via the nitrogen atom and horizontally lying flatly on the surface. The assignment of various scattering types is displayed in Tables 7 and 8.

Table 8. Contributions of various scattering types of pyrazine, horizontally adsorbed on $Ni$ and some other electrodes. Here we specially reveal the scattering via $Q_{\alpha\alpha}$, $d_z$ and other $Q$ and $d$ moments. The sequence of the contributions approximately corresponds to that of the degrees of enhancement.

| Mode symmetry | Wavenumber (cm $^{-1}$) | $(Q-Q)$ | $(d-Q)$ $(Q-d)$ | $(d-d)$ | $Q_{\min or}-Q_{\alpha\alpha}$ $Q_{\alpha\alpha}-Q_{\min or}$ | $Q_{\min or}-d_z$ $d_z-Q_{\min or}$ | $d_{\min or}-Q_{\alpha\alpha}$ $Q_{\alpha\alpha}-d_{\min or}$ | $(d_{\min or}-d_z)$ $(d_z-d_{\min or})$ | $Q_{\alpha\alpha}-Q_{\beta\beta}$ | $d_z-Q_{\alpha\alpha}$ $Q_{\alpha\alpha}-d_z$ | $d_z-d_z$ | Relative intensity |
|---|---|---|---|---|---|---|---|---|---|---|---|---|
| $A_g$ | 1576 | $Q_{xy}Q_{xy}$ $Q_{xz}Q_{xz}$ $Q_{yz}Q_{yz}$ | | $d_xd_x$ $d_yd_y$ | | | | | $(Q_{xx},Q_{yy},Q_{zz})$ $(Q_{xx},Q_{yy},Q_{zz})$ | | $d_zd_z$ | S |
| $A_g$ | 1216 | | | | | | | | | | | S |
| $A_g$ | 1010 | | | | | | | | | | | M |
| $A_g$ | 632 | | | | | | | | | | | S |
| $A_u$ | | | $d_xQ_{yz}$ $d_yQ_{xz}$ | | | $d_zQ_{xy}$ | | | | | | Absent |
| $B_{1g}$ | | $Q_{xz}Q_{yz}$ | | $d_xd_y$ | | | | | | | | Absent |
| $B_{1g}$ | | | | | | | | | | | | Absent |
| $B_{1u}$ | 1045 | | $d_xQ_{xz}$ $d_yQ_{yz}$ | | | | | | | $d_z(Q_{xx},Q_{yy},Q_{zz})$ | | M |
| $B_{2g}$ | 1115 | $Q_{xy},Q_{yz}$ | | | $Q_{xz}(Q_{xx},Q_{yy},Q_{zz})$ | | | $d_xd_z$ | | | | Wv |
| $B_{2u}$ | 795 | | $d_xQ_{xy}$ | | | $d_zQ_{yz}$ | $d_y(Q_{xx},Q_{yy},Q_{zz})$ | | | | | Wv |
| $B_{2u}$ | 450 | | | | | | | | | | | Wv |
| $B_{3u}$ | 1473 | | $d_yQ_{xy}$ | | | $d_zQ_{xz}$ | $d_x(Q_{xx},Q_{yy},Q_{zz})$ | | | | | Wv |

Here the sequence of the scattering types qualitatively corresponds to the degree of enhancement of the contributions. It can be seen that all the pyrazine bands are allowed in the dipole-quadrupole scattering. Moreover the band at $1045 cm^{-1}$, which is forbidden in horizontal adsorption configuration, has significant intensity. This may provide an evidence of the existence of the quadrupole interaction in this system. In the vertical adsorption orientation the adsorbed molecule has $C_{2v}$ symmetry, and the band at $1045 cm^{-1}$ and all other bands correspond to allowed lines in accordance with the selection rules for the contributions (73). Therefore the existence of the forbidden line of pyrazine is not sufficient to make an unambiguous conclusion about the quadrupole interaction in this system. In order to make a more convincing conclusion about the possible existence of the quadrupole interaction, it is necessary to analyze molecules with higher symmetry. Such study can be performed by analyzing of the SER spectra of benzene on a silica supported $Ni$ surface [26]. It should be remind that benzene is a molecule, belonging to the $D_{6h}$ symmetry group and is adsorbed in the planar configuration. The appearance of forbidden lines in particular of the $A_{2u}$ band caused by the $d_z(Q_{xx} + Q_{yy}, Q_{zz})$ type of the scattering in its spectrum demonstrates that the quadrupole light-molecule interaction exists in this system (see Table 9). The existence of the $B_{1u}$ line associated with the $(d_x, d_y)(Q_{xx} - Q_{yy}, Q_{xy})$ type of scattering indicates that there exists a thick layer of adsorbed benzene and all the $d$ moments are the main ones.

Table 9 Part 1. Contributions of various scattering types of benzene adsorbed on silica supported $Ni$. The expression $(d_x, d_y)(Q_{xx} - Q_{yy}, Q_{xy})$ or $(d_x, d_y)(d_x, d_y)$ for example designates, that these combinations of moments contain a contribution, transformed under the irreducible representation of the corresponding line in Table 9. The sequence of contributions approximately corresponds to that of the degrees of enhancement.

| Mode Symmetry | Wavenumber $(cm^{-1})$ | $(Q_{xz}, Q_{yz}\}$ $(Q_{xz}, Q_{yz})$ | $(Q_{xz}, Q_{yz})$ $(Q_{xx} - Q_{yy}, Q_{zz})$ | $(Q_{xx} - Q_{yy}, Q_{xy})$ $(Q_{xx} - Q_{yy}, Q_{xy})$ | Relative intensity |
|---|---|---|---|---|---|
| $E_{2u}$ | 407 | | | | 0.04 |
| $E_{2g}$ | 603 | $(Q_{xz}, Q_{yz}\}$ $(Q_{xz}, Q_{yz})$ | | $(Q_{xx} - Q_{yy}, Q_{xy})$ $(Q_{xx} - Q_{yy}, Q_{xy})$ | 0.08 |
| $B_{2g}$ | 684 | | $(Q_{xz}, Q_{yz})$ $(Q_{xx} - Q_{yy}, Q_{zz})$ | | 0.15 |
| $A_{2u}$ | 776 | | | | 0.09 |
| $E_{1g}$ | 864 | | $(Q_{xz}, Q_{yz})$ $(Q_{xx} - Q_{yy}, Q_{zz})$ | | 0.17 |
| $E_{2u}$ | 969 | | | | 0.04 |
| $A_{1g}$ | 995 | $(Q_{xz}, Q_{yz}\}$ $(Q_{xz}, Q_{yz})$ | | $(Q_{xx} - Q_{yy}, Q_{xy})$ $(Q_{xx} - Q_{yy}, Q_{xy})$ | 1.00 |
| $B_{1u}$ | 1056 | | | | 0.05 |
| $B_{1u}$ | 1120 | | | | 0.02 |
| $E_{2g}$ | 1161 | $(Q_{xz}, Q_{yz}\}$ $(Q_{xz}, Q_{yz})$ | | $(Q_{xx} - Q_{yy}, Q_{xy})$ $(Q_{xx} - Q_{yy}, Q_{xy})$ | 0.10 |
| $E_{2g}$ | 1184 | $(Q_{xz}, Q_{yz}\}$ $(Q_{xz}, Q_{yz})$ | | $(Q_{xx} - Q_{yy}, Q_{xy})$ $(Q_{xx} - Q_{yy}, Q_{xy})$ | 0.32 |
| $E_{1u}$ | 1440 | | | | 0.13 |
| $E_{2g}$ | 1585 | $(Q_{xz}, Q_{yz}\}$ $(Q_{xz}, Q_{yz})$ | | $(Q_{xx} - Q_{yy}, Q_{xy})$ $(Q_{xx} - Q_{yy}, Q_{xy})$ | 0.06 |
| $E_{1u}$ | 3016 | | | | 0.04 |
| $E_{2g}$ | 3040 | $(Q_{xz}, Q_{yz}\}$ $(Q_{xz}, Q_{yz})$ | | $(Q_{xx} - Q_{yy}, Q_{xy})$ $(Q_{xx} - Q_{yy}, Q_{xy})$ | 0.12 |
| $A_{1g}$ | 3060 | $(Q_{xz}, Q_{yz}\}$ $(Q_{xz}, Q_{yz})$ | | $(Q_{xx} - Q_{yy}, Q_{xy})$ $(Q_{xx} - Q_{yy}, Q_{xy})$ | 0.65 |

Table 9 Part 2. Contributions of various scattering types of benzene adsorbed on silica supported $Ni$. The expression $(d_x, d_y)(Q_{xx} - Q_{yy}, Q_{xy})$ or $(d_x, d_y)(d_x, d_y)$ for example designates, that these combinations of moments contain a contribution, transformed under the irreducible representation of the corresponding line in Table 9. The sequence of contributions approximately corresponds to that of the degrees of enhancement.

| Mode Symmetry | Wavenumber $(cm^{-1})$ | $(d_x,d_y)$ $(Q_{xz},Q_{yz})$ | $(d_x,d_y)$ $(Q_{xx}-Q_{yy},Q_{xy})$ | $(Q_{xz},Q_{yz})$ $(Q_{xx}+Q_{yy},Q_{zz})$ | $(Q_{xx}-Q_{yy},Q_{xy})$ $(Q_{xx}+Q_{yy},Q_{zz})$ | $d_z$ $(Q_{xz},Q_{yz})$ | $d_z$ $(Q_{xx}-Q_{yy},Q_{xy})$ | Relative intensity |
|---|---|---|---|---|---|---|---|---|
| $E_{2u}$ | 407 | $(d_x,d_y)$ $(Q_{xz},Q_{yz})$ | | | | | $d_z$ $(Q_{xx}-Q_{yy},Q_{xy})$ | 0.04 |
| $E_{2g}$ | 603 | | | | $(Q_{xx}-Q_{yy},Q_{xy})$ $(Q_{xx}+Q_{yy},Q_{zz})$ | | | 0.08 |
| $B_{2g}$ | 684 | | | | | | | 0.15 |
| $A_{2u}$ | 776 | $(d_x,d_y)$ $(Q_{xz},Q_{yz})$ | | | | | | 0.09 |
| $E_{1g}$ | 864 | | | $(Q_{xz},Q_{yz})$ $(Q_{xx}+Q_{yy},Q_{zz})$ | | | | 0.17 |
| $E_{2u}$ | 969 | $(d_x,d_y)$ $(Q_{xz},Q_{yz})$ | | | | | $d_z$ $(Q_{xx}-Q_{yy},Q_{xy})$ | 0.04 |
| $A_{1g}$ | 995 | | | | | | | 1.00 |
| $B_{1u}$ | 1056 | | $(d_x,d_y)$ $(Q_{xx}-Q_{yy},Q_{xy})$ | | | | | 0.05 |
| $B_{1u}$ | 1120 | | $(d_x,d_y)$ $(Q_{xx}-Q_{yy},Q_{xy})$ | | | | | 0.02 |
| $E_{2g}$ | 1161 | | | | $(Q_{xx}-Q_{yy},Q_{xy})$ $(Q_{xx}+Q_{yy},Q_{zz})$ | | | 0.10 |
| $E_{2g}$ | 1184 | | | | $(Q_{xx}-Q_{yy},Q_{xy})$ $(Q_{xx}+Q_{yy},Q_{zz})$ | | | 0.32 |
| $E_{1u}$ | 1440 | | $(d_x,d_y)$ $(Q_{xx}-Q_{yy},Q_{xy})$ | | | $d_z$ $(Q_{xz},Q_{yz})$ | | 0.13 |
| $E_{2g}$ | 1585 | | | | $(Q_{xx}-Q_{yy},Q_{xy})$ $(Q_{xx}+Q_{yy},Q_{zz})$ | | | 0.06 |
| $E_{1u}$ | 3016 | | | | | $d_z$ $(Q_{xz},Q_{yz})$ | | 0.04 |
| $E_{2g}$ | 3040 | | | | $(Q_{xx}-Q_{yy},Q_{xy})$ $(Q_{xx}+Q_{yy},Q_{zz})$ | | | 0.12 |
| $A_{1g}$ | 3060 | | | | | | | 0.65 |

Table 9 Part 3. Contributions of various scattering types of benzene adsorbed on silica supported $Ni$. The expression $(d_x, d_y)(Q_{xx} - Q_{yy}, Q_{xy})$ or $(d_x, d_y)(d_x, d_y)$ for example designates, that these combinations of moments contain a contribution, transformed under the irreducible representation of the corresponding line in Table 9. The sequence of contributions approximately corresponds to that of the degrees of enhancement.

| Mode Symmetry | Wavenumber $(cm^{-1})$ | $(d_x,d_y)$ $(d_x,d_y)$ | $(d_x,d_y)$ $(Q_{xx}+Q_{yy},Q_{zz})$ | $d_z$ $(d_x,d_y)$ | $(Q_{xx}+Q_{yy},Q_{zz})$ $(Q_{xx}+Q_{yy},Q_{zz})$ | $d_z$ $(Q_{xx}+Q_{yy},Q_{zz})$ | $(d_z,d_z)$ | Relative intensity |
|---|---|---|---|---|---|---|---|---|
| $E_{2u}$ | 407 | | | | | | | 0.04 |
| $E_{2g}$ | 603 | $(d_x,d_y)$ $(d_x,d_y)$ | | | | | | 0.08 |
| $B_{2g}$ | 684 | | | | | | | 0.15 |
| $A_{2u}$ | 776 | | | | | $d_z$ $(Q_{xx}+Q_{yy},Q_{zz})$ | | 0.09 |
| $E_{1g}$ | 864 | | | $d_z$ $(d_x,d_y)$ | | | | 0.17 |
| $E_{2u}$ | 969 | | | | | | | 0.04 |
| $A_{1g}$ | 995 | $(d_x,d_y)$ $(d_x,d_y)$ | | | $(Q_{xx}+Q_{yy},Q_{zz})$ $(Q_{xx}+Q_{yy},Q_{zz})$ | | $(d_z,d_z)$ | 1.00 |
| $B_{1u}$ | 1056 | | | | | | | 0.05 |
| $B_{1u}$ | 1120 | | | | | | | 0.02 |
| $E_{2g}$ | 1161 | $(d_x,d_y)$ $(d_x,d_y)$ | | | | | | 0.10 |
| $E_{2g}$ | 1184 | $(d_x,d_y)$ $(d_x,d_y)$ | | | | | | 0.32 |
| $E_{1u}$ | 1440 | | $(d_x,d_y)$ $(Q_{xx}+Q_{yy},Q_{zz})$ | | | | | 0.13 |
| $E_{2g}$ | 1585 | $(d_x,d_y)$ $(d_x,d_y)$ | | | | | | 0.06 |
| $E_{1u}$ | 3016 | | $(d_x,d_y)$ $(Q_{xx}+Q_{yy},Q_{zz})$ | | | | | 0.04 |
| $E_{2g}$ | 3040 | $(d_x,d_y)$ $(d_x,d_y)$ | | | | | | 0.12 |
| $A_{1g}$ | 3060 | $(d_x,d_y)$ $(d_x,d_y)$ | | | $(Q_{xx}+Q_{yy},Q_{zz})$ $(Q_{xx}+Q_{yy},Q_{zz})$ | | $(d_z,d_z)$ | 0.65 |

It should be noted that only the bands allowed in the dipole-quadrupole approximation appear in the SER spectra of benzene. These bands are $A_{1g}, E_{1g}, E_{2g}, A_{2u}, B_{1u}, E_{1u}$, $E_{2u}$, $B_{2g}$. The fact of appearance of the $B_{1u}$ line was first reported in [40]. Here we should mention the exclusive role of the moments $(Q_{xx} - Q_{yy}, Q_{xy})$ in the enhancement. This result was previously obtained in [40]. In addition it should be noted, that the bands of benzene on the silica supported $Ni$ surface differ significantly from those, for benzene adsorbed on silver. This may result from a stronger chemical interaction for benzene with transition metals compared with $Ag$ , $Cu$ and $Au$. The enhancement factor of the breathing mode is estimated to be $2 \times 10^3$ which is very close to that, obtained in the above-mentioned investigations. However this strong enhancement indicates that the conventional chemical mechanism is inoperative here at all, or its importance is very low. From our point of view the main role in the SER spectra of benzene on $Ni$ belongs to the $d_{x,} d_y$ and $d_z$ moments. Apparently the strongest enhancement of the breathing and totally symmetrical modes $A_{1g}$ ($\nu \sim 995 cm^{-1}$ and $\nu \sim 3060 cm^{-1}$) is due to the $(d_\alpha - d_\alpha)$ types of the scattering. The usually forbidden mode $A_{2u}$ ($\sim 776 cm^{-1}$) has a significantly lower intensity of$\sim 0.09$ compared with the breathing mode $A_{1g}$, $\nu \sim 995 cm^{-1}$, which points out a significantly weaker influence of the quadrupole moments $(Q_{xx} + Q_{yy}, Q_{zz})$ (Table 9). The strong influence of the $d_x$ and $d_y$ moments is manifested in comparatively high intensity of $E_{1u}$ mode $(\nu = 1440 cm^{-1})$ which is mainly due to the $(d_x, d_y)(Q_{xx} + Q_{yy}, Q_{zz})$ scattering. The remain parts $d_z(Q_{xz}, Q_{yz})$ and $(d_x, d_y)(Q_{xx} - Q_{yy}, Q_{xy})$ can not contribute significantly, because they contain the minor quadrupole moments. The strong relative intensity $\sim 0.32$ of the $E_{2g}$ mode $(\nu \sim 1184 cm^{-1})$ is due to the scattering via $d_x$ and $d_y$ dipole moments, such as all another ones, having sufficiently large intensity too. Noteworthy is the increase of the $(Q_{\min or} - Q_{\min or})$ enhancement mechanism. As an example we can mention the relatively large amplitude of enhancement of the $B_{2g}$ mode $(\nu = 684 cm^{-1})$ with the relative intensity of $\sim 0.15$. This effect is apparently associated with the strong distortion of benzene on the silica supported $Ni$. This fact results in weaker "interference" of the wave functions and $Q_{xz}$ and $Q_{yz}$ moments owing to the appearance of a large asymmetry in the electron density and minor quadrupole moments, and to a relative increase of their matrix elements. Another possible reason for the enhancement of this band may be the specific configuration of the benzene molecule, for which the contributions containing $(Q_{xx} - Q_{yy}, Q_{xy})$ strongly affect the SERS spectrum. This fact was mentioned in [40], where the appearance of the $B_{1u}$ and $B_{2u}$ lines was observed for virtually all aromatic molecules with the $D_{6h}$ symmetry. It should be noted that the strong distortion of the benzene molecule is reflected in a large deviation of the spectral lines with ($\nu$ =3016, 1440 and 776 $cm^{-1}$) from their corresponding positions (3080, 1485 and 671 $cm^{-1}$) for the $C_6H_6$ crystalline phase. The final analysis of the SER spectra of benzene on silica-supported $Ni$ demonstrates clearly, that we observe only the lines which are allowed in the dipole-quadrupole approximation and that the dipole interaction in this system apparently stronger, than the quadrupole ones in some mean sense, compared with $Cu$, $Ag$ and $Au$.

The systems analyzed above are based on the mildly rough surface. It will be of particular interest to analyze the intensities of pyridine lines on various metal nanowires [48], which can be considered as the roughest systems at present (Fig. 21).

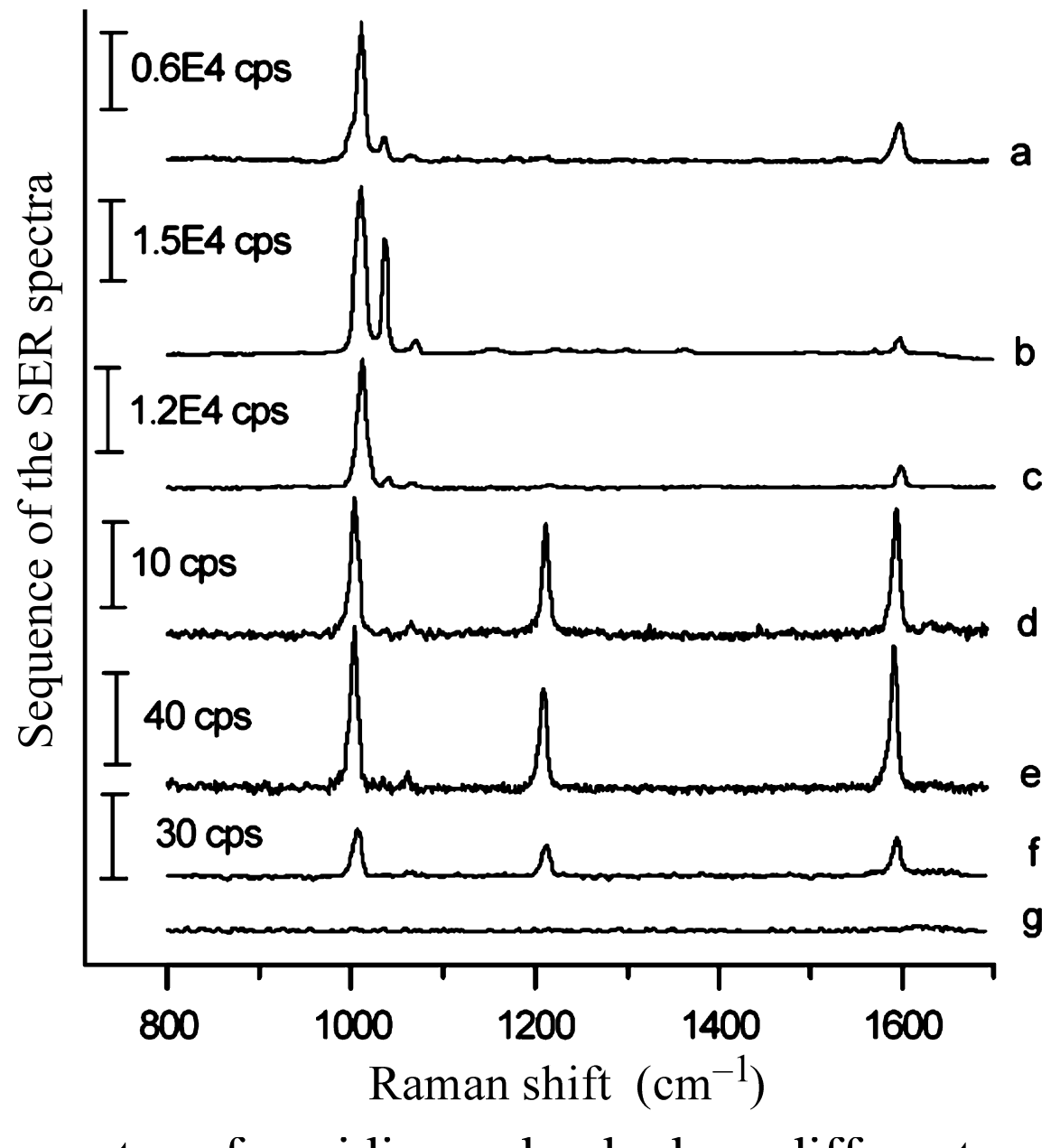

Fig. 21 The SER spectra of pyridine, adsorbed on different metal nanowires, (a) $Au$, (b) $Ag$, (c) $Cu$, (d) $Cd$, (e) $Ni$ and comparison with that, obtained from an electrochemically roughened $Ni$ electrode surface at the potential of –1.2V (f) and around the open circuit potential (g). The solution was 0.01M pyridine and 0.1M KCl.

It can be seen that the relative intensity of the bands at $1593 cm^{-1}$ and at $1215 cm^{-1}$ of pyridine on $Ag$, $Cu$ and $Au$ nanowires is small, compared with the breathing mode, with the enhancement factor $10^5 - 10^6$. For $Ni$ and $Co$ surfaces, the enhancement coefficients are comparable with those for the breathing mode and as small as about $10^3$. These facts confirm that even under such favorable conditions (the strongly rough surface) the enhancement factor for transition metals is still small and there is a strong competition between the breathing and other modes of pyridine. Thus in our opinion SERS on transition metals arises "on the threshold" of appearance of the quadrupole interaction, when the dipole interaction is still stronger, than the quadrupole ones. Therefore the usual rules for SERS regularities, which are valid especially for $Ag$, and a strong roughness degree may be not valid for them. Because the SERS study on transition metal surfaces is still in its infant stage, there are still few sufficient data available for the analysis. However, the same case apparently arises for pyridine on $Ru$, $Rh$ and $Pd$ electrodes [44] which confirm the validity of the above conclusion.

## 19. POSSIBLE REASON OF THE COMPETITION OF THE RAMAN BANDS (INFLUENCE OF THE COMPLEX DIELECTRIC CONSTANT)

As it was mentioned above, the enhancement of the dipole and quadrupole interactions occurs because of heterogeneity of the electric field near the rough metal surface. The heterogeneous field arises from the field reflected by the rough surface and the field emitted by the molecules and also reflected from the surface. This heterogeneity strongly depends on the dielectric properties of the substrate. However, the exact value of the electric field and its derivatives depends on the complex dielectric constant and on the surface configuration which can not be estimated by analytical methods, as mentioned above. Below we analyze the values of the real and imaginary parts of ε', Table 10.

Table 10. The real and imaginary part of the complex dielectric constant of various metals for available values of wavelengths, which are close to the values used in experiment.

| The substance | Wavelength (nm) | $-\mathrm{Re}\,\varepsilon'$ | $\mathrm{Im}\,\varepsilon'$ |
|---|---|---|---|
| Silver | 650 | 17.6 | 0.58 |
| | 600 | 14.1 | 0.45 |
| | 500 | 8.23 | 0.287 |
| Gold | 700 | 15.7 | 1.35 |
| | 600 | 8.77 | 1.37 |
| | 500 | 2.68 | 3.09 |
| Copper | 640 | 7.69 | 1.70 |
| | 520 | 3.71 | 6.99 |
| Nickel | 659 | 14.2 | 17.0 |
| | 521 | 8.27 | 12.7 |
| Palladium | 659 | 16.3 | 15.9 |
| | 521 | 11.1 | 11.6 |
| Cobalt | 659 | 13.2 | 19.2 |
| | 521 | 9.66 | 14.5 |
| Platinum | 640 | 11.1 | 15.7 |

It should be noted that the published data show a very wide scatter for these values. It can be seen however, that for $Ni$, $Pd$, $Co$ and $Pt$, the Im ε' is considerably larger than for $Cu$, $Ag$, and $Au$. The difference between $Ni$ and $Ag$ can be as great as a factor $40$. This may result in a significant decrease in the fields scattered by the surface and emitted by the molecule as well as their derivatives. This decrease may result in a decrease in the enhancement of both the dipole and quadrupole scattering mechanisms. Despite that it is virtually impossible to obtain analytical estimations this may be the most important reason of the relatively small enhancement factors and the insignificant role of the quadrupole enhancement mechanism in the case of transition metals [49]. Thus, we have another limit of the dipole-quadrupole theory in this case, in which the quadrupole interaction is very weak due to the reasons mentioned above.

## 20. SINGLE MOLECULE DETECTION BY THE SERS METHOD AND ITS RELATION TO THE QUADRUPOLE SERS THEORY

As it was pointed out in paragraph 2 there is the enormous SERS effect, which was observed on specially prepared surfaces with a very large roughness [3]. The same effect was observed on relatively large colloidal metal particles with a characteristic size of 80-100 and >100nm under condition of a very small concentration of adsorbed molecules which was approximately equal to the concentration of colloidal particles and the observation conditions corresponded to observation of one or several molecules in the scattering volume [4]. The SER cross-section of such molecules as rodamine 6G, crystal violet, cyanine dye and some others changes from their typical values $10^{-31} - 10^{-29}\ cm^2$ to $10^{-17} - 10^{-16} cm^2$ per molecule that correspond to the enhancement factors up to $10^{14}$. The results in [3] were obtained for the NIR excitation $830nm$, which exclude resonance enhancement for the above molecules. As an evidence of such observations may serve the SER spectra in Fig. 22 and the amplitude of the line 1174 $cm^{-1}$ for the spectra of crystal violet (Fig. 23a).

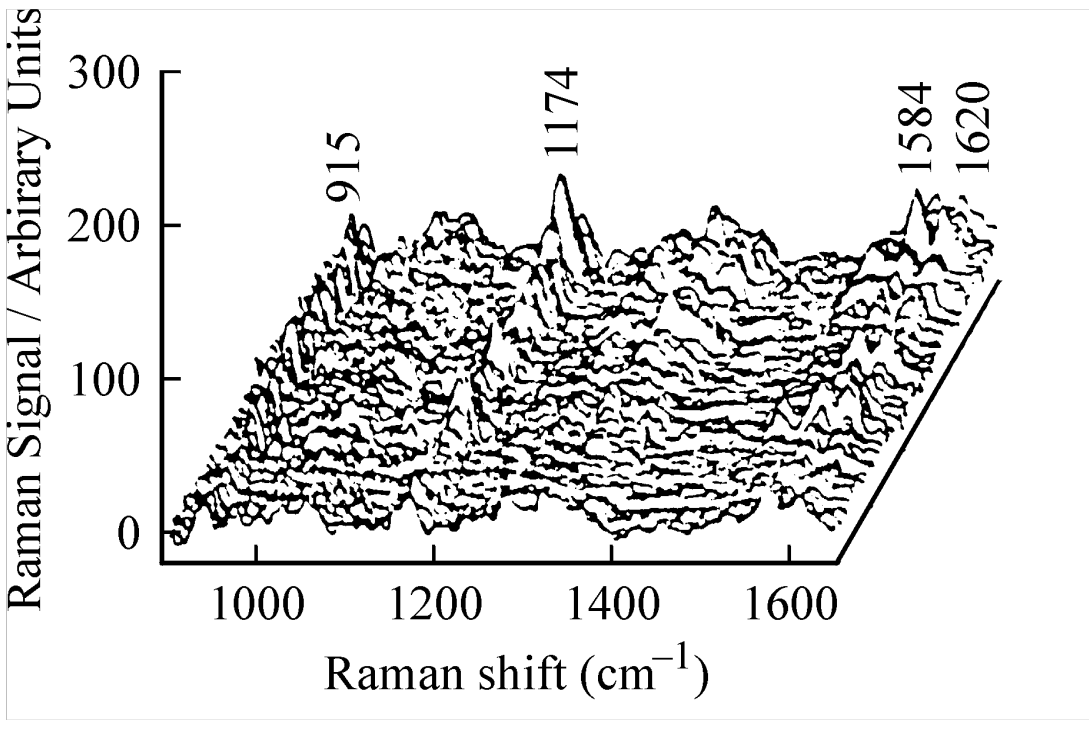

Fig. 22 100 SERS spectra collected from $30pl$ scattering volume containing an average of 0.6 crystal violet molecules, displayed in the time sequence of measurements. Each spectrum is acquired in 1 second.

Figs. 23b and 23c display the background levels without crystal violet molecules and with $10^{14}$ methanol molecules in the scattering volume respectively. The same order of the signals from methanol molecules confirms the above enhancement factor. The spectra of crystal violet are very unstable in the course of time, which correspond to presence of only few molecules in the scattering volume. In addition, statistical analysis of the spectra of methanol, which do not display SERS (Fig. 24a), and the spectra of the line 1174 $cm^{-1}$ of the crystal violet molecules (Fig. 24b and 24c) reveals the Poisson statistics for the Raman signal for the last case, that correspond to appearance of single molecules in the probed volume. The last result demonstrates that single molecules can be detected by the SERS method. Investigations of Emory and Nie [4] performed on colloidal particles revealed that only $0.1-1\%$ of the particles provide the enormous enhancement. A microscopic examination of these

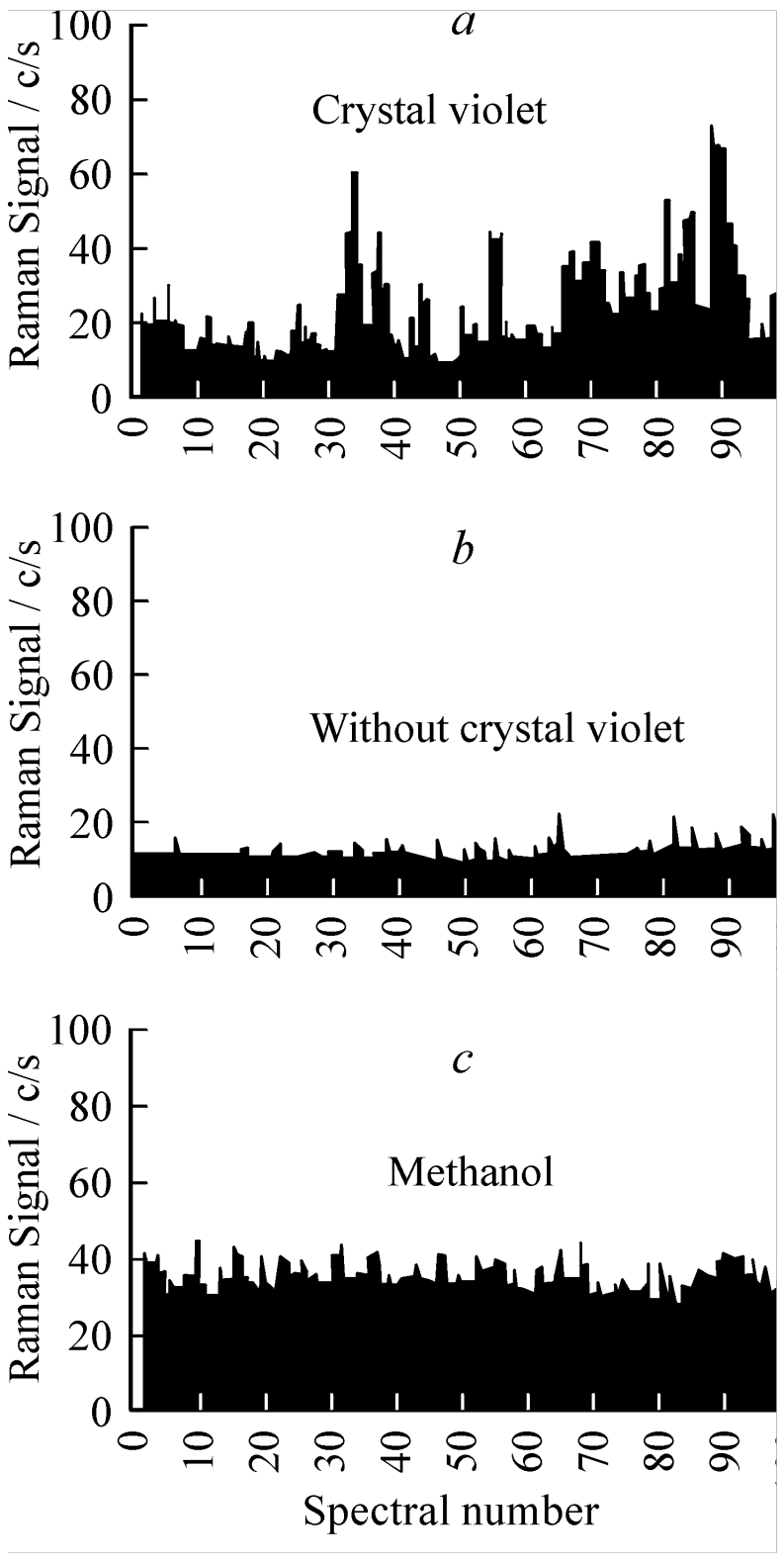

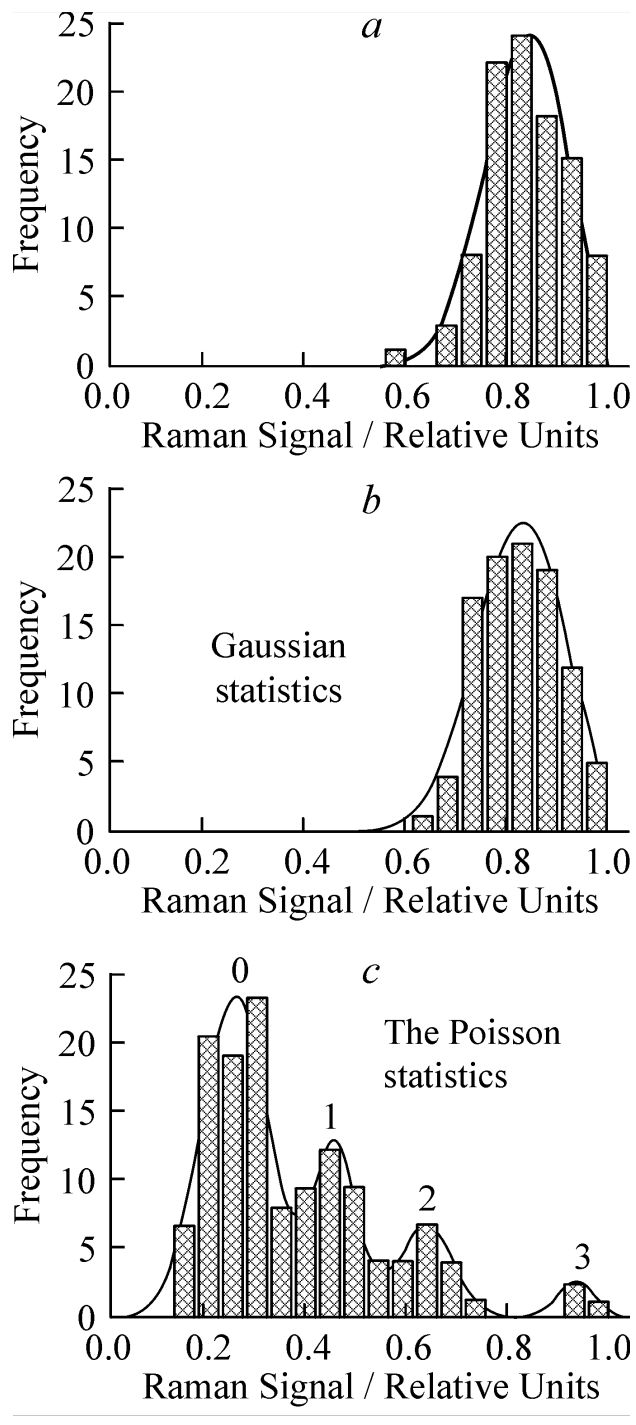

Fig. 23 Fig. 24

Fig. 23 (a) Peak heights of the $1174cm^{-1}$ line for the 100 SERS spectra shown in Fig. 22. (b) Signals measured at $1174cm^{-1}$ for 100 spectra from a sample without crystal violet to establish the background. (c) Peak heights of the Raman line for 100 spectra measured from $3M$ methanol.

Fig. 24 (a) Statistical analysis of 100 "normal" Raman measurements at $1030cm^{-1}$ of $10^{14}$ methanol molecules, (b) Statistical analysis of 100 SERS measurements ($1174cm^{-1}$ Raman line) of six crystal violet molecules in the probed volume. The solid lines are Gaussian fits to the data. (c) Statistical analysis of 100 SERS measurements ($1174cm^{-1}$Raman line) for an average of 0.6 crystal violet molecules in the probed volume. The peaks reflect the probability to find just 0, 1, 2, or 3 molecules in the scattering volume and correspond to the Poisson statistics.

colloidal particles demonstrated, that these “hot” particles have a distinctly faceted shape. Occasionally there were some particles with a rod-like form. The situation for these forms can be analyzed in terms of the quadrupole SERS mechanism, based on the conception of the strong dipole and quadrupole light-molecule interactions, arising in surface fields strongly varying in space [50]. The values of the enhancement coefficients follow approximately from formulas (20, 23 and 48) for such models of the roughness as a finite wedge or a cone, which correspond to the crossings of facets or the rod-like particles. For our estimations we should find the value of $l_1$, for particles, which may differ from the one for the lattices. Its value can be taken to be approximately equal to the height of the finite wedge or cone. It is evident, that when the height $l_1 \rightarrow 0$ the particle dissapears and there is no enhancement. Besides for the volume of $l_1$ radius one can expand the electric field using special functions. We estimated the enhancement coefficient for the molecule of crystal violet placed at the distance $r = a$ from the wedge or cone in the configuration shown in (Fig. 25). We can hope that at this distance our expansion of the light-molecule interaction Hamiltonian, using the dipole and quadrupole terms is valid. Using the approach of paragraph 5 for estimation of mean values of matrix elements, the conventional values of the constants, the values of the oscillator strength $\overline{f}_{nm} = 0.1$ and $\omega_{nm}$ which corespond to the wavelength $\lambda = 830nm$ we can obtain in the current configuration $B = 115$ for the Z-axis and $B = 271$, $a = 1.4nm$ for the Y-axis [50]. The enhancement factor for only the dipole mechanism $G_d$ can be estimated using the formulae (45) and (23) for both the incident and scattered fields using corresponding definitions of the values in (45). Then

$$G_d \sim C_0^4 \left( \frac{l_1}{r} \right)^{4\beta} \tag{80}$$

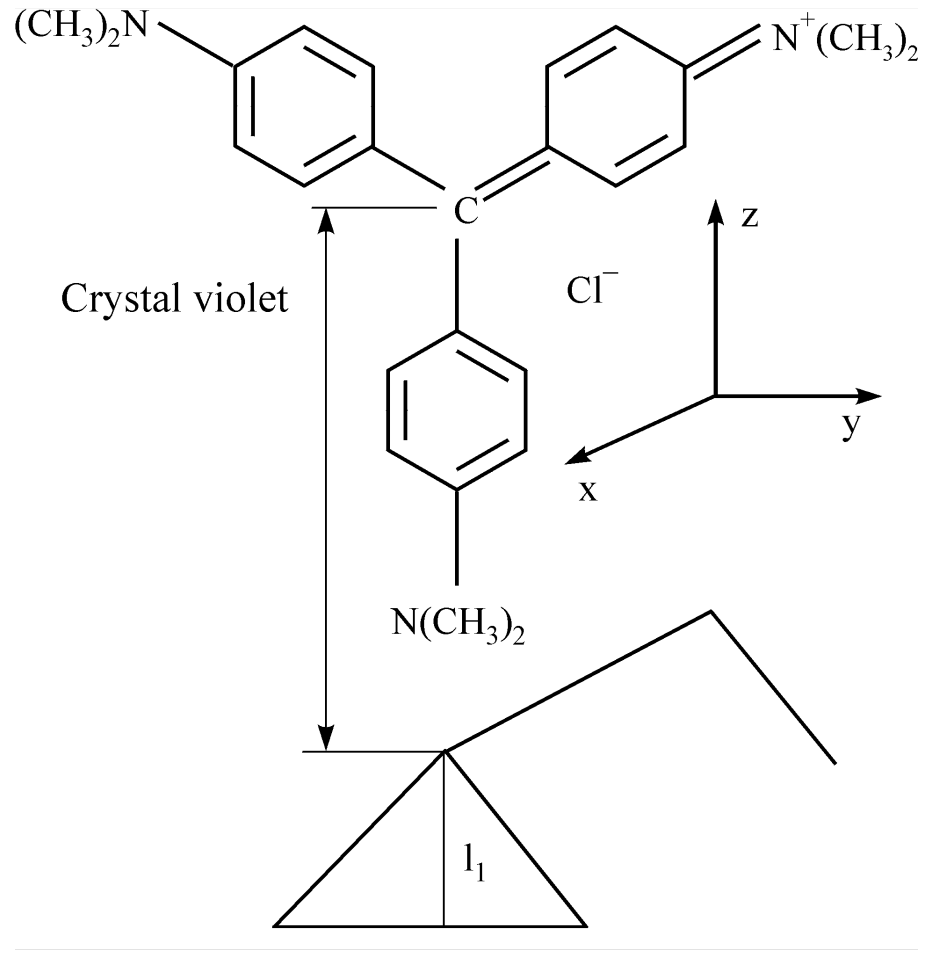

Fig. 25 Configuration of the system: a molecule of crystal violet near the wedge

The relation between the quadrupole and dipole enhancement factors $G_Q = G_0$ and $G_d$ for a rectangular wedge in the Y direction can be estimated using (20) or (23) and (48) in the manner, mentioned above. Then

$$\frac{G_Q}{G_d} \sim \beta^4 \left(\frac{B}{2}\right)^4 \sim 6.4 \times 10^6 \qquad (81)$$

Thus it can be seen, that the dipole approximation is not valid for these models of the rough particles. Estimation of the enhancement coefficients for various models of the particles in accordance with (48) gives the following results. For $l_1 \approx 150 nm$ and $\beta = 1/2,$ which correspond to both a plane, the limiting case of the wedge and a cone with a finite angle at the top, $G_0 = 2.1 \times 10^{11}$. For the limiting case of a tip or a needle with $\beta = 1,$ $G_0 \approx 1.7 \times 10^{16}$. It can be seen, that for various models and parameters values we can obtain estimations of the enhancement coefficients in the range $10^{10} - 10^{16}$ which roughly correspond to those obtained by Kneipp et al. [3]. It is necessary to take into account, that our estimations may give even larger values than those in real systems. When a molecule approaches the top of a wedge or a cone, the strength of the heterogeneous field increases, that may result in a significant increase of the enhancement coefficients. This corresponds to the case, when the full number of expansion terms should be taken into account. Because the field is very close to a singular field, we cannot do this in practice. It is necessary to take into account the precise expression for $\bar{A}$ in the light-molecule interaction Hamiltonian (30). At the same time, the finite value of the conductivity must decrease the enhancement coefficients. Thus our rough estimations lead to the opinion, that the reason for the enormous enhancement is the strong variation in space of the nearly singular electromagnetic fields near intersections of facets or in the vicinity of the tops of rod-like particles. The author clearly understand the roughness of the estimates, but emphasizes the actual possibility of such enormous coefficients.

It should be noted that the integral part of the phenomenon of Single Molecule Detection is blinking of the SERS signal, discovered in the works of Emory and Nie. The signal from a separate molecule changes its intensity in the intermittent regime. The amplitudes of separate signals and the time interval between them depend on many factors. Here we do not describe all experimental features of the blinking phenomenon, since this topic does not refer to the dipole-quadrupole theory directly. However we can give its qualitative explanation.

The electromagnetic field is strongly enhanced in the areas of the tops of large roughness or large particles with strong curvature, as it has been shown above. This field is strongly localized near these points. The number of molecules in experiments on Single Molecule Detection is very low and usually approximately corresponds to the number of colloidal particles. Therefore these molecules diffuse in space and when they move into the regions with the strong electromagnetic field the SERS signal appears. Wherease when they move out from these regions the signal disappears. The intensity and the time width of this signal depends on the path inside the enhanced region and the intensity and configuration of the electromagnetic field near the active site. Besides the signal depends on the temperature of the molecular vibrational system via the temperature of surrounding medium and some other factors. The presented picture is rather complicated. However the above explanation can give some rough notion about the blinking phenomenon. For more correct explanation a carefull analysis of a large number of experimental results, obtained in systems with various conditions is requied.

## 21. CHARGE TRANSFER ENHANCEMENT MECHANISM

For a long time it was considered, that there is no enhancement mechanism on smooth metal single surfaces. However, as it has been mentioned above Campion et al. discovered a system consists of pyrometallic dianhydride PMDA on $Cu(111)$ and $Cu(100)$ [9], which demonstrates the enhancement on an atomically smooth metal surface, for which the electromagnetic contribution can be considered unimportant. This provides strong evidence of a chemical enhancement mechanism operating independently of the electromagnetic mechanism. It was suggested, that this mechanism is just a resonance Raman scattering mechanism via intermediate charge transfer states. Experimental studies demonstrated that the intensity of Raman scattering strongly depends on the wavelength of the incident light (725 and 647 nm) (Fig.26).

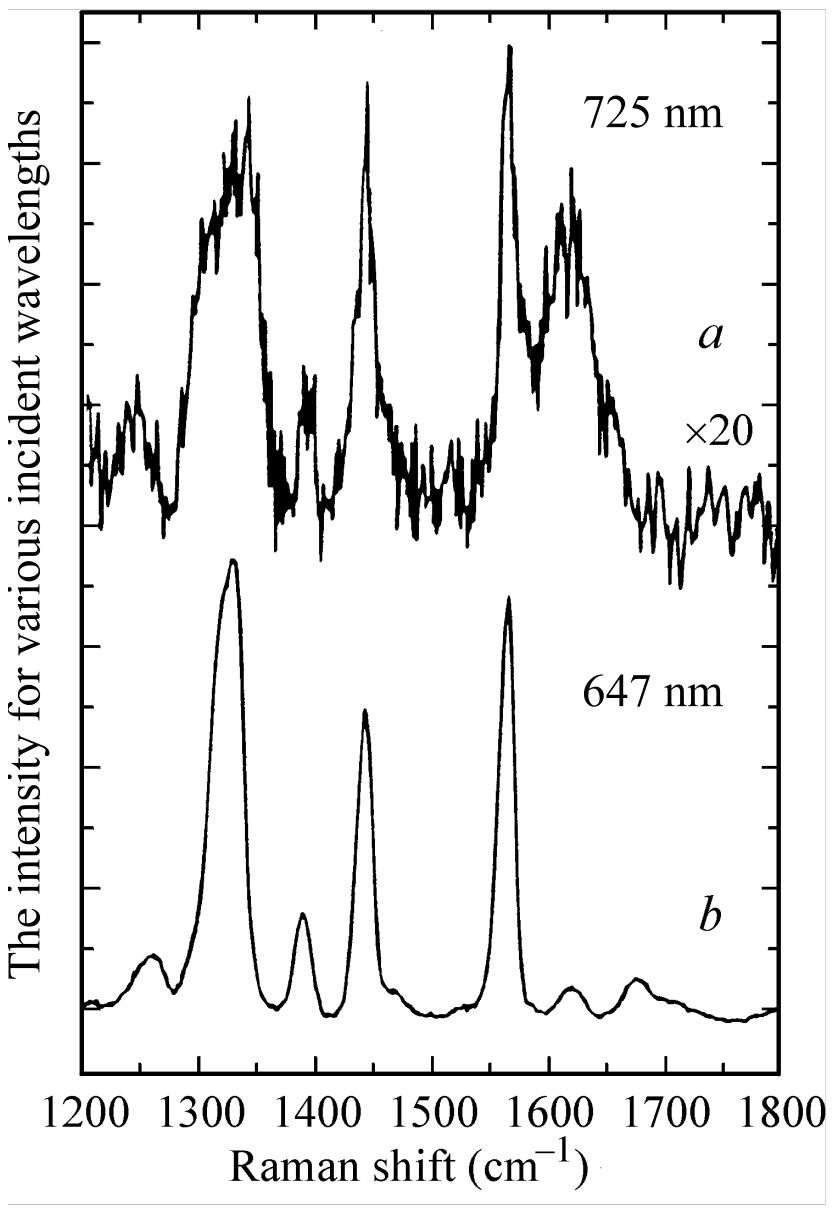

Fig. 26 $TH$ -Polarized Raman spectra of $10A$ of PMDA on $Cu$ (111) at (a) $725nm$ (×20), (b) 647 nm

This points out the resonance character of the enhancement. In addition the results of EELS measurements, demonstrate that there is a peak in the spectrum at~1.9 eV [51], which can be attributed to an electronic excitation of a resonant state (Fig.27). Experimental observations of the enhancement coefficients for different lines give different values. However as it was mentioned above the enhancement coefficients vary approximately within the interval ~ 30-100 (the intensity of~ 2000-1000 counts $s^{-1}w^{-1}$ for $TH$ - polarized excitation for PMDA on the surface and less than 100 counts $s^{-1}w^{-1}$ for PMDA in the volume).

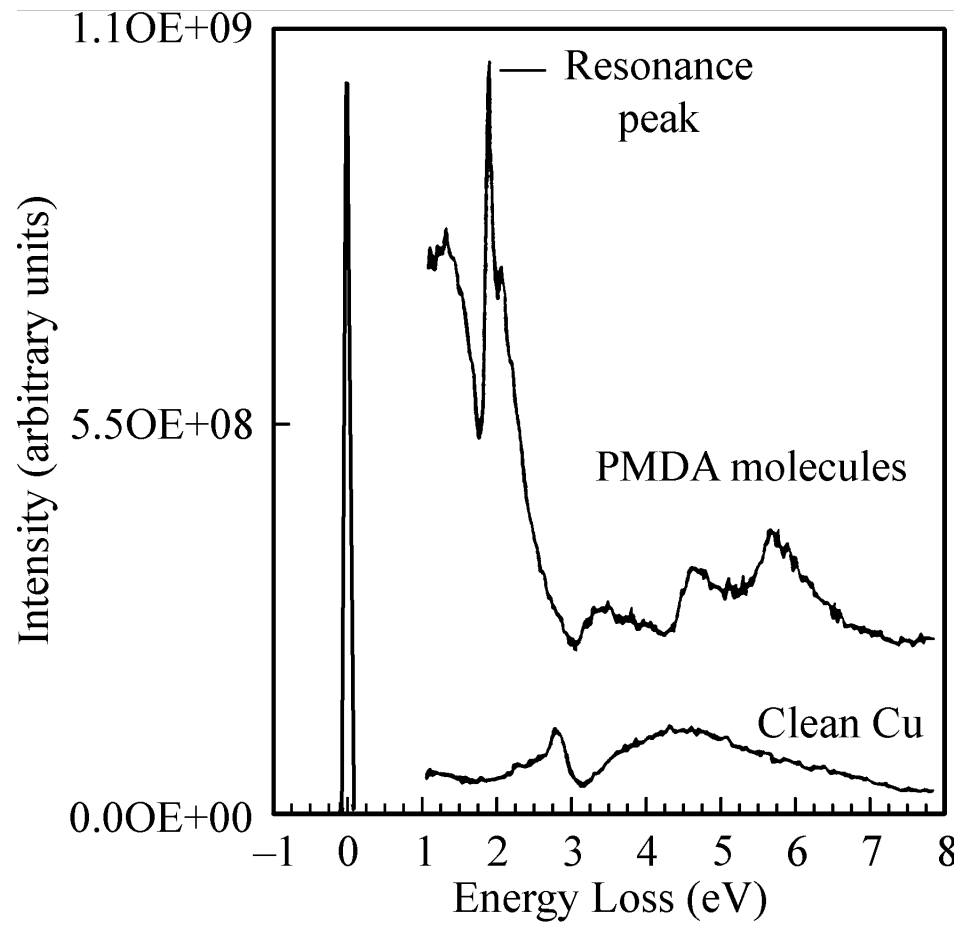

Fig. 27 HREELS spectrum of a $10A$ PMDA thin film adsorbed on $Cu(111)$ at $110K$. The primary beam energy was 11 eV and the spectral resolution was 50 meV. The spectrum demonstrates existence of a strong absorption peak at 1.9 eV, which is attributed to the charge transfer electronic state.

Substitution of PMDA with benzene results in the absence of the enhancement, which points out the chemical mechanism and the absence of the electromagnetic enhancement. As it is stated in literature [9] PMDA is dissociatively chemisorbed by splitting out CO producing a surface carboxylate. For $Cu(111)$ and $Cu(110)$

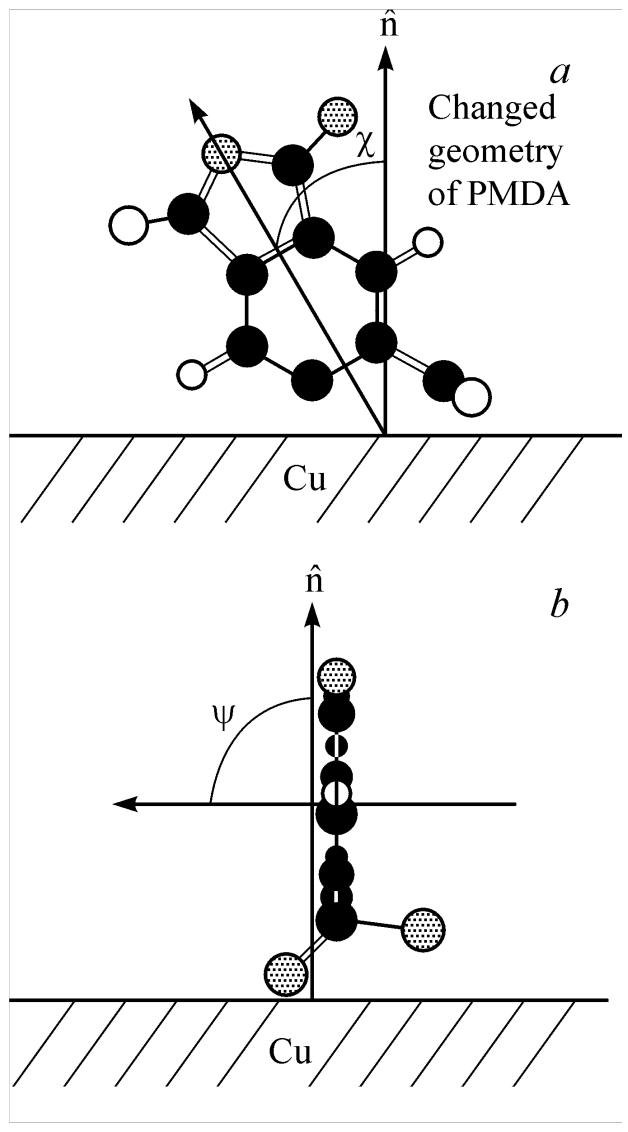

Fig. 28 Adsorption configuration of PMDA on $Cu$ showing the angles (a) $\chi$, between the long axis of the molecule and the surface normal; (b) $\psi$, between the molecule normal and the surface normal; C (black), O (grey), H (white)

the plane of the chemisorbed carboxilate is tilted away with respect to the normal of the surface with its oxygen atoms inequivalently bound to the copper surface, while the plane containing the aromatic moiety of the molecule is oriented perpendicular to the plane of carboxylate (Fig. 28). This configuration allows the unsatisfied valence of the aromatic moiety of the molecule to interact with the surface. It should be noted that the total and relative intensities of the Raman bands depend on the type of the single surface and on the polarization of the incident field (fig 29, 30).

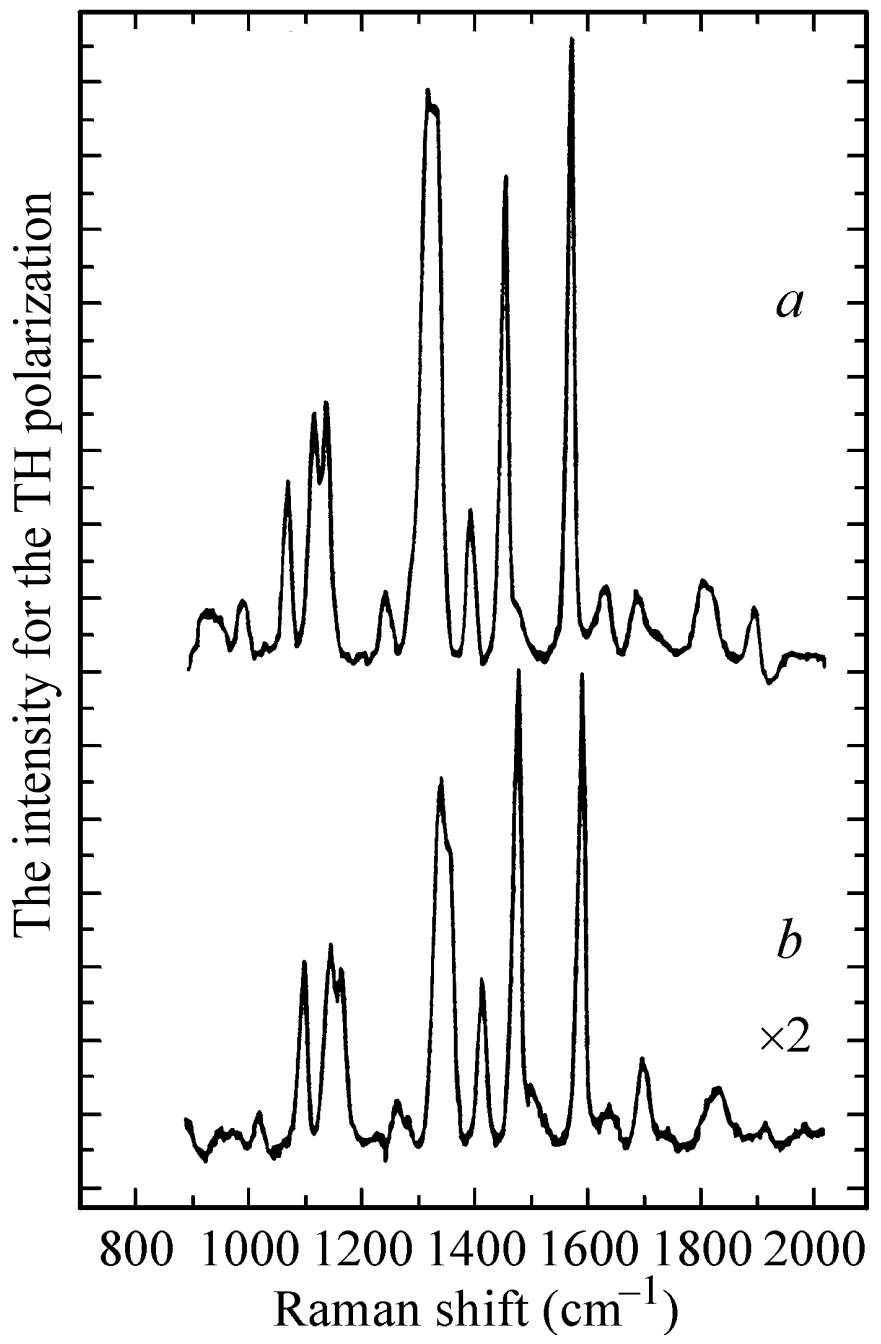

Fig. 29

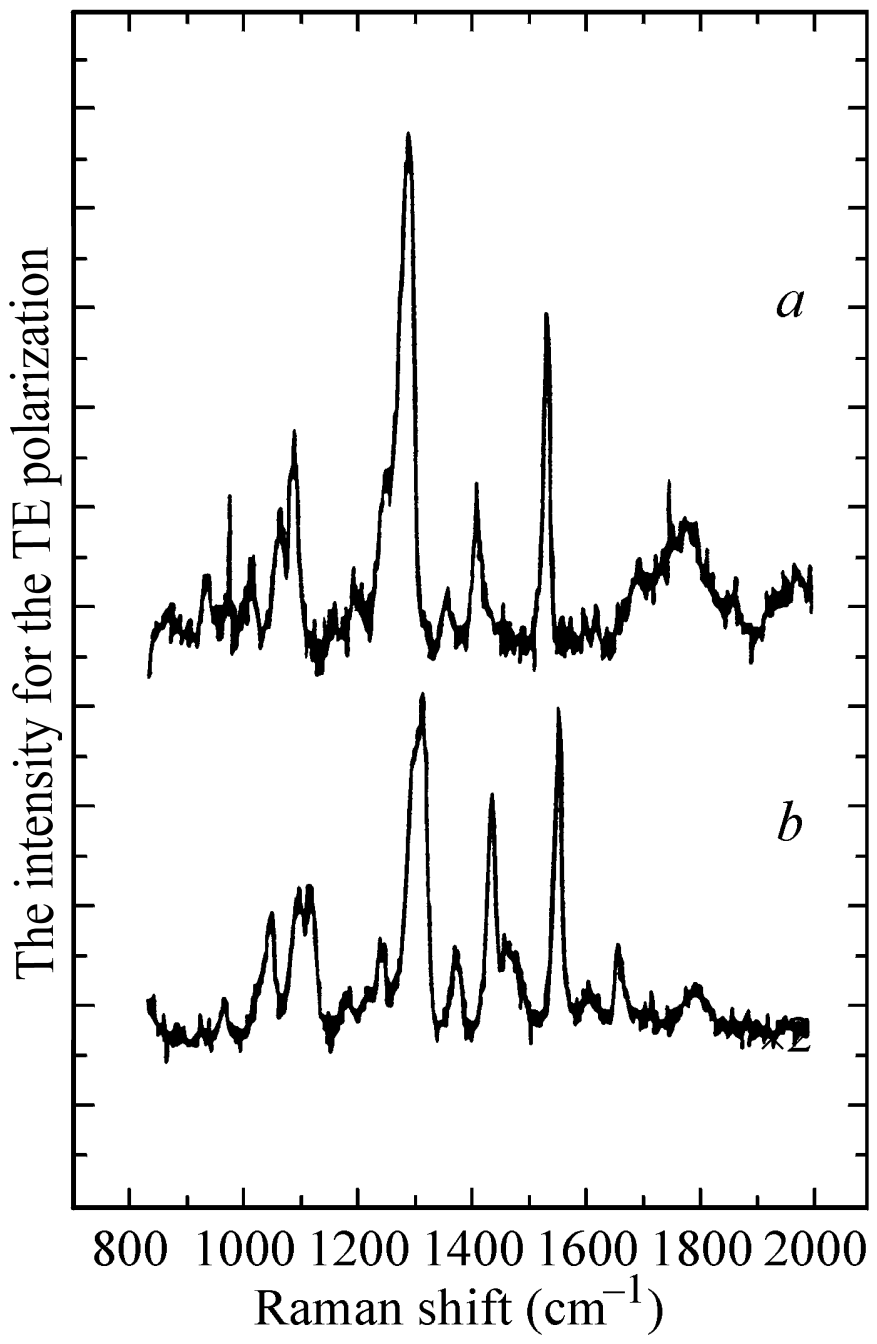

Fig. 30

Fig. 29 *TH*-polarized Raman spectra of 7A of adsorbed PMDA on copper at 647*nm*; PMDA/*Cu* (111), (b) PMDA/*Cu* (100) (×2)

Fig. 30 *TE*-polarized Raman spectra of 7A of adsorbed PMDA on copper at 647*nm*; PMDA/*Cu* (111), (b) PMDA/*Cu* (100)

Usually the peaks observed on $Cu(111)$ have a higher intensity than those on $Cu(100)$. We do not repeat the whole content of [9] concerning these issues, which includes simulations because they are based on very approximate ideas. From our point of view there is indeed a charge transfer mechanism due to some resonant state. However the Raman scattering on this molecule occurs in the surface field, associated with atomic nature of the crystal faces. This field can be expressed as a double Fourier series with periodicity equal to that of the crystal surface. It is evident, that both tangential and normal components of the electric field are excited under illumination with the *TH* and *TE* polarizations. Therefore there is no dramatic change of the Raman intensities in experiment compared with the case of the ideally smooth surface. Besides, since the surface field varies for different faces, this fact results in variation of the absolute and relative intensities of various Raman lines for them. It should be noted that there may be additional changes in the enhancement on atomically rough surfaces due to the changes in bonding compared with single surfaces. It should also be noted that the chemical enhancement, associated with the direct contact of the molecule with the surface is weak (~30-100) as it was mentioned above with respect to the purely electromagnetic enhancement and it is presently observed only for one molecule. Thus the chemical enhancement does not play any important role in SERS, and apparently manifests only in some special cases.

# 22. EXPLANATION OF EXPERIMENTAL PHENOMENA ACCOMPANYING SERS

Basing on the theory presented above we can explain the majority of experimental facts revealed in SERS, in the phenomenon of Single Molecule Detection and other effects from the same point of view. First of all it is role of the surface roughness in both phenomena.

The region of the roughness is a strongly heterogeneous medium. Therefore the electromagnetic field in this region strongly differs from that in a free space and is strongly heterogeneous too. In most cases, when the characteristic size of the roughness is significantly larger than the size of the molecule, we can restrict consideration to the dipole and quadrupole components of light-electrons interaction Hamiltonian (31, 31a). Both the increase of the component of the field, which is perpendicular to the surface and increase of the electromagnetic field derivatives raise the value of the dipole and quadrupole light-molecule interactions. Besides the quadrupole interaction can be more, than the dipole one both due to enhancement of the electric field derivatives and due to the quantum mechanical features of the quadrupole interaction. This accounts for the appearance of the enhancement on rough surfaces, coldly evaporated silver films, island films, colloidal particles, tunnel junctions and other similar objects.

The long-range character of the enhancement in SERS is associated with its electrodynamic nature. The enhancement exists in the whole region of existence of the surface field, which strongly varies in space with the characteristic size equal to the localization size of the surface modes with approximate value $\frac{l_E}{2\pi}$.

The results of the experimental study of the distance dependence of the enhancement in the SERS phenomenon [13] confirm this conclusion (Fig. 2). Estimation of the localization range obtained by the authors of [13] for ordinary SERS gives the value that varies in the interval (2-5) nm and corresponds to the estimate of the mean roughness size of (10-30) nm. This corresponds to our concept concerning the rough structures, used in experiments with ordinary SERS.

As it was noted in [14-16] the maximum of the enhancement is observed on so-called active sites, which disappear after annealing at high temperatures for some type of the roughness, such as coldly evaporated silver films. It is known from [52] that maximum enhancement in SERS occurs near the sharpest roughness. Similarly according to [4, 53] the effect of enormous SERS occurs near the crossing of facets, or near the tops of rod-like particles. Besides as it was demonstrated in [6, 7, 54, 55] adatoms and steps of atomic size do not cause any strong enhancement. Therefore if we consider the sufficiently sharp roughness, that causes appearance of the strong fields and strong field derivatives as the active sites (Fig. 31) then this idea coincides with the conclusion that sites of this kind exist in SERS [14-16].

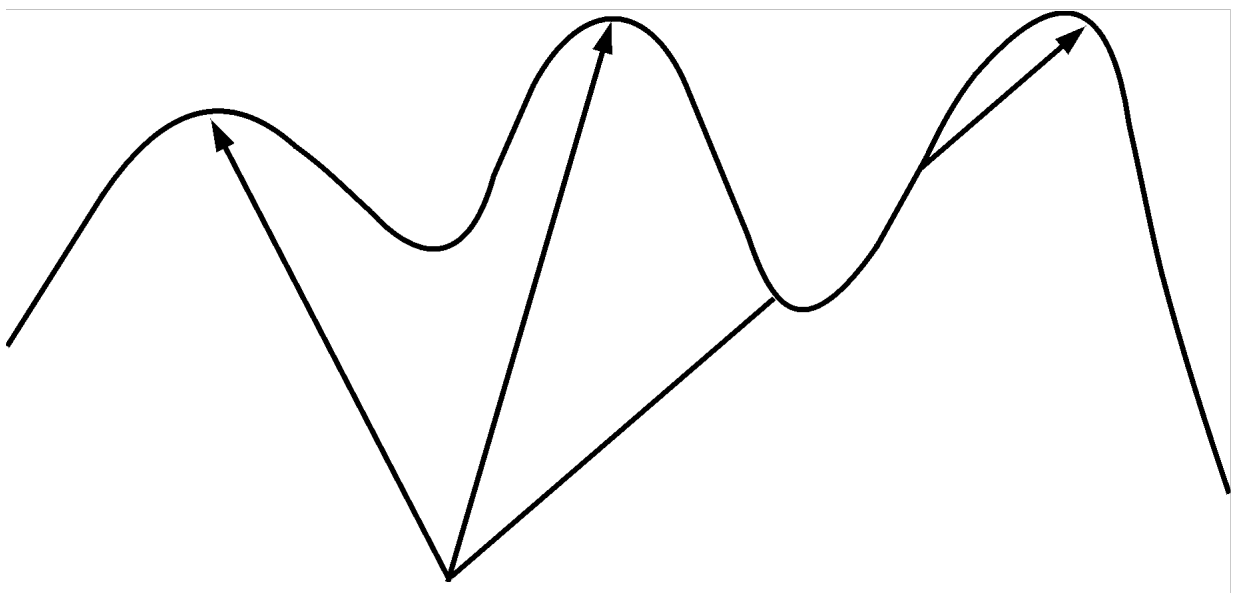

Fig. 31 The active sites

Thus in our opinion the active sites are the sharpest irregularities.

It should be noted that disappearance of the SERS signals after heating of substrates on Fig. 3a and 3b is associated with annealing of these special surfaces and irreversible disappearance of the roughness. Therefore the surface becomes flat and SERS disappears.

The larger the size of these sharp irregularities, the stronger the enhancement, which reaches enormous values in our estimations [50] and is confirmed by the experiments of Kneipp [3] and Emory and Nie [4, 53]. Thus the effect of enormous SERS can be considered as manifestation of the strong enhancement on large sharp irregularities, which increase the efficiency of the active sites.

As it is well known the SERS signal is depolarized. This circumstance is explained by the fact, that the adsorbed molecules are randomly oriented in space and the probabilities of the scattering in the $TE$ and $TH$ polarizations are approximately equal.

The SERS phenomenon was observed not only on metals, but also on the rough surfaces of semiconductors and oxides [5]. The appearance of the enhancement in this case is associated with the same reason – the enhancement of the electric field and their derivatives in the strongly irregular medium. This point of view proves the validity of the proposed mechanism. However the detailed explanation of SERS for such systems, requires more detailed investigations.

As it was pointed out earlier there is a short-range effect in SERS. The ratio of the degrees of enhancement by the first and the second layers is about 100. Now it is clear, that the short range effect is of the electrodynamic rather than the chemical nature [11, 12] and is associated with a very large difference between the enhancement coefficients for the first and the second layers near the active sites, where the electric fields and their derivatives vary very strongly. The corresponding estimations can be made for the roughness of the wedge or cone shape using formula (48) for the enhancement coefficient. Then for the totally symmetric vibrations and $0 < \beta \leq 1/2$ the ratio of enhancement coefficients for the molecules adsorbed in the first and in the second layers (Fig. 32) is given by the value

$$\left(\frac{r_2}{r_1}\right)^{4+4\beta} = (3)^{4\beta+4} \sim 100 - 1000 \tag{82}$$

which is sufficient for explanation of the short range effect in terms of the purely electrodynamic model.

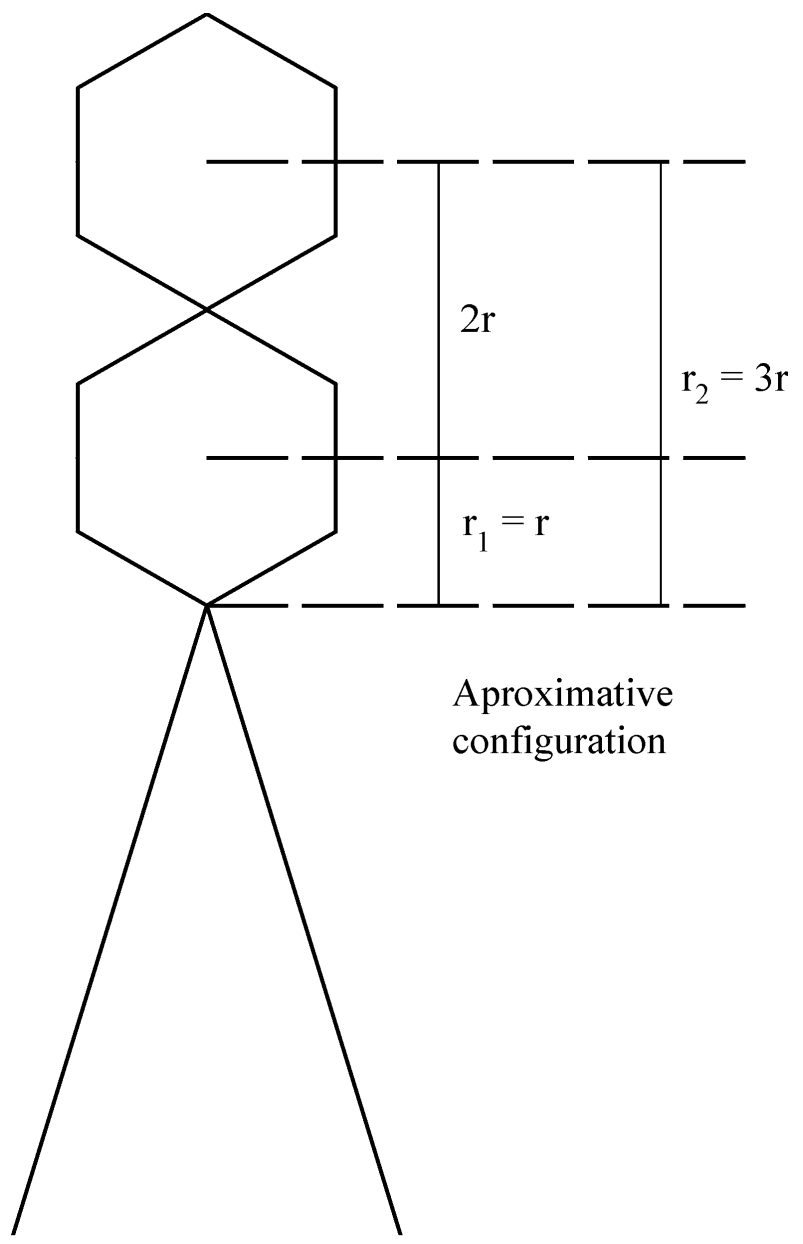

Fig. 32 Molecules adsorbed on the top of the wedge in the first and the second layer

The author clearly understands the roughness of the estimate in this model, however it is apparently sufficient for explanation of the first layer effect by the pure electrodynamic mechanism. It is evident, that the first layer effect must be observed not only for the totally symmetric but for another modes too. Besides the observed first layer effect apparently is not the chemical charge transfer mechanism, because the last one is observed only for PMDA molecule.

The deviation of the frequency dependence from the $(\hbar\omega)^4$ behavior (Fig. 4) is due to variation of the complex dielectric constant with frequency. When we approach the metal plasma frequency the metal dielectric constant decreases to zero and further to unity and the properties of the metal become very close to the properties of vacuum. Therefore the surface field disappears and the enhancement of the fields and their derivatives decreases, thus causing the decrease of the SER signal.

It should be noted that the dependence of the SERS signal on the kind of the metal substrate is determined by the different values of the complex dielectric constant. These values determine $\bar{E}$ and $\partial E_\alpha / \partial x_\alpha$ and hence the enhancement coefficients of various bands.

## 23. CONCLUSION

Thus we have summarized in this review the results concerning explanation of SERS in terms of the strong dipole and quadrupole interactions in the surface fields strongly varying in space. In accordance with selection rules, obtained in the paper for general type of symmetrical molecules, it appears that the dipole-quadrupole theory is consistent with specific features of the SER spectra of such molecules. The theory can account for the following

1. The order of magnitude of the enhancement coefficient for the ordinary and enormous SERS (The Single Molecule Detection by the SERS method) on usual and some special metal surfaces with very strong roughness and on very large colloidal particles.
2. The preferential enhancement of the bands, associated with the totally symmetric vibrations and the strongest enhancement of the breathing mode among them.
3. The appearance of a large number of forbidden lines in molecules with high symmetry.
4. Anomalies of the SER spectra of molecules adsorbed on transition-metal substrates and some other specific systems such as benzene adsorbed on lithium and hexafluorobenzene adsorbed on silver.
5. The slight enhancement of SERS on methane and molecules with cubic symmetry groups, (Electrodynamic forbiddance of the SERS enhancement mechanism).
6. Explanation of the phenomena, accompanying SERS.

It should be noted that the theory presented is valid for surfaces with sufficiently strong roughness, on which the quadrupole interaction can manifest oneself. Consideration of other experimental results and Surface Enhanced processes requires a very careful and detailed analysis of experimental situation. Therefore we restrict consideration in this review only by our own results on the dipole-quadrupole theory and some results, which are spread enough in literature. The critical approach to the latter is valuable and helps to understand clearly the situations arising in observation of SERS under various experimental conditions.

# 24. APPENDIX 1

## WAVEFUNCTIONS OF ARBITRARY MOLECULES

In order to obtain the expression for the SER cross-section it is necessary to know the molecular wavefunctions that take into account the vibrations of nuclei. They can be obtained in the framework of adiabatic perturbation theory [56, 57] by expanding the Hamiltonian $\hat{H}_{mol}$ (26) in the set of the deviations of nuclei from their equilibrium positions. However the direct expansion of (26) has some difficulties. In accordance with the adiabatic perturbation theory $\hat{H}_e$ in (26) can be considered as zero approximation, and $\hat{H}_n + \hat{H}_{e-n}$ as a perturbation. The second term in (28) can be expanded by direct expansion of $1/\left|\bar{R}_{JK}\right|$ in the Taylor series. The $1/\left|\bar{R}_{iJ}\right|$ cannot be expanded directly, because there is a region of radius vectors $\bar{r}_i$ in which the inequality

$$\left|\bar{R}_J^0 - \bar{r}_i\right| > \left|\Delta \bar{R}_J\right| \quad \text{(A1-1)}$$

is not valid.

Therefore let us use the following method to solve (33). Let us try to find the solution in the form

$$\Psi = a^{(0)}\Psi^{(0)} + \Delta\Psi \quad \text{(A1-2)}$$

With the following expression used for the energy

$$\mathrm{E} = \mathrm{E}^{(0)} + \Delta\mathrm{E} \quad \text{(A1-3)}$$

Here $\Psi^{(0)}$ satisfies the equation

$$\hat{H}_e \Psi^{(0)} = E^{(0)} \Psi^{(0)} \qquad \text{(A1-4)}$$

$a^{(0)}$ is some coefficient, $\Delta\Psi$ is an additional part of the molecule wave function, arising due to vibrations, $E^{(0)}$ is the energy of unperturbed Hamiltonian, $\Delta E$ is perturbation of the energy.

Substituting (A1-2) and (A1-3) in (33) we can obtain, taking into account (26)

$$(\hat{H}_e - E^{(0)})\Delta\Psi = -(\hat{H}_n + \hat{H}_{e-n} - \Delta E)(a^{(0)}\Psi^0 + \Delta\Psi) \qquad \text{(A1-5)}$$

Let us designate the eigenfunctions of (A1-4) as $\Psi_n^{(0)}$ and eigenvalues as $E_n^{(0)}$, The additional part $\Delta\Psi_n$ can be expanded in terms of the eigenfunctions of equation (A1-4)

$$\Delta\Psi_n = \sum_{\substack{m \\ m \neq n}} a_{nm} \Psi_m^{(0)} \qquad \text{(A1-6)}$$

Substituting (A1-6) in (A1-5) and projecting on $\Psi_n^{(0)}$ we can obtain the following equation.

$$(\hat{H}_n + \langle n|\hat{H}_{e-n}|n\rangle - \Delta E_n) a_n^{(0)} + \sum_{\substack{m \\ m \neq n}} \langle n|\hat{H}_{e-n}|m\rangle a_{nm} = 0 \qquad \text{(A1-7)}$$

For the projection on $\Psi_l^{(0)}$ ($l \neq n$) we can obtain

$$(E_l^{(0)} - E_n^{(0)}) a_{nl} = -\langle l|\hat{H}_{e-n}|n\rangle a_n^{(0)} - \sum_{\substack{m \\ m \neq n}} \langle l|\hat{H}_n + \hat{H}_{e-n}|m\rangle a_{nm} + \Delta E_n a_{nl}$$

(A1-8)

Let us pass to another variables

$$\Delta\overline{R}_J = \lambda'\overline{U}_J \tag{A1-9}$$

Where

$$\lambda' = \sqrt[4]{\frac{m}{\overline{M}}} \tag{A1-10}$$

is a new parameter of the adiabatic perturbation theory and $\overline{M}$ is a mean mass of nuclei. Equations (A1-7) and (A1-8) can be expanded in powers of this small parameter. Therefore it is necessary to introduce the following expansions for $a_{nl}$ and $\Delta\mathrm{E}_n$

$$a_{nl} = \sum_{k=1}^{\infty} a_{nl}^{(k)} \lambda'^k \tag{A1-11}$$

$$\Delta\mathrm{E}_n = \sum_{k=1}^{\infty} \mathrm{E}_n^{(k)} \lambda'^k \tag{A1-12}$$

The matrix element $\langle n|\widehat{\mathrm{H}}_n|n\rangle = \widehat{\mathrm{H}}_n$ permits a simple analytical expansion, which can be obtained, expanding the functions $1/\left|\overline{R}_{JK}\right|$

$$\widehat{\mathrm{H}}_n = 1/2 \sum_{\substack{JK \\ J\neq K}} \frac{Z_J^* Z_K^* e^2}{\left|\overline{R}_J^0 - \overline{R}_K^0\right|} -$$

$$-1/2\lambda' \sum_{\substack{JK \\ J\neq K}} \frac{e^2 Z_J^* Z_K^* ((\overline{R}_J^0 - \overline{R}_K^0)(\Delta\overline{U}_J - \Delta\overline{U}_K))}{\left|\overline{R}_J^0 - \overline{R}_K^0\right|^3} +$$

$$\lambda'^2 \left[ \begin{array}{l} -\dfrac{\hbar^2}{2m}\sum_J \dfrac{\overline{M}}{M_J}\Delta_{\overline{U}_J} - 1/4\sum_{JK} \dfrac{e^2 Z_J^* Z_K^* \left|\Delta\overline{U}_J - \Delta\overline{U}_K\right|^2}{\left|\overline{R}_J^0 - \overline{R}_K^0\right|^3} + \\ 3/2 \sum_{\substack{JK \\ J\neq K}} \dfrac{e^2 Z^* Z^* ((\overline{R}_J^0 - \overline{R}_K^0)(\Delta\overline{U}_J - \Delta\overline{U}_K))^2}{\left|\overline{R}_J^0 - \overline{R}_K^0\right|^5} \end{array} \right] \tag{A1-13}$$

From expression (29) we can see, that the expansion of matrix elements $\langle l|\hat{\mathrm{H}}_{e-n}|n\rangle$ begins from the first power of $\Delta\bar{R}$ and hence from the first power of $\lambda'$. Then

$$\langle l|\hat{\mathrm{H}}_{e-n}|n\rangle = \lambda' \sum_{J\alpha} \frac{\partial\langle l|\hat{\mathrm{H}}_{e-n}|n\rangle}{\partial X_{J\alpha}} \Delta U_{J\alpha} + $$
$$+ \lambda'^2 \sum_{\substack{JK \\ \alpha\beta}} 1/2 \frac{\partial\langle l|\hat{\mathrm{H}}_{e-n}|n\rangle}{\partial X_{J\alpha}\partial X_{K\beta}} \Delta U_{J\alpha}\Delta U_{K\beta} \cdots\cdots \qquad \text{(A1-14)}$$

The terms which are linear in deviations, can be approximated in the following manner. Let us expand formally the terms $1/|\bar{R}_{iJ}|$ in expression (29) in powers of $\Delta X_{J\alpha}$ and keep only the terms, which are linear in $\Delta X_{J\alpha}$. Then we can write an approximate relation

$$\frac{\partial\langle l|\hat{\mathrm{H}}_{e-n}|n\rangle}{\partial X_{J\alpha}} \cong \sum_i \int \frac{e^2 Z_J^*}{\left|\bar{R}_{iJ}^0\right|^3} (X_{J\alpha}^0 - x_{i\alpha}) \Psi_l^* \Psi_n \times d^3 r_1..d^3 r_N \qquad \text{(A1-15)}$$

The following condition should be noted, because the nuclei are in equilibrium positions

$$-\sum_K \frac{e^2 Z_J^* Z_K^* (X_{J\alpha}^0 - X_{K\alpha}^0)}{\left|\bar{R}_J^0 - \bar{R}_K^0\right|^3} +$$
$$+\sum_i \int \frac{|\Psi_n|^2 e^2 Z_J^* (X_{J\alpha}^0 - x_{i\alpha})}{\left|\bar{R}_{iJ}^0\right|^3} d^3 r_1 \ldots d^3 r_N = 0 \qquad \text{(A1-16)}$$

Using expansions (A1-11 –A1-14), substituting them in (A1-7) and (A1-8) we can obtain relations for determination of $a_n^{(0)}$, $a_{nl}^{(k)}$ and $\mathrm{E}_n^{(k)}$. From expansion (A1-7) for the first power of $\lambda^{'}$ we can obtain the expression for the energy

$$\mathrm{E}_n^{(1)} = 0 \tag{A1-17}$$

and from (A1-8) the coefficients

$$a_{nl}^{(1)} = \frac{1}{(\mathrm{E}_n^{(0)} - \mathrm{E}_l^{(0)})} \sum_{J\alpha} \frac{\partial \langle l | \widehat{\mathrm{H}}_{e-n} | n \rangle}{\partial X_{J\alpha}} \Delta U_{J\alpha} a_n^{(0)} \tag{A1-18}$$

From expansion (A1-7) for the second power of $\lambda^{'}$ we can obtain an equation for determination $a_n^{(0)}$

$$\Bigg[ -\frac{\hbar^2}{2} \sum_J \frac{1}{M_j} \Delta_{\overline{R}_J} - 1/4 \sum_{\substack{JK \\ J \neq K}} \frac{e^2 Z_J^* Z_K^* \left| \Delta \overline{R}_J - \Delta \overline{R}_K \right|^2}{\left| \overline{R}_J^0 - \overline{R}_K^0 \right|^3} +$$

$$+ \frac{3}{2} \sum_{\substack{JK \\ J \neq K}} \frac{e^2 Z_J^* Z_K^* ((\overline{R}_J^0 - \overline{R}_K^0)(\Delta \overline{R}_J - \Delta \overline{R}_K))^2}{\left| \overline{R}_J^0 - \overline{R}_K^0 \right|^5} +$$

$$+ \sum_{\substack{l \\ l \neq n}} \sum_{\substack{JK \\ \alpha\beta}} \frac{1}{(\mathrm{E}_n^{(0)} - \mathrm{E}_l^{(0)})} \frac{\partial \langle l | \widehat{\mathrm{H}}_{e-n} | n \rangle}{\partial X_{J\alpha}} \times \frac{\partial \langle n | \widehat{\mathrm{H}}_{e-n} | l \rangle}{\partial X_{K\beta}} \Delta X_{J\alpha} \Delta X_{K\beta} +$$

$$+ \sum_{\substack{JK \\ \alpha\beta}} \frac{1}{2} \frac{\partial \langle n | \widehat{\mathrm{H}}_{e-n} | n \rangle}{\partial X_{J\alpha} \partial X_{K\beta}} \Delta X_{J\alpha} \Delta X_{K\beta} \Bigg] a_n^{(0)} = \lambda^{'2} \mathrm{E}_n^{(2)} a_n^{(0)} \tag{A1-19}$$

All the terms, except the first in (A1-19) are parts of the potential. Equation (A1-19) gives a solution for $a_n^{(0)}$ as a function of deviations from the equilibrium positions. The corresponding equation in normal coordinates $\xi_s$ with $s$ normal vibration is the following

$$\left[-\frac{\hbar^2}{2}\sum_s \frac{\partial^2}{\partial \xi_s^2} + \sum_s \frac{\omega_s^2 \xi_s^2}{2}\right] a_n^{(0)} = \lambda'^2 \mathrm{E}_n^{(2)} a_n^{(0)} \qquad \text{(A1-20)}$$

with the solution

$$a_n^{(0)} = \alpha_{\overline{V}} = \prod_s N_s H_{V_s}\left(\sqrt{\frac{\omega_s}{\hbar}}\xi_s\right)\exp\left(-\frac{\omega_s \xi_s^2}{2\hbar}\right) \qquad \text{(A1-21)}$$

Here $V_s$ is a quantum number of normal vibration. The set of $V_s$ can be designated as a vector $\overline{V}$. $H_{V_s}$ is a Hermitian polynomial of the $V_s$ power, $N_s$ is a normalization constant. The energy of vibrations

$$E_{\overline{V}} = \lambda'^2 \mathrm{E}_n^{(2)} = \sum_s (V_s + 1/2)\hbar\omega_s \qquad \text{(A1-22)}$$

The values of deviations are determined by the following formulas.

$$\begin{aligned} \Delta \overline{R}_J &= \sum_s \xi_s \overline{R}_{Js} \\ \Delta X_{J\alpha} &= \sum_s \xi_s \overline{X}_{Js\alpha} \end{aligned} \qquad \text{(A1-23)}$$

$\overline{R}_{Js}$ and $\overline{X}_{Js\alpha}$ are the displacement vector and its $\alpha$ projection of the $J$ nucleus in the $s$ vibration mode. Substituting (A1-23) in (A1-18) one can rewrite the expressions for the coefficients $a_{nl}^{(1)}$ in the following form

$$\lambda' a_{nl}^{(1)} = \frac{\sum_s \sqrt{\frac{\omega_s}{\hbar}}\xi_s R_{nls}}{(\mathrm{E}_n^{(0)} - \mathrm{E}_l^{(0)})} a_n^{(0)} \qquad \text{(A1-24)}$$

where

$$R_{nls} = \sum_{J\alpha} \sqrt{\frac{\hbar}{\omega_s}} \overline{X}_{Js\alpha} \frac{\partial \langle l | \hat{\mathrm{H}}_{e-n} | n \rangle}{\partial X_{J\alpha}} \tag{A1-25}$$

Then the general form of the wave function in the first approximation can be written as

$$\Psi_{n\overline{V}} = \left[ \Psi_n^{(0)} + \sum_{\substack{l \\ l \neq n}} \frac{\sum_s R_{nls} \sqrt{\frac{\omega_s}{\hbar}} \xi_s \Psi_l^{(0)}}{(\mathrm{E}_n^{(0)} - \mathrm{E}_l^{(0)})} \right] \alpha_{\overline{V}} \exp - (i \mathrm{E}_{n\overline{V}} t) / \hbar \tag{A1-26}$$

where

$$\mathrm{E}_{n\overline{V}} = \mathrm{E}_n^{(0)} + \sum_s \hbar \omega_s (V_s + 1/2) \tag{A1-27}$$

The term of the Hamiltonian (29) describes interaction of electrons with molecular vibrations or molecular phonons. Owing to the vibrations the potential of the nuclei changes by the value of deformation potential. Therefore we can regard the expansion of matrix elements $\langle 1 | \hat{\mathrm{H}}_{e-n} | n \rangle$ in (A1-14) as matrix elements of deformation potential, which cause electron transitions in the electron shell. The addition to the wave function in (A1-26) is due to deformation potential in a linear approximation. In the zero approximation the round state of the molecule is described by the electron function $\Psi_n^{(0)}$ with the energy $\mathrm{E}_n^{(0)}$ and the wave function of nuclei $\alpha_{\overline{V}}$ with the energy (A1-22). Owing to the deformation potential the electron shell can absorb or emit a molecular phonon (M. ph.). These processes arise due to the second term in (A1-26), when the only nonzero terms in matrix elements of vibrational quantum transitions are those with the change of one of the vibrational quantum numbers on one unit.

To calculate the SER cross-section, we need expressions for the virtual electronic and vibrational states. Their expressions can be obtained as those for the wave function of the ground state and can be written by changing of indices in corresponding expressions.

$$\Psi_{m\bar{V}} = \left[ \Psi_m^{(0)} + \sum_{\substack{k \\ k \neq m}} \frac{\sum_s R_{mks} \sqrt{\frac{\omega_s}{\hbar}} \xi_s \Psi_k^{(0)}}{(\mathrm{E}_m^{(0)} - \mathrm{E}_k^{(0)})} \right] \alpha_{\bar{V}} \exp - (i\mathrm{E}_{m\bar{V}} t)/\hbar \qquad \text{(A1-28)}$$

Here we neglect by changes of the frequencies and vibrational wave functions for the virtual states compared with the ground state. The expression for the coefficients $R_{mks}$ is the following

$$R_{mks} = \sum_{J\alpha} \sqrt{\frac{\hbar}{\omega_s}} \bar{X}_{Js\alpha} \frac{\partial \langle k | \hat{\mathrm{H}}_{e-n} | m \rangle}{\partial X_{J\alpha}} \qquad \text{(A1-29)}$$

# 25. APPENDIX 2

# THE TABLES OF IRREDUCIBLE REPRESENTATIONS OF SOME POINT GROUPS

Below one can find the tables of irreducible representations of some point groups. In the upper line of each table there are designations of the group and of the elements of symmetry. The other lines contain designations of irreducible representations, their characters and the dipole and quadrupole moments, transforming after corresponding irreducible representation.

Table 11. IRREDUCIBLE REPRESENTATIONS OF THE $C_2$, $C_s$ AND $C_i$ GROUPS

| $C_2$ | $C_1$ | $C_2$ | |
|---|---|---|---|
| $A$ | 1 | 1 | $d_z$, $Q_{xx}$, $Q_{yy}$, $Q_{zz}$, $Q_{xy}$ |
| $B$ | 1 | -1 | $d_x$, $d_y$, $Q_{xz}$, $Q_{yz}$ |
| $C_s$ | $C_1$ | $\sigma_h$ | |
| $A'$ | 1 | 1 | $d_x$, $d_y$, $Q_{xx}$, $Q_{yy}$, $Q_{zz}$, $Q_{xy}$ |
| $A''$ | 1 | -1 | $d_z$, $Q_{xz}$, $Q_{yz}$ |
| $C_i$ | $C_1$ | I | |
| $A_g$ | 1 | 1 | $Q_{xx}$, $Q_{yy}$, $Q_{zz}$, $Q_{xy}$, $Q_{xz}$, $Q_{yz}$ |
| $A_u$ | 1 | -1 | $d_x$, $d_y$, $d_z$ |

The main quadrupole moments in these groups are $Q_1 = Q_{xx}$, $Q_2 = Q_{yy}$ and $Q_3 = Q_{zz}$

Table 12. IRREDUCIBLE REPRESENTATIONS OF THE $C_{2v}$ AND $C_{2h}$ GROUPS.

| $C_{2v}$ | $C_1$ | $C_2(z)$ | $\sigma_v(xz)$ | $\sigma_v(yz)$ | |
|---|---|---|---|---|---|
| $A_1$ | 1 | 1 | 1 | 1 | $d_z, Q_{xx}, Q_{yy}, Q_{zz},$ |
| $A_2$ | 1 | 1 | -1 | -1 | $Q_{xy}$ |
| $B_1$ | 1 | -1 | 1 | -1 | $d_x, Q_{xz}$ |
| $B_2$ | 1 | -1 | -1 | 1 | $d_y, Q_{yz}$ |
| $C_{2h}$ | $C_1$ | $C_2(z)$ | $\sigma_h(xy)$ | I | |
| $A_g$ | 1 | 1 | 1 | 1 | $Q_{xx}, Q_{yy}, Q_{zz}, Q_{xy}$ |
| $A_u$ | 1 | 1 | -1 | -1 | $d_z,$ |
| $B_g$ | 1 | -1 | -1 | 1 | $Q_{xz}, Q_{yz}$ |
| $B_u$ | 1 | -1 | 1 | -1 | $d_x, d_y$ |

The main quadrupole moments in these groups are $Q_1 = Q_{xx}$, $Q_2 = Q_{yy}$ and $Q_3 = Q_{zz}$

Table 13. IRREDUCIBLE REPRESENTATIONS OF THE $D_{2h}$ GROUP

| $D_{2h}$ | $C_1$ | $\sigma(xy)$ | $\sigma(xz)$ | $\sigma(yz)$ | $I$ | $C_2(z)$ | $C_2(y)$ | $C_2(x)$ | |
|---|---|---|---|---|---|---|---|---|---|
| $A_g$ | 1 | 1 | 1 | 1 | 1 | 1 | 1 | 1 | $Q_{xx}, Q_{yy}, Q_{zz},$ |
| $A_u$ | 1 | -1 | -1 | -1 | -1 | 1 | 1 | 1 | |
| $B_{1g}$ | 1 | 1 | -1 | -1 | 1 | 1 | -1 | -1 | $Q_{xy}$ |
| $B_{1u}$ | 1 | -1 | 1 | 1 | -1 | 1 | -1 | -1 | $d_z$ |
| $B_{2g}$ | 1 | -1 | 1 | -1 | 1 | -1 | 1 | -1 | $Q_{xz}$ |
| $B_{2u}$ | 1 | 1 | -1 | 1 | -1 | -1 | 1 | -1 | $d_y$ |
| $B_{3g}$ | 1 | -1 | -1 | 1 | 1 | -1 | -1 | 1 | $Q_{yz}$ |
| $B_{3u}$ | 1 | 1 | 1 | -1 | -1 | -1 | -1 | 1 | $d_x$ |

The main quadrupole moments in this group are $Q_1 = Q_{xx}, Q_2 = Q_{yy}$ and $Q_3 = Q_{zz}$

Table 14. IRREDUCIBLE REPRESENTATIONS OF THE $C_{3v}$ GROUP

| $C_{3v}$ | $C_1$ | $2C_3(z)$ | $3\sigma_v$ | |
|---|---|---|---|---|
| $A_1$ | 1 | 1 | 1 | $d_z, Q_{xx}+Q_{yy}, Q_{zz}$ |
| $A_2$ | 1 | 1 | -1 | |
| $E$ | 2 | -1 | 0 | $(d_x, d_y), (Q_{xx}-Q_{yy}, Q_{xy}), (Q_{yz}, Q_{xz})$ |

The main quadrupole moments in this group are $Q_1 = Q_{xx} + Q_{yy}$, $Q_2 = Q_{zz}$

Table 15. IRREDUCIBLE REPRESENTATIONS OF THE $C_{3h}$ GROUP

| $C_{3h}$ | $C_1$ | $C_3$ | $\sigma_h$ | $S_3$ | |
|---|---|---|---|---|---|
| $A'$ | 1 | 1 | 1 | 1 | $Q_{xx}+Q_{yy}, Q_{zz}$ |
| $A''$ | 1 | 1 | -1 | -1 | $d_z$ |
| $E'$ | 2 | -1 | 2 | -1 | $(d_x, d_y), (Q_{xx}-Q_{yy}, Q_{xy})$ |
| $E''$ | 2 | -1 | -2 | 2 | $Q_{yz}, Q_{xz}$ |

The main quadrupole moments in this group are $Q_1 = Q_{xx} + Q_{yy}, Q_2 = Q_{zz}$

Table 16. IRREDUCIBLE REPRESENTATIONS OF THE $C_{6h}$ GROUP

| $C_{6h}$ | $C_1$ | $2C_6$ | $2C_3$ | $C_2''$ | $\sigma_h$ | $2S_6$ | $2S_3$ | I | |
|---|---|---|---|---|---|---|---|---|---|
| $A_g$ | 1 | 1 | 1 | 1 | 1 | 1 | 1 | 1 | $Q_{xx}+Q_{yy}, Q_{zz}$ |
| $A_u$ | 1 | 1 | 1 | 1 | -1 | -1 | -1 | -1 | $d_z$ |
| $B_g$ | 1 | -1 | 1 | -1 | -1 | 1 | -1 | 1 | |
| $B_u$ | 1 | -1 | 1 | -1 | 1 | -1 | 1 | -1 | |
| $E_{1g}$ | 2 | 1 | -1 | -2 | -2 | -1 | 1 | 2 | $(Q_{yz}, Q_{xz})$ |
| $E_{1u}$ | 2 | 1 | -1 | -2 | 2 | 1 | -1 | -2 | $(d_x, d_y)$ |
| $E_{2g}$ | 2 | -1 | -1 | 2 | 2 | -1 | -1 | 2 | $(Q_{xx}-Q_{yy}, Q_{xy})$ |
| $E_{2u}$ | 2 | -1 | -1 | 2 | -2 | 1 | 1 | -2 | |

The main quadrupole moments in this group are $Q_1 = Q_{xx} + Q_{yy}, Q_2 = Q_{zz}$

Table 17. IRREDUCIBLE REPRESENTATIONS OF THE $D_{2d}$ GROUP

| $D_{2d}$ | $C_1$ | $2S_4(z)$ | $C_2$ | $2C_2'$ | $2\sigma_d$ | |
|---|---|---|---|---|---|---|
| $A_1$ | 1 | 1 | 1 | 1 | 1 | $Q_{xx}+Q_{yy},Q_{zz}$ |
| $A_2$ | 1 | 1 | 1 | -1 | -1 | |
| $B_1$ | 1 | -1 | 1 | 1 | -1 | $Q_{xx}-Q_{yy}$ |
| $B_2$ | 1 | -1 | 1 | -1 | 1 | $d_z,Q_{xy}$ |
| $E$ | 2 | 0 | -2 | 0 | 0 | $(d_x,d_y),(Q_{xz},Q_{yz})$ |

The main quadrupole moments in this group are $Q_1=Q_{xx}+Q_{yy},Q_2=Q_{zz}$.

Table 18. IRREDUCIBLE REPRESENTATIONS OF THE $D_{3d}$ GROUP

| $D_{3d}$ | $C_1$ | $2S_6(z)$ | $2C_3$ | I | $3C_2$ | $3\sigma_d$ | |
|---|---|---|---|---|---|---|---|
| $A_{1g}$ | 1 | 1 | 1 | 1 | 1 | 1 | $Q_{xx}+Q_{yy},Q_{zz}$ |
| $A_{1u}$ | 1 | -1 | 1 | -1 | 1 | -1 | |
| $A_{2g}$ | 1 | 1 | 1 | 1 | -1 | -1 | |
| $A_{2u}$ | 1 | -1 | 1 | -1 | -1 | 1 | $d_z$ |
| $E_g$ | 2 | -1 | -1 | 2 | 0 | 0 | $(Q_{xx}-Q_{yy},Q_{xy}),(Q_{xz},Q_{yz})$ |
| $E_u$ | 2 | 1 | -1 | -2 | 0 | 0 | $(d_x,d_y)$ |

The main quadrupole moments in this group are $Q_1=Q_{xx}+Q_{yy},Q_2=Q_{zz}$.

Table 19. IRREDUCIBLE REPRESENTATIONS OF THE $D_{4d}$ GROUP

| $D_{4d}$ | $C_1$ | $2S_8(z)$ | $2S_8^2$ | $2S_8^3$ | $S_8^4$ | $4C_2$ | $4\sigma_d$ | |
|---|---|---|---|---|---|---|---|---|
| $A_1$ | 1 | 1 | 1 | 1 | 1 | 1 | 1 | $Q_{xx}+Q_{yy}, Q_{zz}$ |
| $A_2$ | 1 | 1 | 1 | 1 | 1 | -1 | -1 | |
| $B_1$ | 1 | -1 | 1 | -1 | 1 | 1 | -1 | |
| $B_2$ | 1 | -1 | 1 | -1 | 1 | -1 | 1 | $d_z$ |
| $E_1$ | 2 | $\sqrt{2}$ | 0 | $-\sqrt{2}$ | -2 | 0 | 0 | $(d_x, d_y)$ |
| $E_2$ | 2 | 0 | -2 | 0 | 2 | 0 | 0 | $(Q_{xx}-Q_{yy}, Q_{xy})$ |
| $E_3$ | 2 | $-\sqrt{2}$ | 0 | $\sqrt{2}$ | -2 | 0 | 0 | $(Q_{yz}, Q_{xz})$ |

The main quadrupole moments in this group are $Q_1 = Q_{xx} + Q_{yy}, Q_2 = Q_{zz}$.

Table 20. IRREDUCIBLE REPRESENTATIONS OF THE $D_{3h}$ GROUP

| $D_{3h}$ | $C_1$ | $2S_3$ | $2C_3(z)$ | $\sigma_h$ | $3C_2$ | $3\sigma_v$ | |
|---|---|---|---|---|---|---|---|
| $A_1^{'}$ | 1 | 1 | 1 | 1 | 1 | 1 | $Q_{xx}+Q_{yy}, Q_{zz}$ |
| $A_1^{''}$ | 1 | -1 | 1 | -1 | 1 | -1 | |
| $A_2^{'}$ | 1 | 1 | 1 | 1 | -1 | -1 | |
| $A_2^{''}$ | 1 | -1 | 1 | -1 | -1 | 1 | $d_z$ |
| $E^{'}$ | 2 | -1 | -1 | 2 | 0 | 0 | $(d_x, d_y), (Q_{xx}-Q_{yy}, Q_{xy})$ |
| $E^{''}$ | 2 | 1 | -1 | -2 | 0 | 0 | $(Q_{xz}, Q_{yz})$ |

The main quadrupole moments in this group are $Q_1 = Q_{xx} + Q_{yy}$, $Q_2 = Q_{zz}$.

## Table 21. IRREDUCIBLE REPRESENTATIONS OF THE $D_{4h}$ GROUP

| $D_{4h}$ | $C_1$ | $2C_4(z)$ | $C_2$ | $2C_2^{'}$ | $2C_2^{''}$ | $\sigma_h$ | $2\sigma_v$ | $2\sigma_d$ | $2S_4$ | I | |
|---|---|---|---|---|---|---|---|---|---|---|---|
| $A_{1g}$ | 1 | 1 | 1 | 1 | 1 | 1 | 1 | 1 | 1 | 1 | $Q_{xx}+Q_{yy}, Q_{zz}$ |
| $A_{1u}$ | 1 | 1 | 1 | 1 | 1 | -1 | -1 | -1 | -1 | -1 | |
| $A_{2g}$ | 1 | 1 | 1 | -1 | -1 | 1 | -1 | -1 | 1 | 1 | |
| $A_{2u}$ | 1 | 1 | 1 | -1 | -1 | -1 | 1 | 1 | -1 | -1 | $d_z$ |
| $B_{1g}$ | 1 | -1 | 1 | 1 | -1 | 1 | 1 | -1 | -1 | 1 | $Q_{xx}-Q_{yy},$ |
| $B_{1u}$ | 1 | -1 | 1 | 1 | -1 | -1 | -1 | 1 | 1 | -1 | |
| $B_{2g}$ | 1 | -1 | 1 | -1 | 1 | 1 | -1 | 1 | -1 | 1 | $Q_{xy}$ |
| $B_{2u}$ | 1 | -1 | 1 | -1 | 1 | -1 | 1 | -1 | 1 | -1 | |
| $E_g$ | 2 | 0 | -2 | 0 | 0 | -2 | 0 | 0 | 0 | 2 | $(Q_{xz}, Q_{yz})$ |
| $E_u$ | 2 | 0 | -2 | 0 | 0 | 2 | 0 | 0 | 0 | -2 | $(d_x, d_y)$ |

The main quadrupole moments in this group are $Q_1 = Q_{xx} + Q_{yy}, Q_2 = Q_{zz}$.

## Table 22. IRREDUCIBLE REPRESENTATIONS OF THE $D_{5h}$ GROUP

| $D_{5h}$ | $C_1$ | $2C_5$ | $2C_5^2$ | $5C_2$ | $\sigma_h$ | $2S_5$ | $2S_5^3$ | $5\sigma_v$ | |
|---|---|---|---|---|---|---|---|---|---|
| $A_1^{'}$ | 1 | 1 | 1 | 1 | 1 | 1 | 1 | 1 | $(Q_{xx}+Q_{yy}), Q_{zz}$ |
| $A_1^{''}$ | 1 | 1 | 1 | 1 | -1 | -1 | -1 | -1 | |
| $A_2^{'}$ | 1 | 1 | 1 | -1 | 1 | 1 | 1 | -1 | |
| $A_2^{''}$ | 1 | 1 | 1 | -1 | -1 | -1 | -1 | 1 | $d_z$ |
| $E_1^{'}$ | 2 | $2\cos 72^0$ | $2\cos 144^0$ | 0 | 2 | $2\cos 72^0$ | $2\cos 144^0$ | 0 | $(d_x, d_y)$ |
| $E_1^{''}$ | 2 | $2\cos 72^0$ | $2\cos 144^0$ | 0 | -2 | $-2\cos 72^0$ | $-2\cos 144^0$ | 0 | $(Q_{xz}, Q_{yz})$ |
| $E_2^{'}$ | 2 | $2\cos 144^0$ | $2\cos 72^0$ | 0 | 2 | $2\cos 144^0$ | $2\cos 72^0$ | 0 | $(Q_{xx}-Q_{yy}, Q_{xy})$ |
| $E_2^{''}$ | 2 | $2\cos 144^0$ | $2\cos 72^0$ | 0 | -2 | $-2\cos 144^0$ | $-2\cos 72^0$ | 0 | |

The main quadrupole moments in this group are $Q_1 = Q_{xx} + Q_{yy}, Q_2 = Q_{zz}$.

Table 23. IRREDUCIBLE REPRESENTATIONS OF THE $D_{6h}$ GROUP.

| $D_{6h}$ | $C_1$ | $2C_6(z)$ | $2C_3$ | $C_2$ | $3C_2^{'}$ | $3C_2^{''}$ | $I$ | $\sigma_h$ | $3\sigma_v$ | $3\sigma_d$ | $2S_6$ | $2S_3$ | |
|---|---|---|---|---|---|---|---|---|---|---|---|---|---|
| $A_{1g}$ | 1 | 1 | 1 | 1 | 1 | 1 | 1 | 1 | 1 | 1 | 1 | 1 | $(Q_{xx}+Q_{yy}), Q_{zz}$ |
| $A_{1u}$ | 1 | 1 | 1 | 1 | 1 | 1 | -1 | -1 | -1 | -1 | -1 | -1 | |
| $A_{2g}$ | 1 | 1 | 1 | 1 | -1 | -1 | 1 | 1 | -1 | -1 | 1 | 1 | |
| $A_{2u}$ | 1 | 1 | 1 | 1 | -1 | -1 | -1 | -1 | 1 | 1 | -1 | -1 | $d_z$ |
| $B_{1g}$ | 1 | -1 | 1 | -1 | 1 | 1 | 1 | -1 | -1 | 1 | 1 | -1 | |
| $B_{1u}$ | 1 | -1 | 1 | -1 | 1 | 1 | -1 | 1 | 1 | -1 | -1 | 1 | |
| $B_{2g}$ | 1 | -1 | 1 | -1 | -1 | -1 | 1 | -1 | 1 | -1 | 1 | -1 | |
| $B_{2u}$ | 1 | -1 | 1 | -1 | -1 | -1 | -1 | 1 | -1 | 1 | -1 | 1 | |
| $E_{1g}$ | 2 | 1 | -1 | -2 | 0 | 0 | 2 | -2 | 0 | 0 | -1 | 1 | $(Q_{xz}, Q_{yz})$ |
| $E_{1u}$ | 2 | 1 | -1 | -2 | 0 | 0 | -2 | 2 | 0 | 0 | 1 | -1 | $(d_x, d_y)$ |
| $E_{2g}$ | 2 | -1 | -1 | 2 | 0 | 0 | 2 | 2 | 0 | 0 | -1 | -1 | $(Q_{xx}-Q_{yy}, Q_{xy})$ |
| $E_{2u}$ | 2 | -1 | -1 | 2 | 0 | 0 | -2 | -2 | 0 | 0 | 1 | 1 | |

The main quadrupole moments in this group are $Q_1 = Q_{xx} + Q_{yy}, Q_2 = Q_{zz}$.

Table 24. IRREDUCIBLE REPRESENTATIONS OF THE $T_d$ GROUP

| $T_d$ | $C_1$ | $8C_3$ | $3C_2$ | $6S_4$ | $6\sigma_d$ | |
|---|---|---|---|---|---|---|
| $A_1$ | 1 | 1 | 1 | 1 | 1 | $Q_{xx}+Q_{yy}+Q_{zz}$ |
| $A_2$ | 1 | 1 | 1 | -1 | -1 | |
| $E$ | 2 | -1 | 2 | 0 | 0 | $(Q_{xx}+Q_{yy}-2Q_{zz}, Q_{xx}-Q_{yy})$ |
| $F_1$ | 3 | 0 | -1 | 1 | -1 | |
| $F_2$ | 3 | 0 | -1 | -1 | 1 | $(d_x, d_y, d_z), (Q_{xy}, Q_{xz}, Q_{yz})$ |

The main quadrupole moment in this group is $Q_1 = Q_{xx} + Q_{yy} + Q_{zz}$.

Table 25. IRREDUCIBLE REPRESENTATIONS OF THE $O_h$ GROUP

| $O_h$ | $C_1$ | $8C_3$ | $3C_2$ | $6C_4$ | $6C_2$ | $I$ | $6S_4$ | $8S_6$ | $3\sigma_h$ | $6\sigma_d$ | |
|---|---|---|---|---|---|---|---|---|---|---|---|
| $A_{1g}$ | 1 | 1 | 1 | 1 | 1 | 1 | 1 | 1 | 1 | 1 | $Q_{xx}+Q_{yy}+Q_{zz}$ |
| $A_{1u}$ | 1 | 1 | 1 | 1 | 1 | -1 | -1 | -1 | -1 | -1 | |
| $A_{2g}$ | 1 | 1 | 1 | -1 | -1 | 1 | -1 | 1 | 1 | -1 | |
| $A_{2u}$ | 1 | 1 | 1 | -1 | -1 | -1 | 1 | -1 | -1 | 1 | |
| $E_g$ | 2 | -1 | 2 | 0 | 0 | 2 | 0 | -1 | 2 | 0 | $(Q_{xx}+Q_{yy}-2Q_{zz}, Q_{xx}-Q_{yy})$ |
| $E_u$ | 2 | -1 | 2 | 0 | 0 | -2 | 0 | 1 | -2 | 0 | |
| $F_{1g}$ | 3 | 0 | -1 | 1 | -1 | 3 | 1 | 0 | -1 | -1 | |
| $F_{1u}$ | 3 | 0 | -1 | 1 | -1 | -3 | -1 | 0 | 1 | 1 | $(d_x, d_y, d_z)$ |
| $F_{2g}$ | 3 | 0 | -1 | -1 | 1 | 3 | -1 | 0 | -1 | 1 | $(Q_{xy}, Q_{xz}, Q_{yz})$ |
| $F_{2u}$ | 3 | 0 | -1 | -1 | 1 | -3 | 1 | 0 | 1 | -1 | |

The main quadrupole moment in this group is $Q_1 = Q_{xx} + Q_{yy} + Q_{zz}$.